\documentclass[aps,prc,twocolumn,longbibliography,reprint,superscriptaddress,amsmath,nofootinbib]{revtex4-1}
\usepackage[colorlinks=true, urlcolor=blue,linkcolor=blue, anchorcolor=blue, citecolor=red]{hyperref}
\usepackage[utf8]{inputenc}
\usepackage{natbib}
\usepackage{graphicx}
\usepackage{xcolor}
\usepackage{amsmath}
\usepackage{bm}

\graphicspath{{fig/}}

\newcommand{\taur}{\tau_R}
\newcommand{\btau}{\bar{\tau}}
\newcommand{\PL}{\mathcal{P}_L}
\newcommand{\PT}{\mathcal{P}_T}
\newcommand{\PLbar}{\overline{\mathcal{P}}_L}
\newcommand{\PTbar}{\overline{\mathcal{P}}_T}

\newcommand{\bpi}{\widetilde{\pi}}
\newcommand{\bPi}{\widetilde{\Pi}}

\begin{document}
\preprint{}

\title{On non-conformal kinetic theory and hydrodynamics for Bjorken flow}

\author{Sunil Jaiswal}
\email{sunil.jaiswal@tifr.res.in}
\affiliation{Department of Nuclear and Atomic Physics, Tata Institute of Fundamental Research, Mumbai 400005, India}
\author{Chandrodoy Chattopadhyay}
\email{chattopadhyay.31@osu.edu}
\affiliation{Department of Physics, The Ohio State University, Columbus, Ohio 43210-1117, USA}
\author{Lipei Du}
\email{du.458@osu.edu}
\affiliation{Department of Physics, The Ohio State University, Columbus, Ohio 43210-1117, USA}
\author{Ulrich Heinz}
\email{heinz.9@osu.edu}
\affiliation{Department of Physics, The Ohio State University, Columbus, Ohio 43210-1117, USA}
\author{Subrata Pal}
\email{spal@tifr.res.in}
\affiliation{Department of Nuclear and Atomic Physics, Tata Institute of Fundamental Research, Mumbai 400005, India}
\date{\today}

\begin{abstract}
Using and comparing kinetic theory and second-order Chapman-Enskog hydrodynamics, we study the non-conformal dynamics of a system undergoing Bjorken expansion. We use the concept of `free-streaming fixed lines' for scaled shear and bulk stresses in non-conformal kinetic theory and hydrodynamics, and show that these `fixed lines' behave as early-time attractors and repellors of the evolution. In the conformal limit, the free-streaming fixed lines reduce to the well-known fixed points of conformal Bjorken dynamics. A new fixed point in the free streaming regime is identified which lies at the intersection of these fixed lines. Contrary to the conformal scenario, both kinetic theory and hydrodynamics predict the absence of attractor behavior in the normalised shear stress channel. In kinetic theory a far-off-equilibrium attractor is found for the normalised effective longitudinal pressure, driven by rapid longitudinal expansion. Second-order viscous hydrodynamics fails to accurately describe this attractor. From a thorough analysis of the free-streaming dynamics in Chapman-Enskog hydrodynamics we conclude that this failure results from an inaccurate approximation of the fixed lines and a related incorrect description of the nature of the fixed point. A modified anisotropic hydrodynamic description is presented that provides excellent agreement with kinetic theory results and reproduces the far-from-equilibrium attractor for the scaled longitudinal pressure.
\end{abstract}
\maketitle
%
\section{Introduction}
\vspace*{-2mm}

Causal relativistic dissipative hydrodynamics has been surprisingly successful in describing final-state observables in ultra-relativistic collisions not only among heavy nuclei but also between light nuclei \cite{Romatschke:2007mq, Song:2007ux, Song:2008si, Schenke:2010nt, Heinz:2013th} where the medium is generally expected to be very far from local thermal equilibrium. This has led to a resurgence of interest in understanding the domain of applicability of modern formulations of fluid dynamics \cite{Heller:2011ju, Heller:2013fn,  Heller:2015dha, Kurkela:2015qoa, Blaizot:2017lht, Romatschke:2017vte, Spalinski:2017mel,  Strickland:2017kux, Romatschke:2017acs, Behtash:2017wqg, Blaizot:2017ucy, Denicol:2018pak, Kurkela:2018wud, Mazeliauskas:2018yef, Behtash:2019txb, Heinz:2019dbd, Blaizot:2019scw, Behtash:2020vqk, Denicol:2020eij, Blaizot:2020gql} (see also the reviews \cite{Florkowski:2017olj, Romatschke:2017ejr, Berges:2020fwq}). It is well known that traditional hydrodynamics, formulated as an order-by-order expansion of the energy-momentum tensor and conserved charges in gradients of temperature, chemical potentials and fluid flow velocity, ceases to be a valid description once the system is far from local equilibrium. In fact, these traditional approaches, whose simplest example is the first-order relativistic Navier-Stokes theory \cite{Eckart:1940zz, Landau:1987}, are plagued by acausality: they give rise to superluminal propagation of high-momentum (ultra-violet) modes of the medium, which in turn is closely related to numerical instability of the resulting partial differential equations of such hydrodynamic formulations \cite{Hiscock:1983zz, Hiscock:1985zz}. 

One way to circumvent this problem is to promote the dissipative fluxes to independent dynamical degrees of freedom of the system whose evolution is governed by relaxation-type equations. This approach is taken in the second-order theories of M\"uller, Israel and Stewart (MIS) \cite{Muller:1967zza, Israel:1976tn, Israel:1979wp} where each dissipative quantity relaxes to its Navier-Stokes limit on a microscopic time scale controlled by its corresponding relaxation time. The latter plays the role of ultraviolet regulators for the high-momentum modes and restore causality and stability of the formalism. Somewhat unexpectedly, however, they are found to also enlarge its domain of applicability.

A quantitative way of testing the domain of applicability of an effective theory is to compare its predictions with an underlying theory that is known to govern the microscopic dynamics of the system. It is commonly assumed that the medium formed in ultra-relativistic heavy-ion collisions admits a quasi-particle description in terms of a distribution function whose evolution is described by Boltzmann-like equations \cite{Bass:1998ca, Arnold:2000dr, Arnold:2003zc, Lin:2004en, Xu:2004mz, Xu:2007aa, El:2007vg}. Using kinetic theory as the underlying theory of matter and integrating out the particle degrees of freedom gives rise to an infinite hierarchy of equations for different moments of the distribution function. Some of these moments (e.g. the energy-momentum tensor $T^{\mu\nu}$ and conserved (net-baryon) charge current $N^\mu$) evolve according to hydrodynamic conservation laws while the remaining higher-order moments are `non-hydrodynamic' in nature. Constructing hydrodynamics from kinetic theory amounts to postulating a truncation scheme for an infinite hierarchy of coupled moment equations. Several such truncation schemes have been proposed which led to the development of second- and third-order hydrodynamics \cite{Muronga:2001zk, Baier:2007ix, Bhattacharyya:2007vjd, Denicol:2012cn, Jaiswal:2013npa, Jaiswal:2013vta, Florkowski:2015lra, Grozdanov:2015kqa}, as well as the recently developed anisotropic hydrodynamics \cite{Florkowski:2010cf, Martinez:2010sc, Martinez:2012tu, Bazow:2013ifa, Strickland:2014pga, Florkowski:2014bba, Molnar:2016vvu, Molnar:2016gwq, Alqahtani:2017mhy, McNelis:2018jho, Nopoush:2019vqc}.\footnote{%
    The latter is particularly well suited to describe the large anisotropies in momentum fluxes that are characteristic of the rapid, dominantly longitudinal early-time expansion of the quark-gluon plasma created in relativistic heavy-ion collisions.
} 
All of these causal relaxation-type theories are similar in spirit to the original M\"uller-Israel-Stewart equations, the essential difference being the transport coefficients reflecting the microscopic dynamical properties of the system in each case.     
Several studies comparing results of the above-mentioned hydrodynamic theories to conformal kinetic theories in flow profiles amenable to analytic treatment (such as Bjorken \cite{Bjorken:1982qr} and Gubser flows \cite{Gubser:2010ze}), have revealed a surprising success of hydrodynamics in providing a near-accurate description of the system's macroscopic dynamics even at very early times when the medium is very far from local equilibrium \cite{Denicol:2014xca, Denicol:2014tha, Behtash:2017wqg, Martinez:2017ibh, Strickland:2017kux, Romatschke:2017vte, Romatschke:2017ejr, Behtash:2018moe, Behtash:2019qtk}. A few studies have extended numerically this line of inquiry to non-conformal systems and/or three-dimensionally expanding systems \cite{Romatschke:2017acs, Kurkela:2019set, Dore:2020jye, Ambrus:2021sjg, Du:2021fok}, with mixed results some of which seem to lend further support to this general conclusion while others raise doubts. A key feature that emerged from these studies is that hydrodynamics is governed by a far-from-equilibrium attractor \cite{Heller:2016rtz, Romatschke:2017vte, Florkowski:2017olj, Blaizot:2017ucy, Heller:2018qvh, Dash:2020zqx} to which different initializations of normalised dissipative quantities decay either exponentially at low Knudsen numbers over a time-scale controlled by the relaxation-time, or even faster via power law at high Knudsen-numbers \cite{Jaiswal:2019cju, Kurkela:2019set}. In fact, causal hydrodynamic theories undergoing Bjorken expansion not only exhibit an attractor which is in excellent agreement with the kinetic theory attractor for corresponding dissipative quantities, but it also accurately describes the decay of different initial conditions to the attractor \cite{Chattopadhyay:2018apf}. Although hydrodynamics provide no information on the evolution of higher-order moments of the distribution function, which in Boltzmann kinetic theory are also governed by attractors \cite{Strickland:2018ayk, Strickland:2019hff, Almaalol:2020rnu}, it may thus seem reasonable to treat hydrodynamics as a substitute for kinetic theory as long as we are interested in describing only the evolution of the low-order hydrodynamic moments, i.e. the energy-momentum tensor and conserved charge currents. 

However, almost all of these comparisons of hydrodynamics with kinetic theory have focused on conformal systems with vanishing bulk viscous pressure. A recent comparison of hydrodynamics with boost-invariant kinetic theory for non-conformal systems \cite{Chattopadhyay:2021ive} has shown the former to yield a much less accurate description of the latter than was previously observed for conformal systems. Not only does second-order non-conformal hydrodynamics \cite{Denicol:2014vaa, Jaiswal:2014isa} fail to provide an accurate description of massive kinetic theory in the regime of large Knudsen number, but it is also found \cite{Chattopadhyay:2021ive} to be unable to reproduce the attractor that characterizes the underlying microscopic theory \cite{Romatschke:2017acs, Florkowski:2017jnz}. This seems to contradict the earlier conclusion from conformal studies that hydrodynamics is applicable even far away from equilibrium. To shed light on this puzzle we here present an in-depth investigation of the domain of applicability of second-order {\it non-conformal} hydrodynamics by comparing it with kinetic theory of systems of massive particles undergoing longitudinally boost-invariant medium expansion. A more general study of three-dimensionally expanding systems is left for the future. 

The manuscript is organized as follows: In Sec. \ref{sec:kt} we solve the Boltzmann equation in relaxation-time approximation and obtain the evolution of macroscopic quantities for Bjorken expansion. By studying the early-time dynamics we identify attracting and repelling fixed lines in the space of scaled shear and bulk stresses. In Sec. \ref{sec:hydro} we solve second-order hydrodynamics for Bjorken flow and perform a linearised analysis to extract the behavior of the solutions around the free-streaming fixed points. Comparison between results obtained in hydrodynamics and kinetic theory are presented in Sec. \ref{sec:hydrovskt}. An anisotropic hydrodynamic description is presented in Sec. \ref{aHydro} which is in close agreement with kinetic theory even in far-from-equilibrium regimes. Our conclusions are summarized in Sec. \ref{sec:conclusion}.

\vspace*{-2mm}
\section{Kinetic theory} 
\label{sec:kt}
\vspace*{-2mm}

We consider the Boltzmann equation describing the evolution of the single particle phase space distribution function $f(x,p)$ of a gas with particles of constant mass $m$. For simplicity we choose the relaxation time approximation for the collisional kernel \cite{anderson1974relativistic},
\begin{equation}
\label{BE_gen}
    p^{\mu} \partial_\mu f + \Gamma^{\lambda}_{\mu i} p_\lambda p^\mu \frac{\partial f}{\partial p_i}  = - \frac{u\cdot p}{\tau_{R}}\left(f - f_\mathrm{eq}\right),
\end{equation}
where the relaxation time $\tau_R$ is allowed to depend on position $x$ but not on momentum $p$. The equilibrium distribution function for particles with vanishing chemical potential obeying Boltzmann statistics is $f_{\mathrm{eq}} = \exp(-(u \cdot p)/T)$. We use the standard notation $A \cdot B \equiv A^\mu B_\mu$ to denote scalar products of four-vectors. The Christoffel symbols in (\ref{BE_gen}) are given by derivatives of the metric $g^{\mu\nu}$,
\begin{equation}
    \Gamma^{\mu}_{\alpha\beta} = \frac{g^{\mu \lambda}}{2} \left( \partial_\alpha g_{\beta \lambda} + \partial_\beta g_{\alpha \lambda} - \partial_\lambda g_{\alpha \beta} \right). 
\end{equation}
The particles' four-momenta satisfy the on-shell condition $p \cdot p = m^2$. The appearance of macroscopic variables, i.e. the time-like flow velocity $u^\mu(x)$ and effective temperature $T(x)$, in the collisional kernel essentially makes the RTA Boltzmann equation a hybrid model of describing the microscopic dynamics of a weakly coupled gas. The flow velocity and temperature are defined using the so-called Landau matching condition which is a re-statement of the condition that the RTA collisional kernel satisfies energy-momentum conservation. The energy-momentum tensor $T^{\mu\nu}(x)$ is given by
\begin{equation}\label{Tmunu}
    T^{\mu\nu} = \int dP \, p^\mu \, p^\nu \, f(x,p) \equiv \langle p^\mu p^\nu \rangle, 
\end{equation}
where we use the compact notation $\langle {\cal O}(x,p) \rangle \equiv \int dP \, {\cal O}(x,p) \, f (x,p)$, ${\cal O}(x,p)$ being a generic tensor of particle momenta and space-time. The integration measure is defined as $dP \equiv d^3p/[(2\pi)^3 \sqrt{-g} p^0]$. Using Eqs.~(\ref{BE_gen}) and (\ref{Tmunu}) one finds
\begin{equation}\label{timelike}
    d_\mu T^{\mu\nu} = 0 \quad\implies\quad T^{\mu\nu} u_\nu = \epsilon_\mathrm{eq}(T) \, u^\mu,
\end{equation}
where the functional dependence of the equilibrium energy density on the temperature is
\begin{align} \label{energy}
    \epsilon_\mathrm{eq}(T) & = \int dP \,(u \cdot p)^2 \, f_\mathrm{eq} \equiv \left\langle (u \cdot p)^2 \right\rangle_\mathrm{eq} 
\nonumber \\ 
    & = \frac{3 \, T^4}{\pi^2}\,\left( \frac{z^2}{2} K_2(z)+ \frac{z^3}{6} K_1(z) \right),
\end{align}
with $z=m/T$. Here $K_{n}$ are the modified Bessel functions of the second kind of order $n$. In the limit $z\to 0$ the term in parentheses reduces to 1 and Eq.~(\ref{energy}) reproduces the conformal result $\epsilon_\mathrm{eq}\propto T^4$. Eq.~(\ref{timelike}) defines the effective temperature and flow velocity, with $u^\mu$ and $\epsilon_\mathrm{eq}(T)$ being the time-like eigenvector and eigenvalue of $T^{\mu}_{\nu}$, respectively.

\vspace*{-2mm}
\subsection{Bjorken flow}
\label{sec2a}
\vspace*{-2mm}

In this work we consider a fluid undergoing Bjorken expansion \cite{Bjorken:1982qr}. This is an appropriate description of early-time dynamics of matter formed in ultra-relativistic heavy-ion collisions. Bjorken symmetries enforce homogeneity in the transverse $(x,y)$ plane, boost invariance along the $z$ (longitudinal or beam) direction, and reflection symmetry $z \to -z$. This implies that the flow profile is $v^x =v^y = 0$ and $v^z = z/t$. The symmetries are manifest in Milne coordinate system with proper time $\tau = \sqrt{t^2 - z^2}$ and space-time rapidity $\eta_s = \tanh^{-1}(z/t)$. The Milne metric is given by $g_{\mu\nu} = \mathrm{diag}(1,-1,-1,-\tau^2)$, and the non-vanishing Christoffels are $\Gamma^{\tau}_{\eta\eta} = 1/\tau$, $\Gamma^{\eta}_{\tau\eta} = \Gamma^{\eta}_{\eta\tau} = \tau$. In these coordinates the fluid appears to be static, $u^\mu = (1,0,0,0)$, and all macroscopic quantities depend only on proper time. 

The symmetries of Bjorken expansion significantly constrain the space-time and momentum dependencies of the distribution function $f(x,p)$. Boost invariance along the beam direction implies that $f(x,p)$ can depend only on the longitudinally boost invariant variables $\tau$ and $w \equiv p_\eta = t p^z - z E_p$. Homogeneity and rotational invariance in the transverse plane forbids any dependence on $(x, y)$ and $\phi_p = \tan^{-1}(p^y/p^x)$. Thus, the distribution function depends only on 3 variables: $f(x,p) = f(\tau, p_T, w)$, where $p_T = \sqrt{(p^x)^2 + (p^y)^2}$ \cite{Baym:1984np, Florkowski:2013lya}. The momentum-space measure in terms of boost-invariant variables is $dP = (p_T dp_T d\phi_p dw)/\left[(2\pi)^3 \tau p^\tau \right]$, with $p^\tau = \sqrt{(p_T)^2 + w^2/\tau^2 + m^2}$.

Using the projection operators $u^\mu u^\nu$ and $\Delta^{\mu\nu}=g^{\mu\nu}-u^\mu u^\nu$ parallel and perpendicular to the time direction in the local rest frame, the energy-momentum tensor can be decomposed as follows:
\begin{equation}
    T^{\mu\nu} = \langle p^\mu p^\nu \rangle = \epsilon u^\mu u^\nu - (P + \Pi) \Delta^{\mu\nu} + \pi^{\mu\nu}.
\end{equation}
Here $\epsilon = \left\langle (p^\tau)^2 \right\rangle$ is the energy density in the local rest frame, $P = \frac{1}{3} \left\langle (p_T^2 + (w/\tau)^2 ) \right\rangle_\mathrm{eq}$ is the equilibrium pressure, $\Pi=\frac{1}{3} \left\langle (p_T^2 + (w/\tau)^2) \right\rangle - P$ is the bulk viscous pressure, and the remainder $\pi^{\mu\nu}$ (which is traceless and transverse to $u^\mu$) is the shear stress tensor. Bjorken symmetry considerably simplifies the structure of the latter: instead of a tensor with five independent components in general, it reduces to a diagonal matrix with a single independent component $\pi \equiv \pi^{\eta}_{\eta} = \frac{2}{3\tau^2} \bigl\langle (\frac{w^2}{\tau^2} - p_T^2/2) \bigr\rangle$: $\pi^{\mu\nu} = \mathrm{diag}(0, \pi/2, \pi/2, -\pi/\tau^2)$. As a result the energy-momentum tensor becomes diagonal in Milne coordinates, $T^{\mu\nu} = \mathrm{diag}(\epsilon, \PT, \PT, \PL)$, with effective transverse pressure $\PT \equiv (1/2)\left\langle p_T^2 \right\rangle$ and effective longitudinal pressure $\PL \equiv \left\langle (w/\tau)^2 \right\rangle$. Note that for the central Bjorken cell ($z = 0$), $w/\tau$ is equal to the longitudinal momentum $p^z$. 

Expressed in terms of boost-invariant variables $(\tau, p_T, w)$, the Boltzmann equation (\ref{BE_gen}) takes the simple form
\begin{equation}
\label{BEMilne}
    \frac{\partial f}{\partial \tau} = - \frac{f - f_\mathrm{eq}}{\tau_R},
\end{equation}
where $f_\mathrm{eq} = \exp\bigl(-\sqrt{p_T^2 + w^2/\tau^2 + m^2}/T\bigr)$. The distribution function admits an analytic solution \cite{Florkowski:2013lya, Florkowski:2014sfa}:
\begin{align}
\label{soln}
    f(\tau; p_T, w) &= D(\tau,\tau_0) f_\mathrm{in}(\tau_0; p_T,w)
\nonumber \\ 
    &+ \int_{\tau_0}^{\tau} \, \frac{d\tau'}{\tau_R(\tau')} \, D(\tau,\tau') \, f_\mathrm{eq} (\tau', p_T, w).
\end{align}
Here $f_\mathrm{in}$ is the initial distribution function at proper time $\tau_0$. In this work we take the initial distribution to be given in generalized Romatschke-Strickland form \cite{Romatschke:2003ms}:
\begin{equation}
\label{f_in}
    f_\mathrm{in}(\tau_0; p_T, w) = \frac{1}{\alpha_0} \exp\left(- \frac{\sqrt{p_T^2  + (1{+}\xi_0) w^2/\tau_0^2 + m^2 }}{\Lambda_0} \right).
\end{equation}
It has 3 parameters ($\alpha_0, \xi_0, \Lambda_0$), corresponding to the 3 independent components of $T^{\mu\nu}$ ($\epsilon$, $\PT$, $\PL$). The initial anisotropy in momentum space is parametrised by $\xi_0$, the typical momentum scale is set by $\Lambda_0$, and $\alpha_0$ ensures that all initial conditions have the same initial energy density or, equivalently, the same effective temperature. The damping function
\begin{equation}
    D(\tau_2,\tau_1) = \exp\left(- \int_{\tau_1}^{\tau_2} \frac{d\tau'}{\tau_R(\tau')} \right)
\end{equation}
depends on the scattering rate $1/\tau_{R}$ and controls the rate at which the distribution function loses memory of its initial form. In Milne coordinates Bjorken flow is static; therefore, to know the distribution function at any proper time one only needs to determine the time evolution of the temperature. As mentioned above, this is done via the Landau matching condition, $\epsilon = \left\langle (p^\tau)^2 \right\rangle = \epsilon_\mathrm{eq}(T)$. Using the formal solution (\ref{soln}) for $f$ this yields effectively an integral equation for the temperature \cite{Florkowski:2013lya, Florkowski:2014sfa}:
\begin{align}
\label{e_sol}
    \epsilon_\mathrm{eq}(T) &= D(\tau,\tau_0) \frac{\Lambda_0^4}{4\pi^2 \alpha_0} \tilde{H}_{\epsilon} \left[\frac{\tau_0}{\tau \sqrt{1+\xi_0}}, \frac{m}{\Lambda_0} \right] 
\\\nonumber  
    & + \frac{1}{4\pi^2} \int_{\tau_0}^{\tau}         \frac{d\tau'}{\tau_R(\tau')} D(\tau,\tau') T^4(\tau') \tilde{H}_{\epsilon} \left[\frac{\tau'}{\tau},\frac{m}{T(\tau')}\right],
\end{align}
where
\begin{equation}
    \!\!\!\!
    \tilde{H}_{\epsilon}(y,z) \equiv \int_{0}^{\infty}\!\!du \, u^3 \, \exp\left(-\sqrt{u^2 + z^2}\right)\, H_{\epsilon}\left[y,\frac{z}{u} \right] ,
\end{equation}
with \cite{Florkowski:2014sfa}
\begin{equation}
\nonumber
    H_{\epsilon}(y,z) = y \left( \sqrt{y^2{+}z^2} + \frac{1{+}z^2}{\sqrt{y^2{-}1}} \tanh^{-1} \sqrt{\frac{y^2{-}1}{y^2{+}z^2}} \right).
\end{equation}
Eq.~(\ref{e_sol}) is solved for $T(\tau)$ by numerical iteration.

Before proceeding we point out an important difference between conformal and non-conformal systems in the role played by the initial distribution function for the temperature evolution. The first term on the r.h.s. of Eq.~(\ref{e_sol}) shows that, for fixed $\xi_0$, different choices of $(\alpha_0, \Lambda_0)$ that yield identical initial energy densities, $\epsilon(\tau_0)$, give rise to different energy density evolutions. In other words, for given momentum anisotropy and initial energy density, the evolution of $\epsilon(\tau)$ is sensitive to the momentum scale $\Lambda_0$ that characterizes the initial distribution function. In the conformal $(m{\,=\,}0)$ limit, on the other hand, the second argument of the function $\tilde{H}_{\epsilon}$ drops out,
\begin{align}
\label{e_sol_conf}
    \epsilon_\mathrm{eq}^\mathrm{conf}(T) &= D(\tau,\tau_0) \frac{\Lambda_0^4}{4\pi^2 \alpha_0} \tilde{H}_{\epsilon} \left[\frac{\tau_0}{\tau \sqrt{1+\xi_0}}, 0 \right] 
\\\nonumber 
    & + \frac{1}{4\pi^2} \int_{\tau_0}^{\tau} \frac{d\tau'}{\tau_R(\tau')} D(\tau,\tau') T^4(\tau') \tilde{H}_{\epsilon} \left[\frac{\tau'}{\tau},0 \right],
\end{align}
and, for fixed initial momentum anisotropy $\xi_0$ and energy density $\epsilon_0 \propto \Lambda_0^4/\alpha_0$, the evolution of $\epsilon_\mathrm{eq}^\mathrm{conf}(T) = 3T^4/\pi^2$ is insensitive to the initial momentum scale $\Lambda_0$. In fact, all initial distribution functions of the form $f_\mathrm{in}(\tau_0; p_T,w) = \psi\left[(p_T^2 + (1+\xi_0) w^2/\tau_0^2)/\Lambda_0^2\right]$ that yield identical initial energy densities then give rise to identical $\epsilon(\tau)$. This is because, for $f_\mathrm{in} = \psi$, a change of the functional dependence of $\psi$ on its argument simply results in a rescaling of the function $\tilde{H}_{\epsilon} \left[\frac{\tau_0}{\tau \sqrt{1+\xi_0}}, 0 \right]$ in the first term on the r.h.s. of Eq.~(\ref{e_sol_conf}) by the dimensionless factor
\begin{equation}
    {\cal C} = \frac{1}{4\pi^2} \int_{0}^{\infty} dx \, x^3 \, \psi(x)
\end{equation}
which can be absorbed into the initial energy density.

After obtaining the solution $T(\tau)$ of Eq. (\ref{e_sol}) we use it to calculate the effective transverse and longitudinal pressures:
\begin{align}
\label{PT}
    \PT(\tau) =&  D(\tau,\tau_0) \frac{\Lambda_0^4}{8 \pi^2 \alpha_0} \tilde{H}_{T} \left[\frac{\tau_0}{\tau \sqrt{1+\xi_0}}, \frac{m}{\Lambda_0} \right] 
\\\nonumber 
    & + \frac{1}{8\pi^2}  \int_{\tau_0}^{\tau} \frac{d\tau'}{\tau_R(\tau')} D(\tau,\tau') T^4(\tau') \tilde{H}_{T} \left[\frac{\tau'}{\tau},\frac{m}{T(\tau')}\right], 
\\
\label{PL}
    \PL(\tau) =&  D(\tau,\tau_0) \frac{\Lambda_0^4}{4 \pi^2 \alpha_0} \tilde{H}_{L} \left[\frac{\tau_0}{\tau \sqrt{1+\xi_0}}, \frac{m}{\Lambda_0} \right] 
\\\nonumber  
    & + \frac{1}{4\pi^2}  \int_{\tau_0}^{\tau} \frac{d\tau'}{\tau_R(\tau')} D(\tau,\tau') T^4(\tau') \tilde{H}_{L} \left[\frac{\tau'}{\tau},\frac{m}{T(\tau')}\right]. 
\end{align}
Here the functions $\tilde{H}_{T,L}$ are defined by \cite{Florkowski:2014sfa}
\begin{align}
\nonumber
    \tilde{H}_{T,L}(y,z) &\equiv \int_{0}^{\infty} du \,  u^3  \exp(-\sqrt{u^2 + z^2})\, H_{T,L}\left(y,\frac{z}{u}\right), 
\end{align}
with
\begin{align}
    H_{T}(y,z) =&  \frac{y}{(y^2-1)^{3/2}} \bigg[ - \sqrt{(y^2-1)(y^2+z^2)} 
\nonumber \\    
    & + \left( z^2 + 2y^2 - 1 \right) \tanh^{-1}\sqrt{\frac{y^2-1}{y^2+z^2}} \bigg] ,
\end{align}
\begin{align}
    H_{L}(y,z) =& \frac{y^3}{(y^2-1)^{3/2}} \bigg[ \sqrt{(y^2-1)(y^2+z^2)}  
\nonumber \\
    & -  \left( z^2 + 1 \right) \tanh^{-1}\sqrt{\frac{y^2-1}{y^2+z^2}} \bigg].
\end{align}
From Eqs.~(\ref{PT},\ref{PL}) it is straightforward to obtain the bulk and shear viscous stresses as $\Pi = \frac{1}{3}(\PL + 2 \PT - 3 P)$ and $\pi = \frac{2}{3}(\PT{-}\PL)$.

In \cite{Chattopadhyay:2021ive} it is shown that in kinetic theory for weakly interacting particles the bulk and shear stresses obey certain bounds. From the positivity of the distribution function, $f(x,p) \geq 0$, the positivity of $\PT = \langle p_T^2 \rangle/2\geq 0$ and $\PL=\langle p_z^2 \rangle\geq 0$ follows immediately. So does the positivity of the total isotropic pressure $P + \Pi = \langle |\vec{p}|^2 \rangle/3 =\langle p_T^2 + w^2/\tau^2 \rangle/3 \geq 0$. Moreover, the trace of the energy momentum tensor is positive: $T_\mu^\mu = \epsilon - 3(P+\Pi) =  m^2 \langle 1 \rangle \geq 0$. These inequalities results in the following bounds on $\Pi/P$ and $\pi/P$:
\begin{align}
\label{pi_Pi_bounds}
    & \frac{\Pi}{P} + \frac{\pi}{2P} \geq -1 , \quad \frac{\Pi}{P} - \frac{\pi}{P} \geq -1 \, ,
\nonumber \\ 
    & \frac{\Pi}{P} \geq -1, \qquad \qquad \frac{\Pi}{P} \leq \frac{\epsilon}{3P} - 1. 
\end{align}
Note that, unlike the first three bounds, the last one is a function of $m/T$ and thus moves with the evolving temperature of the system. For our following studies we select a variety of initial conditions within the allowed region spanned by $(\Pi/P, \pi/P)$; Table~\ref{table:IC} identifies them by the colors in which they are plotted in the figures. The parameters $(\alpha_0, \Lambda_0, \xi_0)$ used in the initial distribution (\ref{f_in}) to generate these initial conditions are summarized in Appendix \ref{Initial paramters}.

\begin{table}[h!]
 \begin{center}
 \resizebox{\columnwidth}{!}{
  \begin{tabular}{|c|c|c|c|c|c|c|c|}
   \hline
  &  Blue &  Green  &  Magenta &  Maroon &  Orange &  Black &  Cyan \\
   \hline
    $(\Pi/P)_0$ & 0 & $-0.25$ & $-0.37$ & 0 & 0 & $-0.25$ & $-0.85$ \\
   \hline
    $(\pi/P)_0$ & $-1$ & $-1$ & $-1$ & 0.99 & $-1.8$ & 0 & 0 \\
   \hline
  \end{tabular}}
  \caption{Association of initial conditions $\bigl(\Pi/P)_0,(\pi/P)_0\bigr)$ with the colors of the curves showing their evolution in the figures below.}
  \label{table:IC}
 \end{center}
 \vspace*{-.6cm}
\end{table}

In the following we study the time evolution of the system described by Eq.~(\ref{BEMilne}). Throughout the paper, initial conditions will be set at $\tau_0=0.1$\,fm/$c$, with an initial temperature $T_0=500$\,MeV. For the particle mass we take $m{\,=\,}200$\,MeV. The system thus behaves approximately as a conformally symmetric gas at very early times when $m/T=0.4 < 1$ while non-conformal corrections rapidly increase at $\tau>\tau_0$ since the temperature drops rapidly. For the relaxation time we make the conformal ansatz $\tau_{R}(\tau) = 5 C/T(\tau)$ where we set $C = 10/4\pi$ throughout the paper. This choice corresponds to an initial specific shear viscosity $\eta/s \approx 10/(4\pi) \approx 0.8$. For a massless system this value of $\eta/s$ remains constant throughout the evolution, for a massive system it decreases with time (see Fig.~\ref{Fig_etas_zetas}b below).  

\vspace*{-2mm}
\subsection{Free-streaming dynamics and fixed lines} 
\label{KT_fs}
\vspace*{-2mm}

Previous analyses of conformal dynamics have shown the existence of early-time fixed points for the scaled shear stress at $\pi/P = 1$ and $-2$ \cite{Blaizot:2017ucy}.\footnote{%
    For conformal systems, $-2\leq \pi/P \leq 1$ is the allowed kinetic bound which follows from the positivity of the distribution function ($\PL, \PT \geq 0$).}%
$^,$\footnote{%
    Note that to define equilibrium pressure one has to perform Landau matching which is not a necessity for free-streaming RTA Boltzmann equation ($\tau_R \to \infty$). We still choose to do it, having in mind a physical picture of a gas which interacts, albeit very weakly, such that $\tau_R$ is very large but finite.
    }
They are `fixed' in the sense that if the system is non-interacting, as is approximately the case at very early times when $\tau \ll \tau_R$, and it is initialized with $\pi/P = 1$ or $\pi/P = -2 $, then $\pi/P$ stays fixed at these initial values. Moreover, any other initial condition where $\pi/P$ is not exactly $-2 $ will eventually hit $\pi/P = 1$ if the system is non-interacting. Thus, $\pi/P = 1$ is an attractive fixed point whereas $\pi/P = -2$ is a repulsive one. 

We here perform a similar early-time analysis to identify fixed `points' of non-conformal dynamics. Accordingly, we again first consider a non-interacting system, taking $\tau_R\to\infty$ in the RTA Boltzmann equation. In this limit the solution for the distribution function is simply $f(\tau; p_T, w) = f_\mathrm{in}(p_T,w)$ at all times. Accordingly, the free-streaming solutions for energy density, transverse and longitudinal effective pressures are simply
\begin{align}
    \epsilon^{\mathrm{fs}}(\tau) &= \frac{\Lambda_0^4}{4\pi^2 \alpha_0} \tilde{H}_{\epsilon} \left[\frac{\tau_0}{\tau \sqrt{1+\xi_0}}, \frac{m}{\Lambda_0} \right] , 
\label{e_fs} 
\\ 
    \PT^{\mathrm{fs}}(\tau) &= \frac{\Lambda_0^4}{8\pi^2 \alpha_0} \tilde{H}_{T} \left[\frac{\tau_0}{\tau \sqrt{1+\xi_0}}, \frac{m}{\Lambda_0} \right],
\label{PT_fs}  
\\
    \PL^{\mathrm{fs}}(\tau) &= \frac{\Lambda_0^4}{4\pi^2 \alpha_0} \tilde{H}_{L} \left[\frac{\tau_0}{\tau \sqrt{1+\xi_0}}, \frac{m}{\Lambda_0} \right].
\label{PL_fs}
\end{align}

\begin{figure}[t!]
\begin{center}
 \includegraphics[width=.85\linewidth]{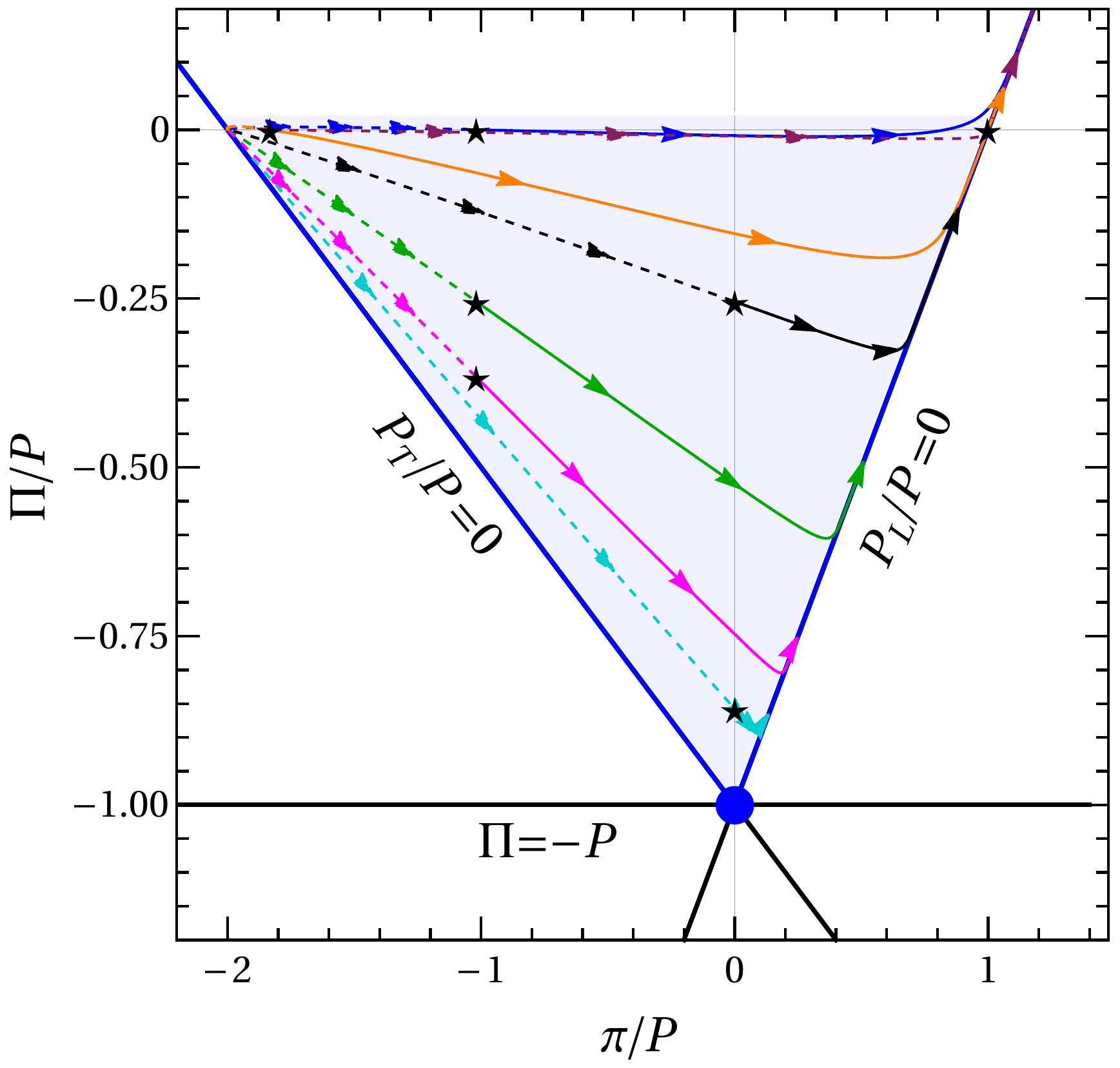}
\end{center}
\vspace*{-6mm}
\caption{
    Evolution of various solutions of free streaming kinetic theory with initial conditions given in Table \ref{table:IC}. The solid curves shows forward evolution ($\tau/\tau_0\geq 1$), while the dashed curves are backward evolution in proper time. 
    \vspace*{-3mm}  }
\label{CM_fs}
\end{figure} 

Fig.~\ref{CM_fs} shows the free-streaming evolution of the normalized shear and bulk viscous stresses for various initial conditions (indicated by black stars). The blue-shaded region shows the bounds imposed by kinetic theory. The fourth bound listed in Eq.~(\ref{pi_Pi_bounds}) (i.e. the upper edge of the shaded region) is taken at the initial time with temperature $T_0 = 0.5$~GeV; it moves dynamically with the evolving temperature (not shown). Arrows indicate the direction of time. All the trajectories are seen to move toward the line $\PL = 0$, as expected for longitudinal free-streaming, corresponding to positive shear stress $\pi\geq0$. The initial slopes of the curves depend on the initial distribution function. After merging with the $\PL{\,=\,}0$ line, all trajectories continue to move upward along that line. As the thermal pressure decreases, the normalized shear stress $\pi/P$ continues to increase. The bulk viscous pressure moves from negative territory (familiar for expanding systems that stay close to the Navier-Stokes limit\footnote{%
    Note, however, that for our choice of $m/T_0$, the Navier-Stokes limit is very close to $\Pi\approx 0$; see Fig.~\ref{CM_exact}. 
}%
) eventually into positive territory (i.e. far away from its Navier-Stokes expectation $\Pi_\mathrm{NS} = -\zeta/\tau$), indicating that the system moves farther and farther away from local thermal equilibrium. On the other hand, if the system is evolved backward in time from the chosen initial conditions (the dashed curves in Fig.~\ref{CM_fs} show backward evolution), all trajectories merge toward the point $(\Pi/P{\,=\,}0,\, \pi/P{\,=\,}{-}2)$ as $\tau \to 0$. This can be deduced analytically from Eqs.~(\ref{e_fs})-(\ref{PL_fs}) using the properties of $\tilde{H}_{\epsilon,T,L}(y,z)$ near $y \to \infty$. In the limit $\tau \to 0$, both $\epsilon^{\mathrm{fs}}$ and $\PL^{\mathrm{fs}}$ diverge as $1/\tau^2$ (with identical proportionality constants) such that $\PL^{\mathrm{fs}}/\epsilon^{\mathrm{fs}} \to 1$. In contrast, $\PT^{\mathrm{fs}}$ diverges slower than $1/\tau^2$ leading to $\PT^{\mathrm{fs}}/\epsilon^{\mathrm{fs}} \to 0$. Also, at early times, $P{\,\approx\,}\epsilon^{\mathrm{fs}}/3$, such that we obtain, $\Pi/P = \frac{1}{3} \left( \PL^{\mathrm{fs}}/P + 2 \PT^{\mathrm{fs}}/P - 3  \right) \to 0$, and $\pi/P = \frac{2}{3} \left( \PT^{\mathrm{fs}}/P - \PL^{\mathrm{fs}}/P  \right) \to -2$. The point $(\Pi/P{\,=\,}0,\, \pi/P{\,=\,}{-}2)$, which matches with the repulsive fixed point of \textit{conformal} dynamics, acts like a stable (attractive) fixed point when trajectories are evolved backward in time. It is thus a \textit{repulsive} fixed point for forward evolution. 

From Fig.~\ref{CM_fs} it is clear that if the system is initialized on the $\PL = 0$ line it will remain on it, irrespective of how far away from local equilibrium this takes the system. For any other initialization the evolution trajectory first takes it to the $\PL = 0$ line, which corresponds to a very specific non-equilibrium state in which the longitudinal momenta of all particles are redshifted to zero by the longitudinal expansion. The system then continues moving toward even more extreme non-equilibrium configurations, characterized by fixed much larger transverse momenta (as initially given to the particles) than the increasingly smaller values that would be required for a thermalized system whose temperature and thermal pressure keeps decreasing.

Obviously, for non-conformal free-streaming systems the line $\PL{\,=\,}0$ acts as an {\it attractive fixed line}. Moreover, any initialisation on the line $\PT{\,=\,}0$, corresponding to vanishing transverse momenta for all particles, will continue to stay on the same line during its evolution.\footnote{%
    This is harder to demonstrate numerically since our initial-state parametrization (\ref{f_in}) does not permit us to put the system precisely on this line with finite values of the parameters ($\Lambda_0,\alpha_0,\xi_0$); however, the systematic pattern of the green, magenta and blue lines in Fig.~\ref{CM_fs} clearly supports this claim.}
Thus, $\PT{\,=\,}0$ acts as another {\it fixed line} of the dynamics. However, any initial condition that deviates slightly from this `fixed line' will move away from it and eventually join the $\PL = 0$ line, identifying the former as a repulsive fixed line of free-streaming Bjorken dynamics.

In the conformal limit $(m = 0)$ the bulk viscous pressure vanishes, and the allowed region in Fig.~\ref{CM_fs} shrinks to a line at $\Pi/P{\,=\,}0$. In that case the corners of the allowed region (corresponding to intersections with the lines $P_L{\,=\,}0$ and $P_T{\,=\,}0$, respectively) yield the two {\it conformal fixed points}, namely the longitudinal and transverse free-streaming fixed points. Note that, at the level of the distribution function, the attractive and repulsive fixed lines map to $f \propto \delta(p_z)$ and $f \propto \delta(p_T)$, respectively. For non-conformal systems, the intersection of the {\it fixed lines} $\PL{\,=\,}0$ and $\PT{\,=\,}0$ results in a {\it fixed point} at $(\Pi/P = -1, \pi/P = 0)$, represented by a blue circle in Fig.~\ref{CM_fs}, reflecting a spherically symmetric distribution function $f \propto \delta^{(3)}(\bm{p}) \propto \delta(|\bm{p}|)/|\bm{p}|^2$. At this fixed point, all particles are essentially condensed at zero momenta. For non-conformal systems this kind of phase-space configuration yields a finite energy density solely due to rest mass energy of the particles while the total isotropic pressure vanishes:
\begin{equation}
\label{pressure:iso}
    P + \Pi = \frac{1}{3} \left\langle |\bm{p}|^2 \right\rangle  \propto \int_{0}^{\infty} \frac{p^4\,dp}{E_p} \, \frac{\delta(p)}{p^2} = 0,
\end{equation}
where $p\equiv|\bm{p}|$. Moreover, due to isotropy of $\delta(|\bm{p}|)$, the shear stress tensor also vanishes at this fixed point. 

\vspace*{-2mm}
\subsection{Kinetic theory with finite scattering rate $1/\tau_R$}
\label{sec2c}
\vspace*{-2mm}

\begin{figure}[t]
\begin{center}
 \includegraphics[width=.85\linewidth]{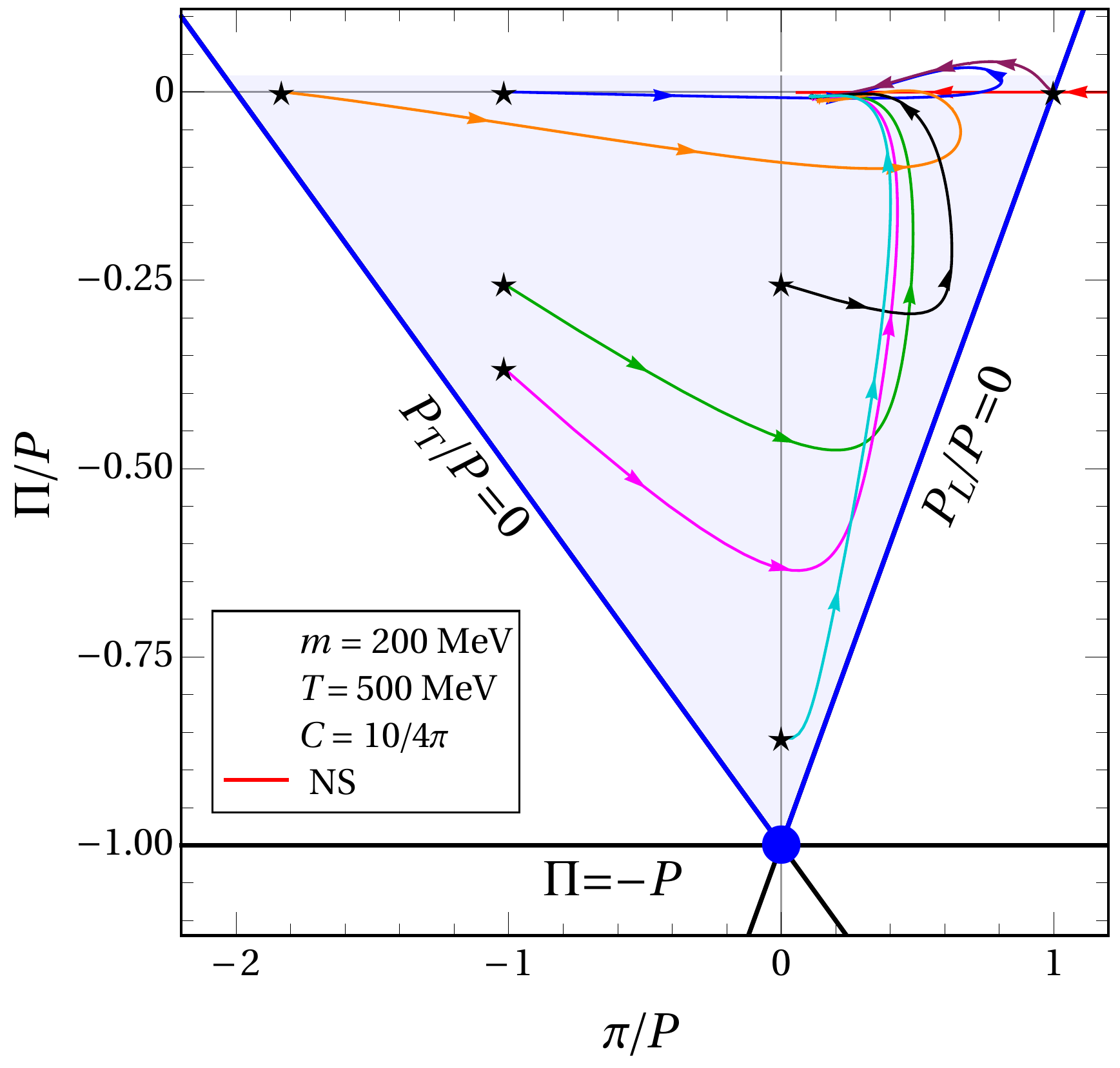}
\end{center}
\vspace*{-6mm}
  \caption{
  Evolution of various solutions of kinetic theory with $\tau_R = 5 C/T$, $C = 10/4\pi$, using initial conditions from Table~\ref{table:IC}. 
  \vspace*{-5mm}  
  \label{CM_exact}}
\end{figure} 

\begin{figure}[t!]
\begin{center}
 \includegraphics[width=.8\linewidth]{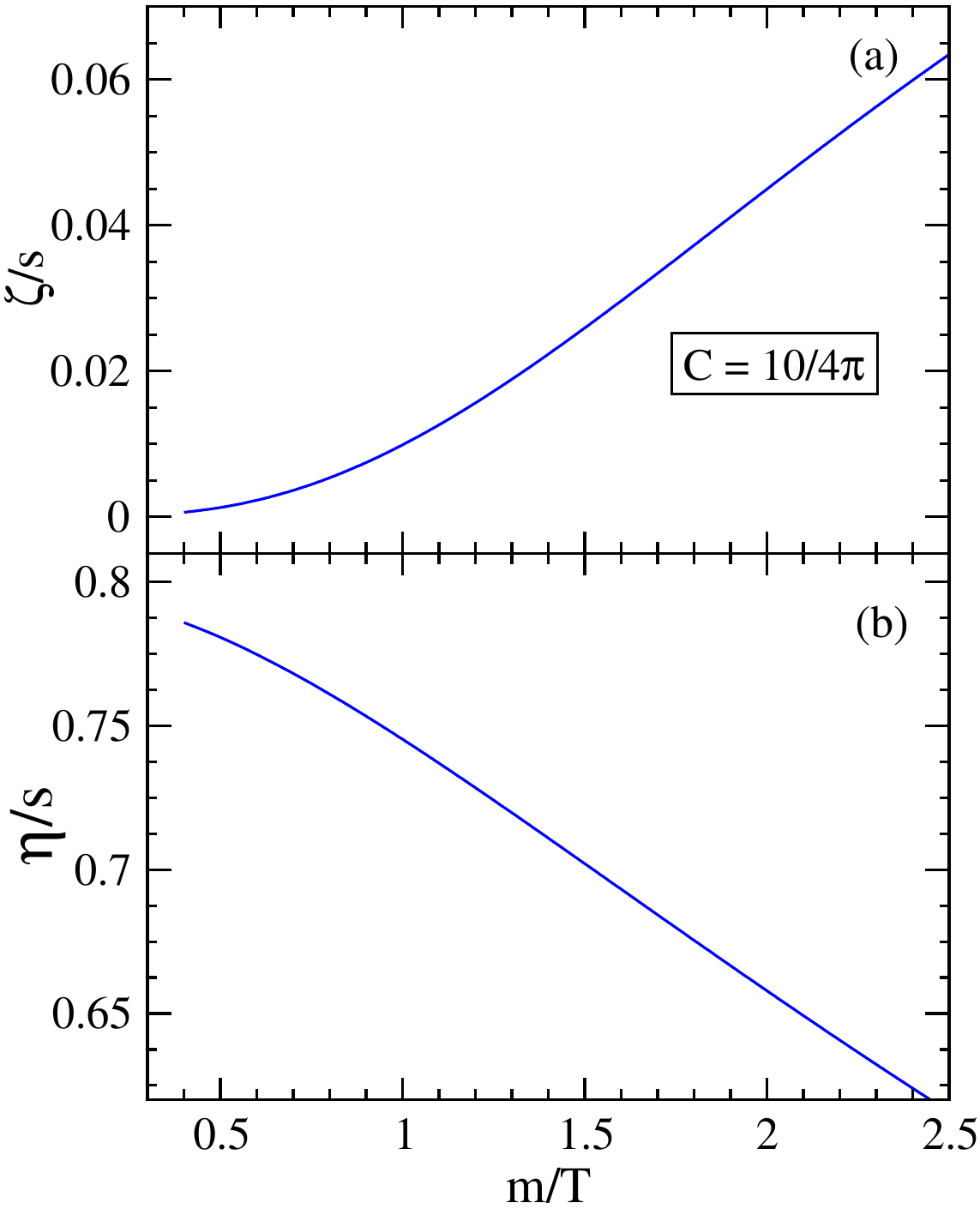}
\end{center}
\vspace*{-6mm}
\caption{
    $m/T$ dependence of (a) $\zeta/s$ and (b) $\eta/s$, for\\ $\tau_R{\,=\,}5C/T$ with $C=10/(4\pi)$. 
  \vspace*{-2mm} 
  }
\label{Fig_etas_zetas}
\end{figure}

Next we study the effect of collisions on this free-streaming fixed-line pattern. We parametrize the relaxation time $\tau_{R}$ as described at the end of Sec.~\ref{sec2a}. Fig.~\ref{CM_exact} shows the evolution trajectories for the same set of initial conditions whose free-streaming evolution was studied in Fig.~\ref{CM_fs}. As an additional red line we added the evolution of the Navier-Stokes values $\pi_\mathrm{NS} = \frac{4}{3}\frac{\eta}{\tau}$, $\Pi_\mathrm{NS}=-\frac{\zeta}{\tau}$, where the shear and bulk viscosities are computed using the standard definitions \cite{Jaiswal:2014isa},
\begin{align}
    \eta(T) &\equiv \frac{\tau_R}{T} \frac{1}{15} \int \frac{d^3p}{(2\pi)^3 \, E_p^2} \, p^4 \, \exp\left(- \sqrt{p^2 + m^2}/T \right), \label{eta}
\\
    \zeta(T) &\equiv \frac{5}{3} \eta(T) - \tau_R\, (\epsilon + P) \, c_s^2, 
\label{zeta}
\end{align}
with the squared speed of sound
\begin{equation}\label{cs2}
    c_s^2 = \frac{\epsilon + P}{3 \epsilon + (3+z^2)P}
\end{equation}
where $z = m/T$. For $\tau_R = 5 C/T$, the dependence of $\zeta/s$ and $\eta/s$ on the scaled particle mass $m/T$ is shown in Fig.~\ref{Fig_etas_zetas}. While the specific shear viscosity $\eta/s$ decreases monotonously with increasing particle mass, the opposite is true for $\zeta/s$. For conformal systems $\eta = (4/5)  \tau_R P$ and $c_s^2 = 1/3$, leading to $\zeta = 0$. For small $m/T$, the ratio of the viscosities satisfies $ \zeta/\eta \approx 75 \left( 1/3 - c_s^2 \right)^2$ \cite{Jaiswal:2014isa}, ensuring positivity of $\zeta$ and the entropy production rate.

As in Fig.~\ref{CM_fs}, the shaded region in Fig.~\ref{CM_exact} delineates the initially allowed region for the normalized shear and bulk viscous stresses at $T_0=500$\,MeV. We emphasize that all the kinetic theory curves satisfy the bounds given by Eqs.~(\ref{pi_Pi_bounds}), staying within the allowed region of $(\Pi/P, \pi/P)$ space throughout their entire evolution.\footnote{%
    While the blue and maroon curves in the upper right corner appear to stray outside the {\it initially} allowed region, we note that the upper limit of the allowed region {\it rises with time} as the temperature drops, and the dynamically evolving kinetic bound $\Pi/P \leq (\epsilon{-}3P)/(3P)$  is always respected.}

Except for the maroon curve, all trajectories in Fig.~\ref{CM_exact} start away from the $\PL{\,=\,}0$ ``free-streaming attractor''. As in the free-streaming case shown in Fig.~\ref{CM_fs}, they initially move towards this attractor, but before they can reach it collisions kick in and drive the system away from zero longitudinal pressure. The same is true for the maroon history where the system is initialized at approximately zero $\PL$ but driven away from that line by collisions. The collisions move the system closer to local thermal equilibrium, and its evolution trajectories eventually merge with the Navier-Stokes line which describes its late-time behavior and eventually takes it to the thermal fixed point $\Pi/P = 0 =\pi/P$.

\begin{figure}[t]
\begin{center}
 \includegraphics[width=.8\linewidth]{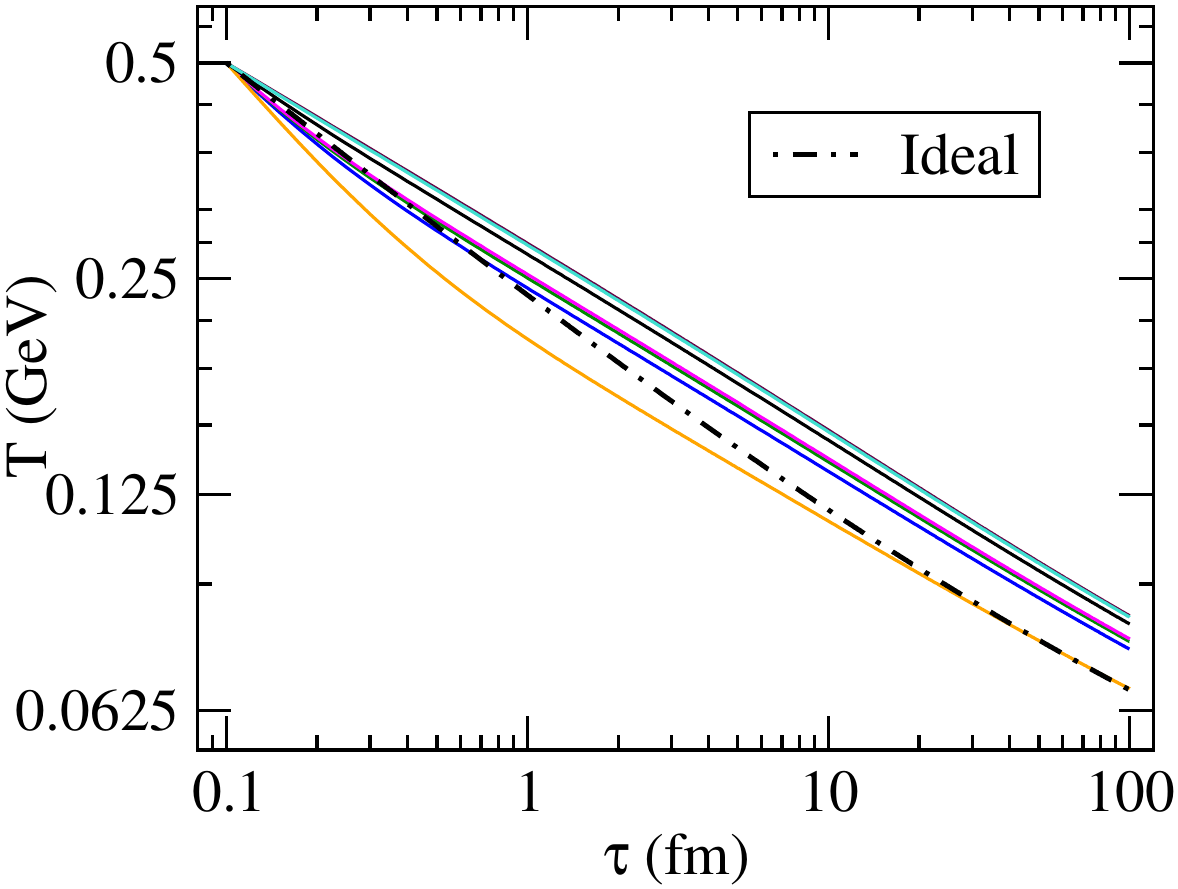}
\end{center}
\vspace*{-6mm}
  \caption{
  {Time evolution of temperature using the initial conditions
  given in Table 1.}
  \vspace*{-2mm}  }
  \label{Fig_temp}
\end{figure}

The corresponding temperature evolution histories are shown in Fig.~\ref{Fig_temp} and compared with ideal fluid dynamics (i.e. perfect local thermal equilibrium).\footnote{%
    Note that for a \textit{conformal} ideal fluid the time evolution for the temperature is a straight line with slope $-1/3$ in a log-log plot. For non-conformal ideal evolution the entropy per unit rapidity is still conserved, $s(\tau) \propto 1/\tau$, but the $z{\,=\,}m/T$ dependence of the entropy density
\begin{equation}
    s(T) = \frac{T^3}{2\pi^2} z^3 K_3(z)
\end{equation}
    causes the corresponding temperature $T(\tau)$ to deviate from the simple $\tau^{-1/3}$ law for a massless ideal fluid.}  
The early-time differences in the slopes of the trajectories are direct manifestations of differences in the amount of viscous heating caused by the different initial shear and bulk viscous stresses. The orange curve corresponds to the largest (positive) initial effective longitudinal pressure, leading to the largest amount of work done against the longitudinal expansion,
\begin{equation}
    \frac{d\epsilon}{d\tau} = - \frac{1}{\tau} \left( \epsilon + P_L \right),
\end{equation}
and thus resulting in the fastest initial drop of the temperature. As the effective longitudinal pressure $P_L$ for the orange curve remains above the thermal pressure throughout the first 0.2\,fm/$c$ (see Fig.~\ref{PL_P_KT}) where the rates of expansion and work done by the longitudinal pressure are highest, its temperature trajectory remains below that of ideal fluid dynamics for an extended period of time: only after about 10\,fm/$c$ has viscous heating by bulk and shear viscous stresses brought the temperature back up to that of the ideal fluid.  

\begin{figure}[t]
\begin{center}
 \includegraphics[width=.8\linewidth]{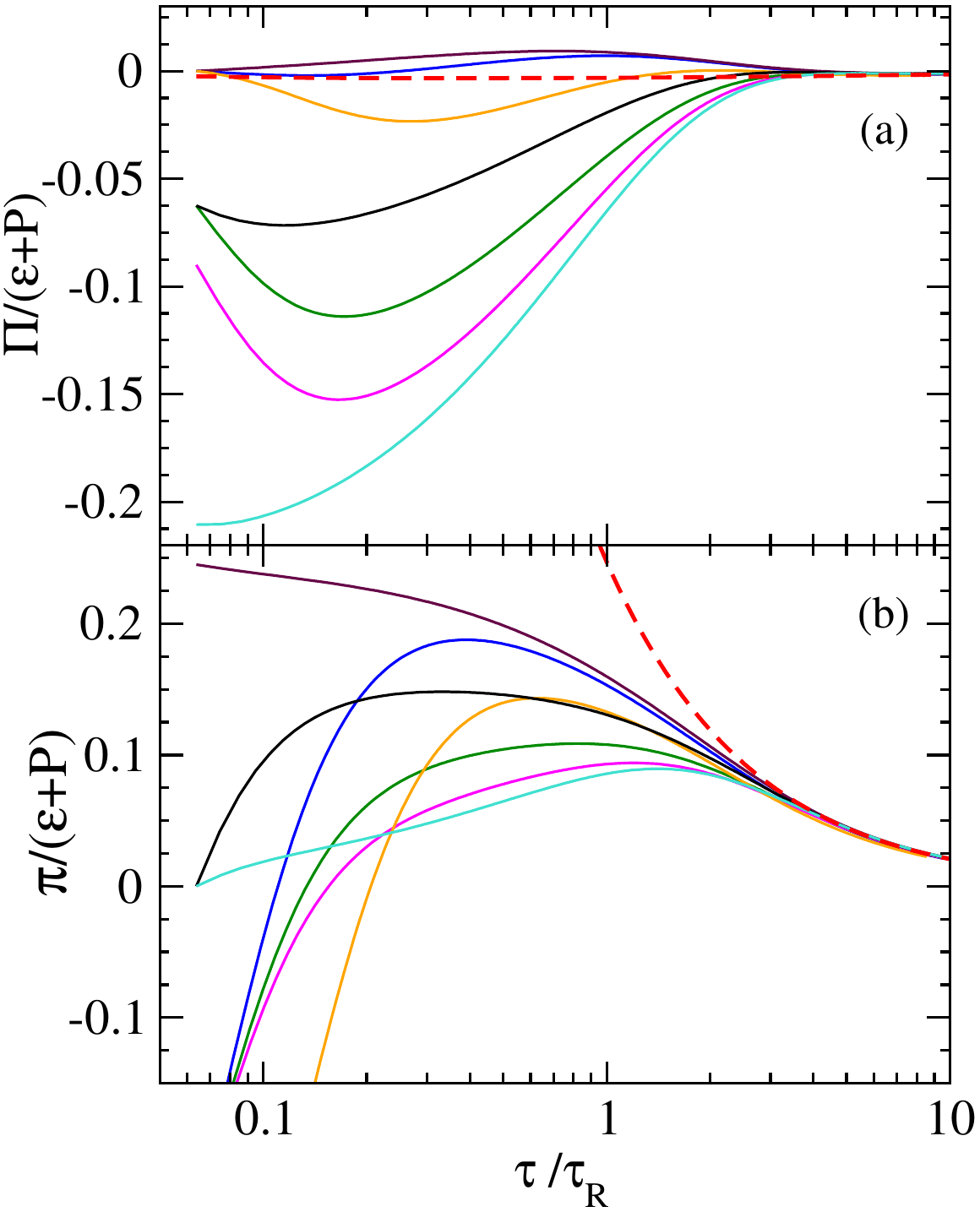}
\end{center}
\vspace*{-6mm}
  \caption{
  {Scaled time evolution of (a) $\Pi/(\epsilon{+}P)$ and (b) $\pi/(\epsilon{+}P)$.}
  \vspace*{-2mm}  }
  \label{fig:bulk_shear}
\end{figure}

The evolution histories for the enthalpy-normalised bulk and shear stresses, $\bPi \equiv \Pi/(\epsilon{+}P)$ and $\bpi \equiv \pi/(\epsilon{+}P)$, as a function of scaled time $\tau/\tau_R$ are shown in Fig.~\ref{fig:bulk_shear}. In panel (a) the maroon and orange curves show that the initial slope of trajectories corresponding to identical initial values for the bulk viscous pressure (here $\bPi_0 = 0$) depends on the initial shear stress associated with the anisotropy of the initial kinetic momentum distribution, i.e. (in this case) on whether the distribution function is initially sharply peaked along $p_z$ ($\bpi_0 \approx 0.24$, maroon) or along $p_T$ ($\bpi_0 \approx - 0.45$, orange). The black and green trajectories illustrate the same point for a nonzero value of the initial bulk stress. The orange, green and magenta curves together illustrate that negative $\bpi_0$ results in large negative initial slopes for $\bPi(\tau/\tau_R)$.  Panel (b) shows that the evolution of sets of curves starting with identical initial normalized shear stress $\bpi_0$ varies significantly under variation of the initial bulk viscous pressure $\bPi_0$ -- see the black and cyan pair of lines with $\bpi_0 = 0$, and the triplet of blue, green and magenta lines with $\bpi_0 = -0.25$. Negative initial $\bPi_0$ is seen to decrease the initial slope of $\bpi$. The blue and orange curves are both characterised by large negative initial shear stresses and small deviations from conformality ($\bPi \approx 0$); they exhibit near-identical early-time slopes, reminiscent of the power-law decay of $\bpi$ towards a hydrodynamic attractor that was previously identified as the characteristic early-time behaviour of weakly-coupled systems in conformal Bjorken expansion \cite{Jaiswal:2019cju, Kurkela:2019set}. Together, the two panels of Fig.~\ref{fig:bulk_shear} provide evidence for strong bulk-shear coupling effects in RTA Boltzmann kinetic theory -- we will return to these in Sec.~\ref{sec:hydro}.

The different trajectories in Fig.~\ref{fig:bulk_shear} for both $\bPi$ and $\bpi$ all merge with the Navier-Stokes trajectory $\bPi_\mathrm{NS} = -(\zeta/s) / (5C) \times (\tau_R/\tau)$, $\bpi_\mathrm{NS} = 4(\eta/s)/(15C) \times (\tau_R/\tau)$ (red dashed line), after $\bar\tau \approx 3$. Owing to weak temperature dependencies of $\eta/s$ and $\zeta/s$ caused by the nonzero particle mass $m$, different curves approach slightly different Navier-Stokes solutions; the red dashed line should therefore rather be shown as a narrow band of red curves. To prevent clutter we here only show the Navier-Stokes solution corresponding to an initial equilibrium distribution function. 

\begin{figure}[t!]
\begin{center}
 \includegraphics[width=.77\linewidth]{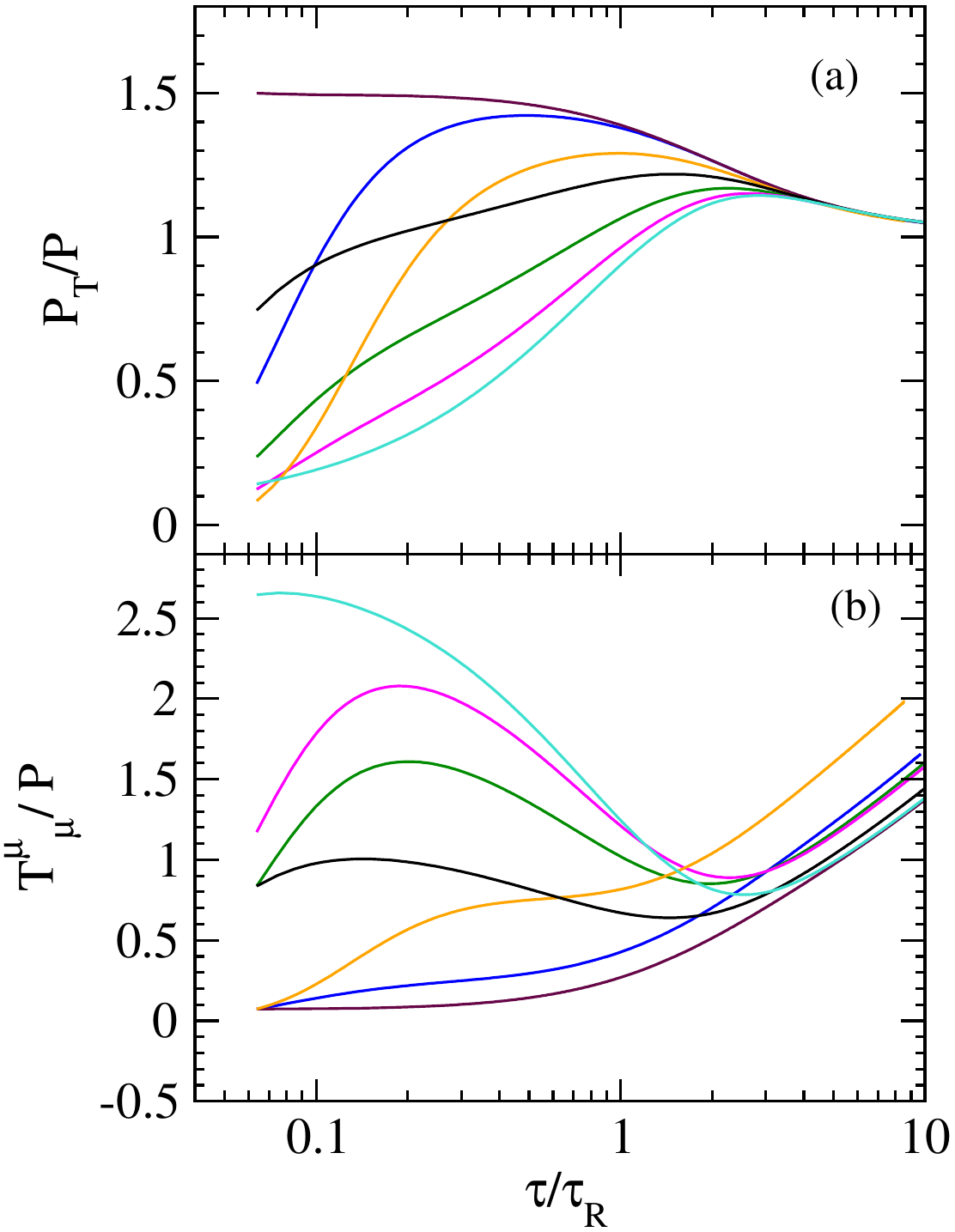}
\end{center}
\vspace*{-6mm}
  \caption{
  {Time evolution of (a) scaled transverse effective pressure, $\PT/P$ and (b) scaled trace, $T^{\mu}_{\mu}/P$.  }
  \vspace*{-2mm}  }
  \label{PT_Trace}
\end{figure} 

Figure~\ref{PT_Trace} shows the time evolution of the transverse pressure $\PT$ and the trace of the energy momentum tensor $T^{\mu}_{\mu} = \epsilon{\,-\,}2 \PT{\,-\,}\PL$, both scaled by the thermal pressure $P$. The substantial difference between the orange and magenta curves in panel (a), in spite of starting from a very similar initial transverse pressure $P_T/P$, shows that the evolution of $\PT/P$ is strongly affected by the initial shear stress $\bpi_0$ or, equivalently, by the initial longitudinal pressure $\PL$. We also note that, similar to panel (b) of Fig.~\ref{fig:bulk_shear}, the orange and blue trajectories, being driven by large shear stress combined with small bulk viscous pressure (approximately conformal dynamics), again exhibit similar early-time slopes. As different trajectories are seen to cross each other, non-conformal dynamics is seen to be quite different from conformal dynamics: there is no sign of attractor-controlled early-time dynamics. Universality is achieved only at late times, after $\btau \approx 3$, when all trajectories join the late-time Navier-Stokes trajectory of dissipative fluid dynamics which ultimately leads the system to local thermal equilibrium, characterized by $\PT/P\to 1$.

The normalized trace of the energy momentum tensor shown in panel (b) is a measure of conformality and its breaking by particle mass effects. While the maroon, blue and orange curves start out with zero initial bulk viscous pressure, the cyan curve features the largest initial bulk stress. While these differences in the initial {\it bulk stress} control the early-time evolution of the trace of $T^{\mu\nu}$, its late-time behaviour is more affected by the initial value of the shear stress $\bpi_0$: Among all the curves in Fig.~\ref{PT_Trace}b, the orange one (which is initialized with the largest negative shear stress $\bpi_0$) shows the strongest breaking of conformality. Once again this indicates strong shear-bulk coupling effects. However, another mechanism comes into play additionally: The generic rise of $T^{\mu}_{\mu}/P$ at late times shared by all evolution trajectories is caused by the non-zero particle mass $m$ whose conformality-breaking effects grow at late times when the temperature $T$ decreases and $m/T$ increases. Similar late-time slopes of the temperature trajectories in Fig.~\ref{Fig_temp} translate into similar late-time slopes for $T^{\mu}_{\mu}/P$ in Fig.~\ref{PT_Trace}b. The fact that the orange curve has the largest $T^{\mu}_{\mu}/P$ at late times reflects, among other contributing factors, the fact that it has the lowest temperature throughout its evolution.

\begin{figure}[!t]
\begin{center}
 \includegraphics[width=0.9\linewidth]{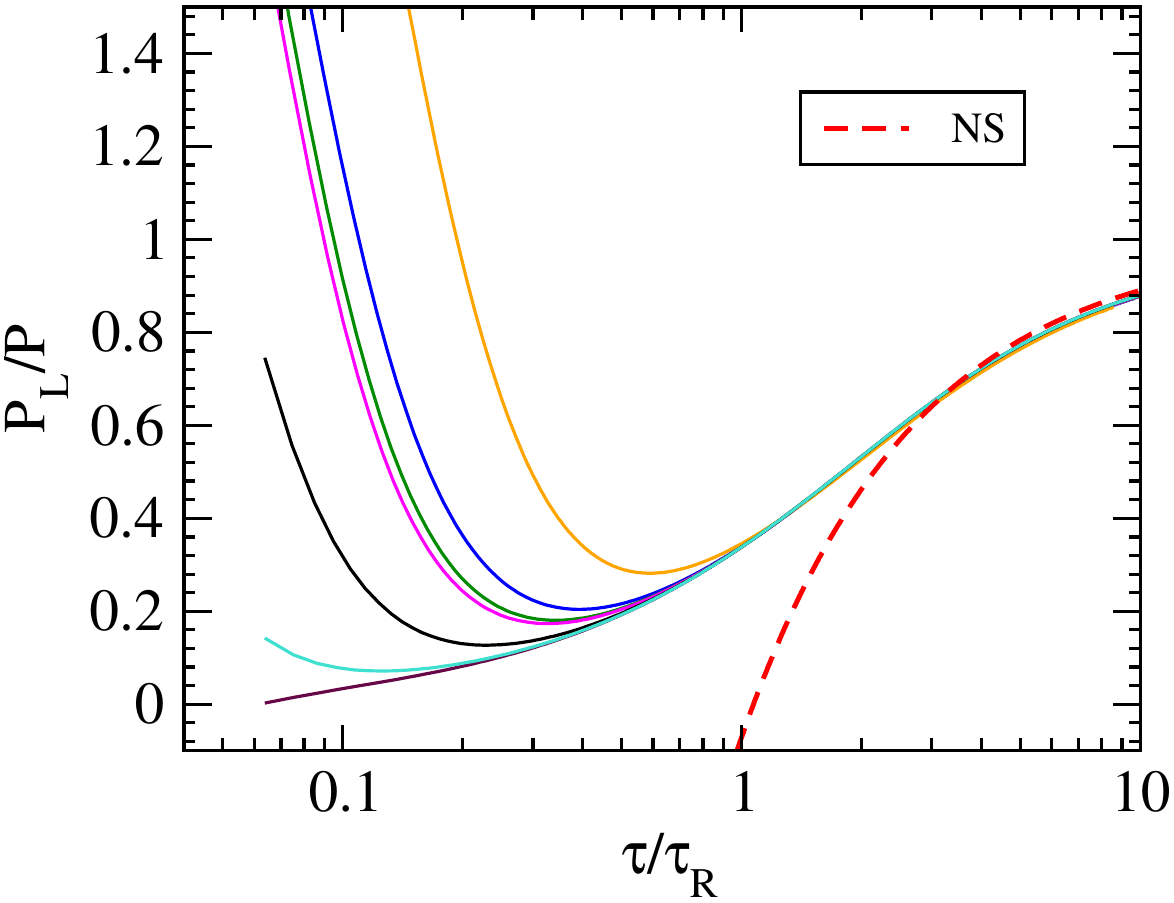}
\end{center}
\vspace*{-6mm}
  \caption{
  Scaled time evolution of $P_L/P$. First-order (Navier-Stokes) result is denoted by red dashed curve (see paragraph below Eq.~(\ref{coeff_rel})).
  \vspace*{-2mm}  }
\label{PL_P_KT}
\end{figure} 

\begin{figure*}[!t]
\begin{center}
 \includegraphics[width=\linewidth]{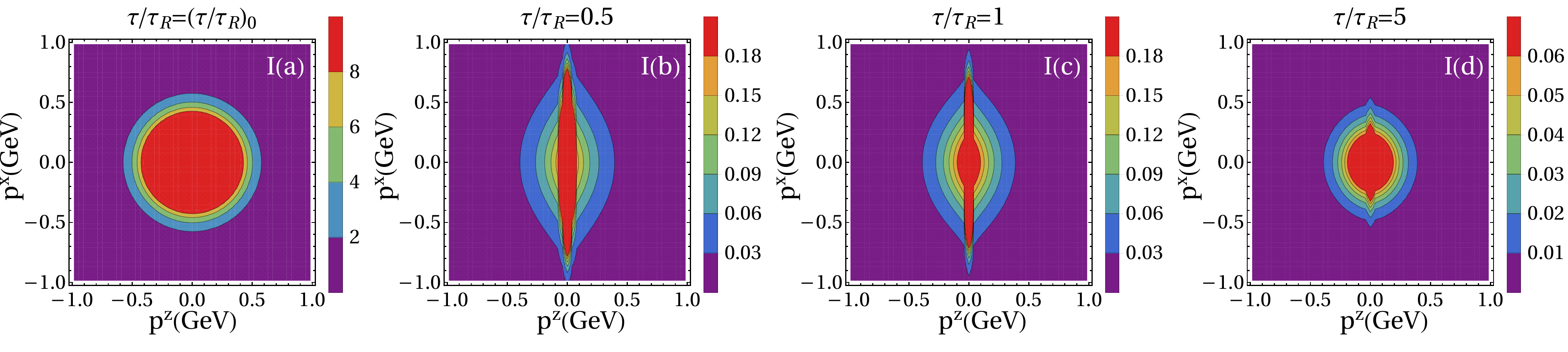}
 \includegraphics[width=\linewidth]{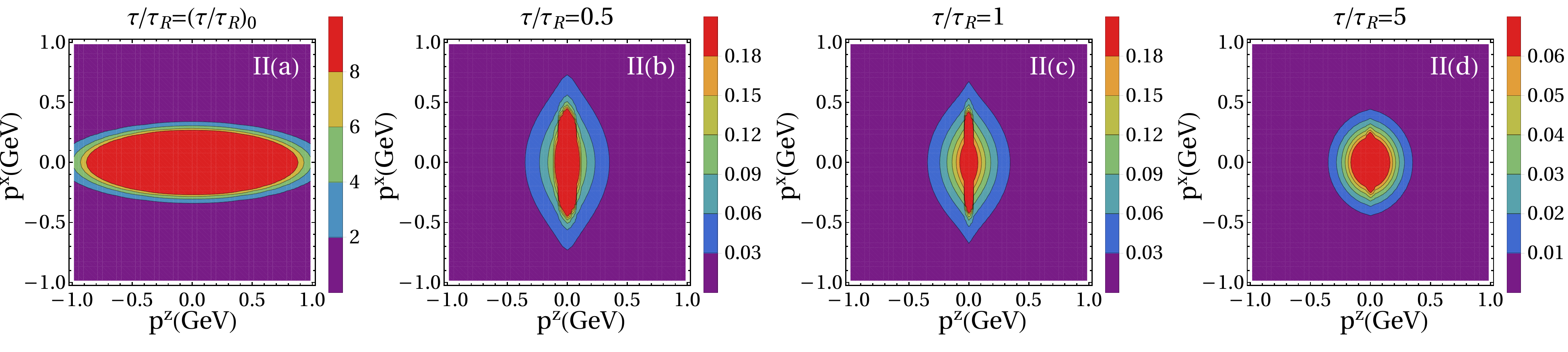}
 \includegraphics[width=\linewidth]{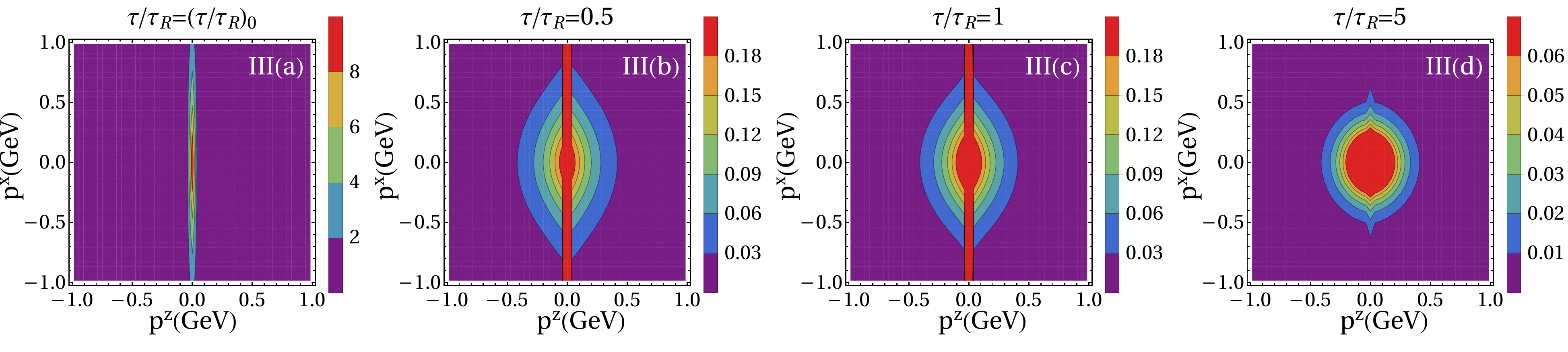}
\end{center}
\vspace*{-6mm}
  \caption{
  Contour plots showing, from left to right, the time evolution of the distribution functions corresponding, from top to bottom, to the initial conditions for the black, green, and maroon trajectories (see Table \ref{table:IC}).
  \vspace*{-2mm}  }
  \label{fig:Contour}
\end{figure*} 

The most convincing early-time, far-off-equilibrium attractor behavior manifests itself in the evolution of the effective longitudinal pressure $\PL/P$, shown in Fig.~\ref{PL_P_KT}. For all initial conditions the evolution trajectories show rapid decay to a universal attracting curve, merging with this ``attractor" already at $\tau/\tau_R < 1$, i.e. before the collision rate starts to exceed the expansion rate. For all practical purposes, the maroon curve, which smoothly connects the universal early-time non-thermal fixed point, $\PL/P \approx 0$,  with the late-time thermal fixed point, $\PL/P = 1$, may be thought of as an attractor. The identical slopes of various trajectories indicate that their decay to the attractor is governed by a power-law at early times. This is similar to what has been observed for $\bpi(\tau/\tau_R)$ for conformal systems where it was demonstrated that, due to rapid expansion, excursions from the free-streaming attractor ($P_L = 0$) at \textit{early} times decay at a scale set by the initialization time, $\tau_0$, and that at those early times the decay rate is essentially \textit{independent} of the relaxation-time \cite{Jaiswal:2019cju, Kurkela:2019set}. 

The presence of an early-time attractor in $\PL/P$ and not separately in the normalized shear or bulk channels can also be understood as follows: As already mentioned, the distribution function at the stable fixed line is sharply peaked along longitudinal momentum: $f = f_{\rm fs} \approx \psi(p_T) \delta(w)$, where $\psi$ is an arbitrary function of transverse coordinates. Accordingly, one has
\begin{align}
    \pi_0 &= \int_p \, \left(  \frac{ (u \cdot p)^2}{3} - m^2  - \frac{w^2}{\tau_0^2} \right)  f_{\rm fs}, \nonumber \\ 
    &=  \int_p \, \left(  \frac{ (u \cdot p)^2}{3} - m^2 \right)  f_{\rm fs} = P_0 + \Pi_0.
\end{align}
Therefore, the free-streaming distribution function does not correspond to universal fixed points for either $\pi/P$ or $\Pi/P$, rather it is $\pi_0/P_0 - \Pi_0/P_0 = 1$, or equivalently, $(\PL)_0/P_0 = 0$ which corresponds to a universal stable fixed point of non-conformal kinetic theory. In fact, using Eq.~\eqref{PL_fs} one can show that the leading order decay of $\PL^{\rm fs}$ at late times goes as $(\PL/P)_{\rm fs} \propto (\tau_0/\tau)^2$. Therefore, all free-streaming solutions of $\PL/P$ rapidly approach zero with a $\tau^{-2}$ fall-off. In contrast, leading order free-streaming solutions for $\pi/P$ and $\Pi/P$ are \textit{independent} of $\tau_0/\tau$, and instead depend on the choice of initial parameter $\Lambda_0$ via the ratio $m/\Lambda_0$; hence, they do not attain a universal limit.

To summarize, our analysis shows that even in non-conformal Bjorken flow, the decay of trajectories to an early-time attractor is predominantly driven by rapid longitudinal expansion of the medium. The evolution of the attractor itself, on the other hand, is affected throughout by the relaxation time, or rather by its competition with the expansion rate, i.e. by the Knudsen number $\tau_R/\tau$ \cite{Chattopadhyay:2019jqj}. We close this section by mentioning that in \cite{Romatschke:2017vte}, Romatschke explored non-conformal attractor behavior in Bjorken flow using kinetic theory for a slightly different quantity, namely, $A_1 \equiv -1 + \pi/(\epsilon +P) - \Pi/(\epsilon+P)$ versus the `inverse gradient strength', $\Gamma \equiv \tau/\gamma_s$ with $\gamma_s = (4\eta/3 + \zeta)/(\epsilon + P)$. Although, the quantity $A_1$ shows universality at late-times $(\tau > \tau_R)$, we show in Appendix \ref{A1_Gamma_appendix} that $A_1$ does not posses a universal (i.e., independent of $m/T_0$) early-time limit, and accordingly does not exhibit early-time attractor behavior once $m/T_0$ is varied. In contrast, the scaled longitudinal pressure continues to show early-time universality even for different choices of the ratio $m/T_0$.  

\vspace*{-2mm}
\subsection{Evolution of the distribution function}
\label{sec2d}
\vspace*{-2mm}

The appearance of an attractor for $\PL/P$ should be a reflection of some universal feature of the evolution of the microscopic distribution of particle momenta. We explore this in Fig.~\ref{fig:Contour} by studying the time evolution of the shape and structure of contour plots of the momentum distributions in longitudinal and transverse momentum, $p^z$ and $p^x$. The top, middle and bottom rows in Fig.~\ref{fig:Contour} correspond to the black, green and maroon trajectories in the previous figures, exploring three different initial conditions for the bulk and shear stresses as listed in Table~\ref{table:IC}. 

The top row features an isotropic initial distribution function $(\pi_0 = 0)$ with a non-equilibrium energy distribution in the local rest frame ($\Pi \neq 0$). The initial momentum distribution for the middle row is elongated along the $p^z$ direction, corresponding to a negative shear stress $\bpi_0 \approx -0.25$. The initial distribution in the bottom row is sharply peaked in $p^z$ around zero and very wide in $p^x$, corresponding to a large positive initial shear stress $\bpi_0 \approx 0.24$.

For typical thermal $p^x$ values, say $p^x{\,\sim\,}T_0{\,=\,}0.5$\,GeV, the $p^z$ dependencies for these rather different initial distribution functions are seen to become qualitatively similar already at early times, $\tau/\tau_R \sim 0.5$, when the expansion rate still exceeds the microscopic scattering rate. Rapid longitudinal expansion shrinks the top and middle initial $p^z$ distributions whereas the bottom one (which initially has almost zero longitudinal pressure, $\mathcal{P}_{L,0} \approx 0$) cannot be shrunk any further. At the same time, microscopic interactions begin to generate a second, isotropic component which populates the low-momentum region in similar ways in all three rows. This was observed previously \cite{Kurkela:2018vqr, Strickland:2018ayk} and has given rise to recent proposals for improved hydrodynamic approximations that take this feature into account \cite{McNelis:2020jrn, Alalawi:2020zbx}. Among the different macroscopic quantities considered, the effective longitudinal pressure $\PL \equiv \langle (p^z)^2 \rangle$ is most sensitive to the $p^z$ profile of the distribution function and essentially captures this early time universal behavior. 

\begin{figure*}[!tbp]
\begin{center}
 \includegraphics[width=\linewidth]{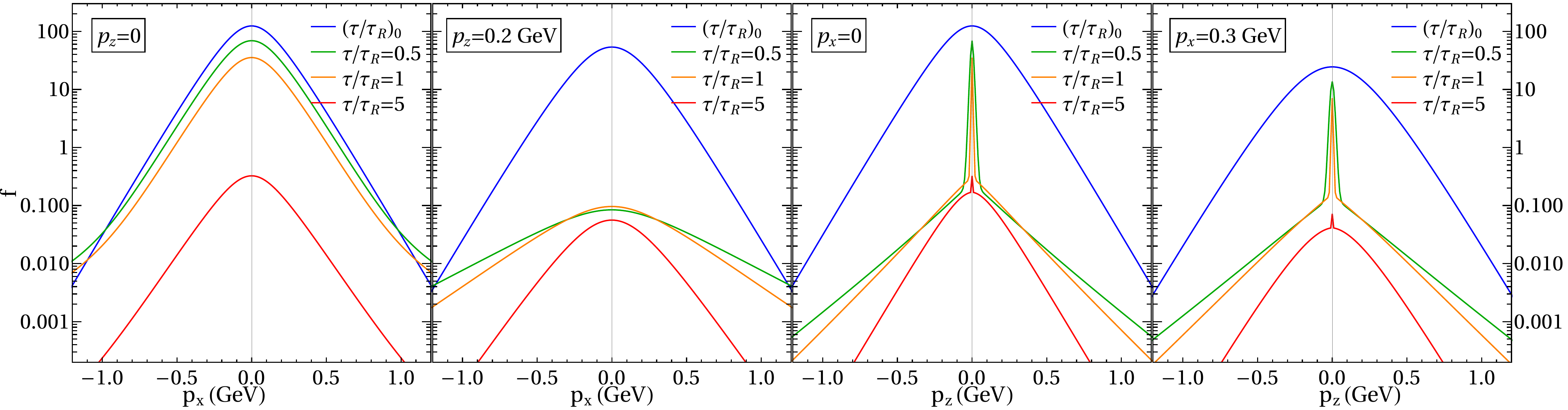}
 \includegraphics[width=\linewidth]{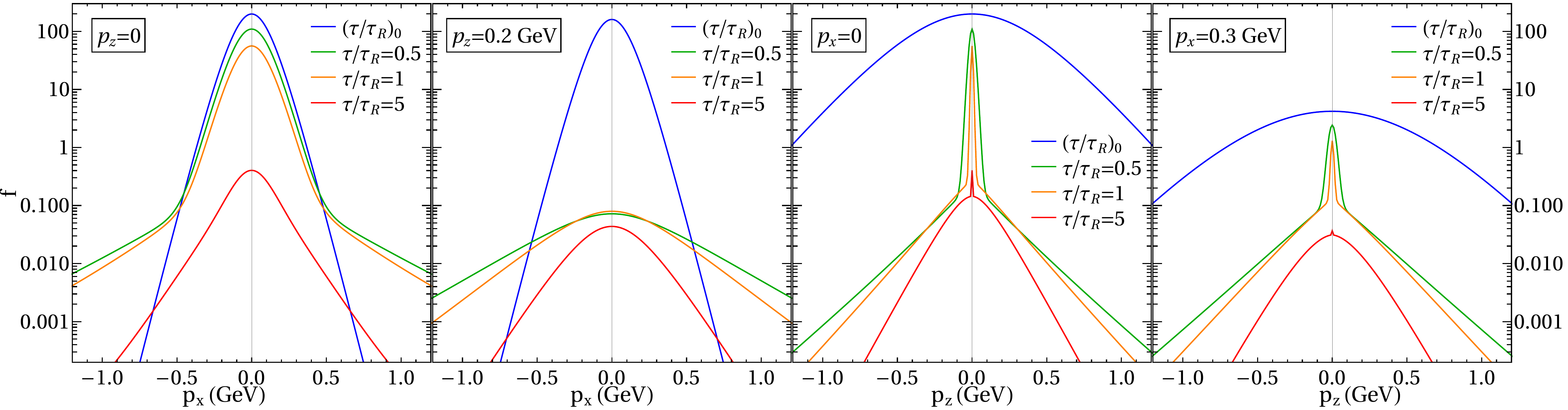}
 \includegraphics[width=\linewidth]{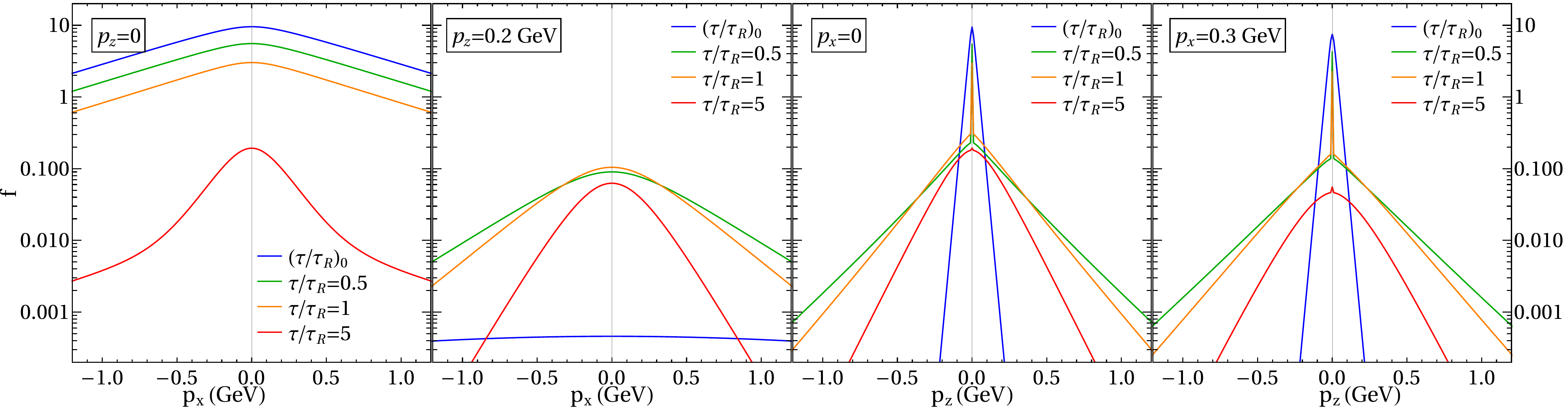}
\end{center}
\vspace*{-6mm}
  \caption{
  Evolution of the distribution function at different momentum slices. The rows here correspond to those in Fig.~\ref{fig:Contour} and show, respectively, the momentum dependencies of $f(\tau;p_T,p^z)$ corresponding to the black, green, and maroon trajectories in Table~\ref{table:IC} at various times $\tau/\tau_R$. The left two columns show $p^x$ distributions at fixed $p^z=0$ and 0.2\,GeV/$c$, the right two columns show $p^z$ distributions at fixed $p^x=0$ and 0.3\,GeV/$c$.
  \vspace*{-2mm}  }
  \label{fig:f_pt_pz}
\end{figure*} 

With increasing $\tau/\tau_R$ the expansion rate decreases and the system begins to thermalize, leading to a near-isotropic distribution around $\tau = 5 \tau_R$. The sharp red bars near $p^z \approx 0$ in the bottom and, less prominently, in the top rows, which arise from the initial $p^z$ distributions getting red-shifted to $p^z\simeq 0$ by longitudinal expansion, eventually get damped away by the damping function $D(\tau,\tau_0)$. The middle row (corresponding to the green curves in previous figures) is seen to isotropize faster than the top and bottom rows: Since its initial distribution is strongly elongated along $p_z$, the rapid early longitudinal expansion does not shrink it as much in $p_z$ as for the top and bottom distributions before the thermalizing effects of interactions take over. 

The contour plots shown in Fig.~\ref{fig:Contour} show these features with limited resolution. For a more detailed and quantitative view we follow Ref.~\cite{Strickland:2018ayk} and provide in Fig.~\ref{fig:f_pt_pz} horizontal and vertical cuts through the panels shown in Fig.~\ref{fig:Contour}, through the center ($p^z=0$ and $p^x=0$) in columns 1 and 3 and off-center in columns 2 and 4 of Fig.~\ref{fig:f_pt_pz}, respectively. The top, middle, and bottom rows in Figs.~\ref{fig:Contour} and \ref{fig:f_pt_pz} correspond to each other. We invite the reader to reread the preceding two paragraphs and follow the discussion along in Fig.~\ref{fig:f_pt_pz}.

\begin{figure*}[t!]
\begin{center}
 \includegraphics[width=\linewidth]{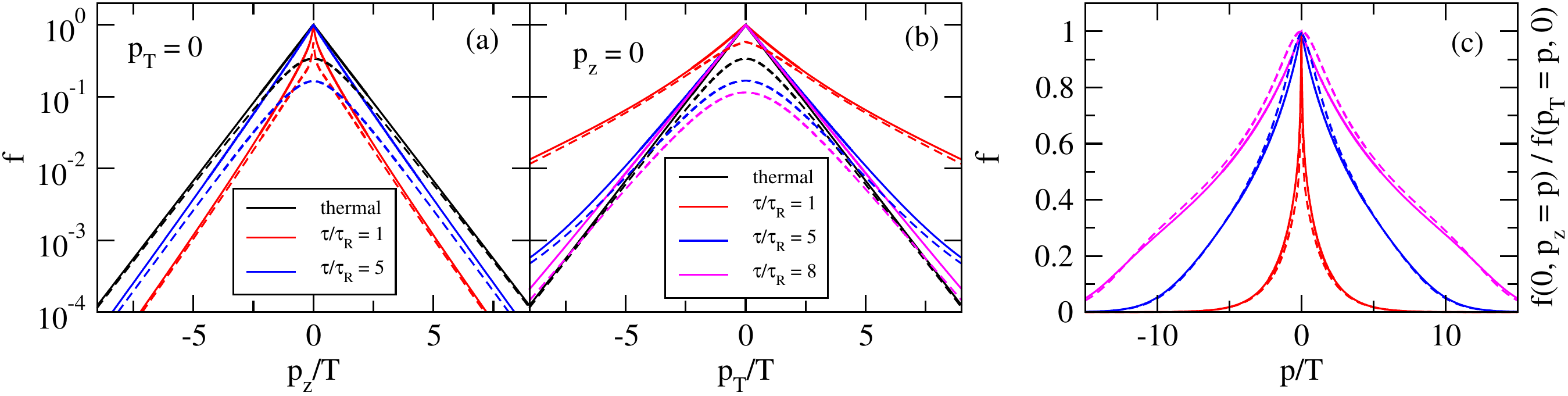}
\end{center}
\vspace*{-6mm}
  \caption{
  {Normalised momentum dependence of (a) $f(\tau; p_T = 0, p_z)$, (b) $f(\tau; p_T, p_z = 0)$, and (c) the ratio $f(\tau; p_T = 0, p_z = p)/f(\tau; p_T = p, p_z = 0)$, obtained using conformal (solid lines) and non-conformal (dashed lines) dynamics at various scaled times.}
  \vspace*{-2mm}  }
  \label{fig:f_momenta}
\end{figure*} 

We close this subsection with a brief discussion of features specifically associated with the breaking of conformal symmetry by the particle mass $m$, by comparing our results with those presented in Ref.~\cite{Strickland:2018ayk} for the conformal case.\footnote{\label{fn6}%
    During the preparation of this manuscript an error was discovered in the code used to generate Figs.~9-11 in Ref.~\cite{Strickland:2018ayk} for conformal theories. We thank Mike Strickland for helping us resolve this issue. We checked that the results shown in Fig.~\ref{fig:f_momenta} are in perfect agreement with those from Strickland's corrected code.
    }  
Following the analysis in \cite{Strickland:2018ayk} we compare in Fig.~\ref{fig:f_momenta} snapshots at three different times of the longitudinal (panel (a)) and transverse (panel (b)) momentum distributions, as well as their ratio in panel (c), for an expanding gas of massless ($m{\,=\,}0$, solid lines) and massive ($m{\,=\,}0.2$\,GeV, dashed lines) particles, for identical initial conditions $f_{\mathrm{in}}(p_T, w) = \exp\bigl(- \sqrt{p_T^2 + w^2/\tau_0^2}/\Lambda_0\bigr)$  with $\Lambda_0 = 0.5$\,GeV at $\tau_0 = 0.1$\,fm/$c$.\footnote{%
    This corresponds to an initial temperature $T_0{\,=\,}0.5$\,GeV for the conformal system and  $T_0 \approx 0.503$\,GeV for the non-conformal gas.} 
For comparison we also include in panels (a) and (b) as black solid and dashed lines the conformal and non-conformal thermal equilibrium distributions. Plotted as a function of $p/T$, the conformal equilibrium distribution does not evolve with time whereas the non-conformal one does; its peak at $p{\,=\,}0$ decreases with time whereas at large momenta $p\gg m$ it merges with the time-independent conformal equilibrium distribution. To avoid clutter we plot the thermal equilibrium distribution for the massive case for only one of the three times shown, i.e. at $\tau/\tau_R = 1$. The colored curves show the solutions of the RTA Boltzmann equation at $\tau/\tau_R = 1$ (red), 5 (blue) and 8 (magenta). At large $p/T$ where mass effects are negligible, panels (a,b) demonstrate clearly that for both massless and massive systems the distribution functions approach their respective thermal equilibrium distributions as $\tau/\tau_R$ increases. We have checked that the same holds true also at small values of $p/T$ although this is not obvious from Fig.~\ref{fig:f_momenta} for the non-conformal case for which we plotted the thermal distribution only for $\tau/\tau_R = 1$ (black dashed). 

We note that in the massless case with the above choice of initial conditions we have $f_\mathrm{in}(0, 0) = f_\mathrm{eq}(\tau'; 0 ,0 ) = 1 $ such that Eq.~(\ref{soln}) gives
\begin{equation}
\label{f_0_0}
    f(\tau; 0, 0) = D(\tau, \tau_0) + \int_{\tau_0}^{\tau} \frac{d\tau’}{\tau_R (\tau')} \, D(\tau, \tau') .
\end{equation}
With $\partial D(\tau,\tau')/\partial \tau' = D(\tau,\tau')/\tau_R(\tau')$ it follows that for a massless gas $f(\tau;0,0){\,=\,}D(\tau,\tau){\,=\,}1$ at all times. The same does not remain true for $m\ne 0$; this explains the different behavior near $p/T=0$ of the solid and dashed lines in Figs.~\ref{fig:f_momenta}a,b.

In Fig.~\ref{fig:f_momenta}c we plot the ratio of the distribution functions taken along the transverse and longitudinal momentum slices: $f(\tau; p_T{=}0, p_z{=}p)/ f(\tau; p_T{=}p, p_z{=}0)$. By definition, this ratio always equals 1 at $p/T = 0$; for massless particles it has a cusp at $p/T=0$ whereas for massive particles its slope vanishes at $p/T=0$. For a static thermalized system this ratio would be 1 at {\it all} values of $p/T$. We see that the latter limit is never reached in a system undergoing Bjorken expansion: While the ratio decreases less steeply with increasing $p/T$ as time increases, it never becomes flat as expected in static equilibrium.\footnote{%
    Apparently contradictory results shown in Fig.~11 of Ref.~\cite{Strickland:2018ayk} were caused by the coding error discussed in footnote~\ref{fn6}.}
%

\vspace*{-.2cm}
\section{Relativistic dissipative hydrodynamics in Bjorken flow}
\label{sec:hydro}
\vspace*{-.2cm}

In this section, we shall focus on a hydrodynamic description of the non-conformal boost-invariant system we have been studying thus far, with the goal of comparing the hydrodynamic evolution with the underlying kinetic theory. The hydrodynamic equations that we use consist of terms up to second-order in velocity gradients and have been obtained by solving the RTA Boltzmann equation approximately in a Chapman-Enskog (CE) like iterative series around local equilibrium \cite{Chapman:1970}. The corresponding energy-momentum conservation equations together with bulk and shear evolution equations of second-order CE hydrodynamics for a system undergoing Bjorken expansion are \cite{Denicol:2014vaa, Jaiswal:2014isa}:
\begin{align}
\label{eng_bj}
    \frac{d\epsilon}{d\tau} &= -\frac{1}{\tau}\left(\epsilon + P + \Pi -\pi\right) \, ,  
\\ 
\label{bulk_bj}
    \frac{d\Pi}{d\tau} + \frac{\Pi}{\tau_\Pi} &= -\frac{\beta_\Pi}{\tau} - \delta_{\Pi\Pi}\frac{\Pi}{\tau}
    +\lambda_{\Pi\pi}\frac{\pi}{\tau} \, ,  
\\ 
\label{shear_bj}
    \frac{d\pi}{d\tau} + \frac{\pi}{\tau_\pi} &= \frac{4}{3}\frac{\beta_\pi}{\tau} - \left( \frac{1}{3}\tau_{\pi\pi}
    +\delta_{\pi\pi}\right)\frac{\pi}{\tau} + \frac{2}{3}\lambda_{\pi\Pi}\frac{\Pi}{\tau} \, . 
\end{align}
Here $\beta_\pi \equiv \eta/\tau_\pi, \, \beta_\Pi \equiv \zeta/\tau_\Pi$ are first-order transport coefficients related to the shear and bulk viscosities, and the second-order transport coefficients $\delta_{\Pi\Pi}, \,  \lambda_{\Pi\pi}, \,  \delta_{\pi\pi}, \tau_{\pi\pi}$ and $\lambda_{\pi\Pi}$ are unitless functions of temperature $T$ and mass $m$ as given in Ref. \cite{Jaiswal:2014isa}. We note  $\beta_\Pi = \frac{5}{3}\beta_\pi -(\epsilon+P)c_s^2$. The transport coefficients $(\eta, \zeta)$ and the squared speed of sound are given by Eqs.~(\ref{eta}-\ref{cs2}). Also, from the exact form of the  coefficients the following relations can be deduced:\footnote{%
    These relations between the transport coefficients hold irrespective of whether the hydrodynamic equations are obtained in the 14-moment \cite{Denicol:2014vaa} or the Chapman-Enskog approximation \cite{Jaiswal:2014isa}, even though the coefficients derived with the two methods differ.}
\begin{align}
\label{coeff_rel}
\delta_{\Pi\Pi}\,=\,& \frac{5}{6} \lambda_{\pi\Pi} -c_s^2 , \, \quad 
\lambda_{\Pi\pi}\,=\, \delta_{\pi\pi} - (1+c_s^2) \, ,
\nonumber \\
\tau_{\pi\pi}\,=\,& \frac{6}{7} (2 \delta_{\pi\pi} -1) \, . 
\end{align}
Since there are three relations between the five coefficients, only two of them are linearly independent. 

First-order Navier-Stokes (NS) theory is obtained by solving Eq.~(\ref{eng_bj}) together with the first-order constituent equations $\Pi=-\zeta/\tau$, $\pi=4\eta/(3\tau)$.

\vspace*{-2mm}
\subsection{Early-time dynamics}
\label{dim_eq}
\vspace*{-2mm}

In the following we shall focus on the early-time features of second-order hydrodynamics to explore the structure of fixed points and fixed lines, in analogy to the analysis performed for kinetic theory in the previous section. For easier comparison we rewrite Eqs.~(\ref{eng_bj}-\ref{shear_bj}) using dimensionless variables, $\PLbar \equiv \PL/P, \PTbar \equiv \PT/P$ and $z \equiv m/T$:
\begin{align}
\label{eq_z1}
    &\frac{dz}{d\tau} = - \frac{1}{\tau} \left( \frac{P}{d\epsilon/dz} \right)  \left( \frac{\epsilon}{P} + \PLbar \right),
\\ 
\label{eq_plbar1}
    &\frac{d\PLbar}{d\tau} + \frac{\PLbar{-}1}{\tau_R}
    = \frac{1}{\tau} \left( a_1 + a_2 \PLbar +c_s^2 \PLbar^2 + a_3 \PTbar \right)\!,
\\ 
\label{eq_ptbar1}
    &\frac{d\PTbar}{d\tau} +\frac{\PTbar{-}1}{\tau_R} 
    = \frac{1}{\tau} \left( b_1 + b_2 \PTbar + c_s^2 \PLbar \PTbar + b_3 \PLbar \right)\!.
\end{align}
Here $\epsilon = \frac{3 T^4}{\pi^2}\,\bigl(\frac{z^2}{2} K_2(z)+\frac{z^3}{6} K_1(z) \bigr)$ is the equilibrium energy density and $P=\frac{T^4}{\pi^2} \bigl(\frac{z^2}{2} K_2(z) \bigr)$ the thermal pressure. The coefficients $a_i$ and $b_i$ in the above equations are expressed below in terms of two independent second-order transport coefficients, $\tau_{\pi \pi}$ and $\lambda_{\pi\Pi}$, along with the first-order ones: 
{\small
\begin{align}
\label{coeff:01}
&a_1 = \frac{3}{2} \lambda_{\pi\Pi} - c_s^2 \left( 1 + \frac{\epsilon}{P} \right) -\frac{\beta_\Pi}{P} -\frac{4}{3} \frac{\beta_\pi}{P} \, ,
\nonumber \\
&a_2 = c_s^2 \frac{\epsilon}{P} - \tau_{\pi\pi} - \frac{\lambda_{\pi\Pi}}{2}  \, ,
\nonumber \\
&a_3 = \tau_{\pi\pi} - \lambda_{\pi\Pi} \, ,
\nonumber \\
&b_1 =  \frac{\lambda_{\pi\Pi}}{2} - c_s^2 \left( 1 + \frac{\epsilon}{P} \right) -\frac{\beta_\Pi}{P} +\frac{2}{3} \frac{\beta_\pi}{P} \, ,
\nonumber \\
&b_2 = c_s^2 \frac{\epsilon}{P} - \frac{1}{2} + \frac{\tau_{\pi\pi}}{12} - \frac{\lambda_{\pi\Pi}}{3} \, ,
\nonumber \\
&b_3 = \frac{1}{2} - \frac{\tau_{\pi\pi}}{12} - \frac{\lambda_{\pi\Pi}}{6} \, .
\end{align}
}

As we did for kinetic theory in Sec.~\ref{KT_fs}, we now analyze the early-time dynamics described by these equations in the limit $\tau/\tau_R\to0$, by setting the relaxation time $\tau_R$ to infinity. The equations can be further simplified by assuming $T \gg m$ as $\tau \to 0$ and taking the $z \to 0$ limit of the transport coefficients, $\tau_{\pi\pi} \to 10/7$, $\lambda_{\pi\Pi} \to 6/5$, $\beta_\Pi \to 0$ and $\beta_\pi \to 4P/5$, in  \eqref{coeff:01}. Eqs.~(\ref{eq_plbar1},\ref{eq_ptbar1}) then simplify to
\begin{align}
\label{plbar_limit1}
 \frac{d\PLbar}{d\tau} &=
	 \frac{1}{\tau} \!\left(\! -\frac{3}{5} -\frac{36}{35}\PLbar +\frac{1}{3} \PLbar^2 + \frac{8}{35}\PTbar \!\right) ,
 \\ \label{ptbar_limit1}
 \frac{d\PTbar}{d \tau} &=
	 \frac{1}{\tau} \!\left(\! -\frac{1}{5} +\frac{23}{105}\PTbar +\frac{1}{3} \PLbar \PTbar + \frac{19}{105}\PLbar \!\right) .
\end{align}
Note that taking the limit $z \to 0$ in the transport coefficients decouples the $\PLbar$ and $\PTbar$ equations from the $z$ evolution \eqref{eq_z1}. This is analogous to the conformal hydrodynamics of Bjorken flow, where the scaled time evolution of $\pi/P$ decouples completely from the corresponding temperature evolution.

Fixed points of Eqs.~(\ref{plbar_limit1},\ref{ptbar_limit1}) are obtained by setting the derivatives on the l.h.s. of the equations to zero:
\begin{align}
&-\frac{3}{5} -\frac{36}{35}\PLbar +\frac{1}{3} \PLbar^2 + \frac{8}{35}\PTbar  = 0, \\
&-\frac{1}{5} +\frac{23}{105}\PTbar +\frac{1}{3} \PLbar \PTbar + \frac{19}{105}\PLbar = 0.
\end{align}
The resulting cubic equations for $\PLbar$ and $\PTbar$ have three pairs of real solutions:
\begin{align}
\label{fp:non_conf}
\bigl\{\PLbar^*, \PTbar^*\bigr\} = \{-0.212,1.61\};\ 
                       \{3.64,-0.32\};\
                       \{-1, -3.33\},
\end{align}
corresponding to
\begin{align}\label{fp:non_conf_2}
    \left\{\frac{\Pi^*}{P}, \frac{\pi^*}{P}\right\} 
    = \{0,1.214\};\  \{0,-2.64\};\  \{-3.56,-1.56\}.
\end{align}
Recall that in kinetic theory we had only one fixed point; here, for hydrodynamics in the $m/T \to 0$ limit, we obtain three of them. The first two fixed points are precisely the longitudinal and transverse free-streaming fixed points of second-order {\it conformal} hydrodynamics \cite{Denicol:2012cn, Jaiswal:2019cju}; they arise from taking the $z \to 0$ limit of the transport coefficients. In this limit the first-order (Navier-Stokes) term $\beta_\Pi$ and bulk-shear-coupling coefficient $\lambda_{\Pi \pi}$ in the equation (\ref{bulk_bj}) for the bulk viscous pressure both drop out, and the equation for $\Pi$ thus has the solution $\Pi = 0$, telling us that the dynamics remains conformal \textit{at all times}. 

The three hydrodynamic fixed points are crude approximations of the coordinates of the three corners of the kinetically allowed region in the $(\Pi/P,\pi/P)$ plane shown in Fig.~\ref{CM_fs}. The third fixed point is absent in conformal hydrodynamics and thus is a feature of {\it non-conformal} hydrodynamics. We show below (see Fig.~\ref{fig:streamplot}) that this third fixed point exhibits attractor behavior while the second one is an unstable fixed point (or ``repellor'') and the first one acts as a saddle point. We emphasize that all three hydrodynamic fixed points lie {\it outside the bounds} of the region allowed by kinetic theory. The first two fixed points\footnote{%
    For the conformal case these agree exactly with those obtained from second-order hydrodynamics: $\pi/P \simeq 1.214$ and ${-}2.64$.}
break the bound on the scaled {\it shear} stress\footnote{%
    This is similar to what happens in the conformal case where second-order hydrodynamics yields stable and unstable fixed points at $\pi^*/(\epsilon{+}P) \simeq 0.3$ and ${-}0.66$, respectively, whereas positivity of both $\PL^*$ and $\PT^*$ requires $\pi^*/(\epsilon{+}P)$ to lie in the interval $[-0.5,0.25]$.}
while the third one (corresponding to $\frac{\Pi^*}{P} = -3.56, \frac{\pi^*}{P} = -1.56$) breaks the bound $\frac{\Pi}{P}\geq -1$ for the scaled {\it bulk} viscous pressure. 

We next discuss the convergent resp. divergent evolution of initial conditions near the various fixed points.

\vspace*{-2mm}
\subsection{Convergence of initial conditions in the free streaming regime}
\label{sec:early_conv}
\vspace*{-2mm}

To explore the existence of convergent dynamics
at early times ($z \approx 0, \tau/\tau_R \ll 1$) we define deviations from the fixed points,
\begin{equation}
\delta_L = \PLbar - \PLbar^* \, , \qquad \delta_T = \PTbar - \PTbar^* \, ,
\end{equation}
where $\PLbar^*$ and $\PTbar^*$ represent any of the three fixed point values in Eq.~(\ref{fp:non_conf}).
With these substitutions, and ignoring terms that are nonlinear in $\delta_{L,T}$, Eqs.~(\ref{eq_plbar1},\ref{eq_ptbar1}) simplify in the free streaming regime $\taur \to \infty$ to
\begin{align}
\label{eq:deltaL}
\frac{d\delta_L}{d\tau}  &= \frac{1}{\tau}( q_1\, \delta_L + a_3\, \delta_T) ,
\\ \label{eq:deltaT}
\frac{d\delta_T}{d\tau}  &= \frac{1}{\tau}( q_2\, \delta_T + q_3 \, \delta_L) ,
\end{align}
where the coefficients $q_1\,,\,q_2\,,\,q_3$ are given by
\begin{align}
q_1 &= a_2 + 2 c_s^2\, \PLbar^* ,
\nonumber \\
q_2 &= b_2 + c_s^2\,\PLbar^* ,
\nonumber \\ 
q_3 &= b_3 + c_s^2\,\PTbar^* .
\end{align}
The solutions of Eqs.~(\ref{eq:deltaL},\ref{eq:deltaT}) are:
\begin{align}
\delta_L &= c_1\, \tau^{(q_1+q_2+q_4)/2} + c_2\, \tau^{(q_1+q_2-q_4)/2} ,
\\
\delta_T &= c_1 \left(\frac{q_2-q_1+q_4}{2a_3}\right) \tau^{(q_1+q_2+q_4)/2} 
    \nonumber \\ 
    & \qquad + c_2 \left(\frac{q_2-q_1-q_4}{2a_3}\right) \tau^{(q_1+q_2-q_4)/2} .
\end{align}
where $q_4\,=\,\sqrt{(q_1-q_2)^2\,+\,4a_3\,q_3}$ and $c_1, c_2$ are integration constants specified by the initial conditions for $\delta_L$ and $\delta_T$. The solutions corresponding to the three sets of fixed points obtained in \eqref{fp:non_conf} (in $z\to 0$ limit) are
\begin{align}
\label{delta_sol}
\{-0.212,1.61 \} :& \quad \delta_{L} = c_1\, \tau^{0.26} + c_2\, \tau^{-1.28}  \nonumber \\ 
 &  \quad  \delta_{T} = 6.27\, c_1\, \tau^{0.26} - 0.5\, c_2\, \tau^{-1.28} 
\nonumber \\
\{3.64,  -0.32 \} :& \quad \delta_{L} = c_1\, \tau^{1.55} + c_2\, \tau^{1.28}  \\ 
 &  \quad  \delta_{T} =  0.65\, c_1\, \tau^{1.55} + -0.5\, c_2\, \tau^{1.28}
\nonumber \\    
\{-1, -3.33 \} :& \quad \delta_{L} =  c_1\, \tau^{-0.26} + c_2\, \tau^{-1.55}  \nonumber \\ 
 &  \quad  \delta_{T} =6.27\, c_1\, \tau^{-0.26} + 0.65\, c_2\, \tau^{-1.55}  \nonumber
\end{align}

From the first solution in Eqs.~\eqref{delta_sol} it is clear that the first fixed point, $\{\PL^*,\PT^*\}=\{-0.212,1.61\}$, behaves as an attractor only for initial deviations that satisfy $c_1 = 0$. For all other initial conditions it acts as a repellor for both $\delta_L$ and $\delta_T$, as the growth term with coefficient $c_1$ becomes dominant with time.\footnote{%
    Note that the growth and decay rates differ by a factor 5.} 
We shall accordingly label the first fixed point as a saddle point. -- The second solution in Eqs.~\eqref{delta_sol} has all trajectories diverging from the second fixed point $\{3.64,-0.32\}$, unless initialised with exactly $c_1 = c_2 = 0$. This identifies the second fixed point as a repellor. -- The third solution corresponds to the new fixed point $\{-1, -3.33\}$ that is absent in conformal hydrodynamics. Sufficiently small deviations from this fixed point all decay with a superposition of two power laws whose powers differ by a factor $\sim 6$. This third fixed point (whose position has been calculated above in the small-mass limit  $z{\,\to\,}0$ for the transport coefficients) thus is an attracting fixed point of non-conformal hydrodynamics.

\begin{figure}[!t]
\begin{center}
 \includegraphics[width=.85\linewidth]{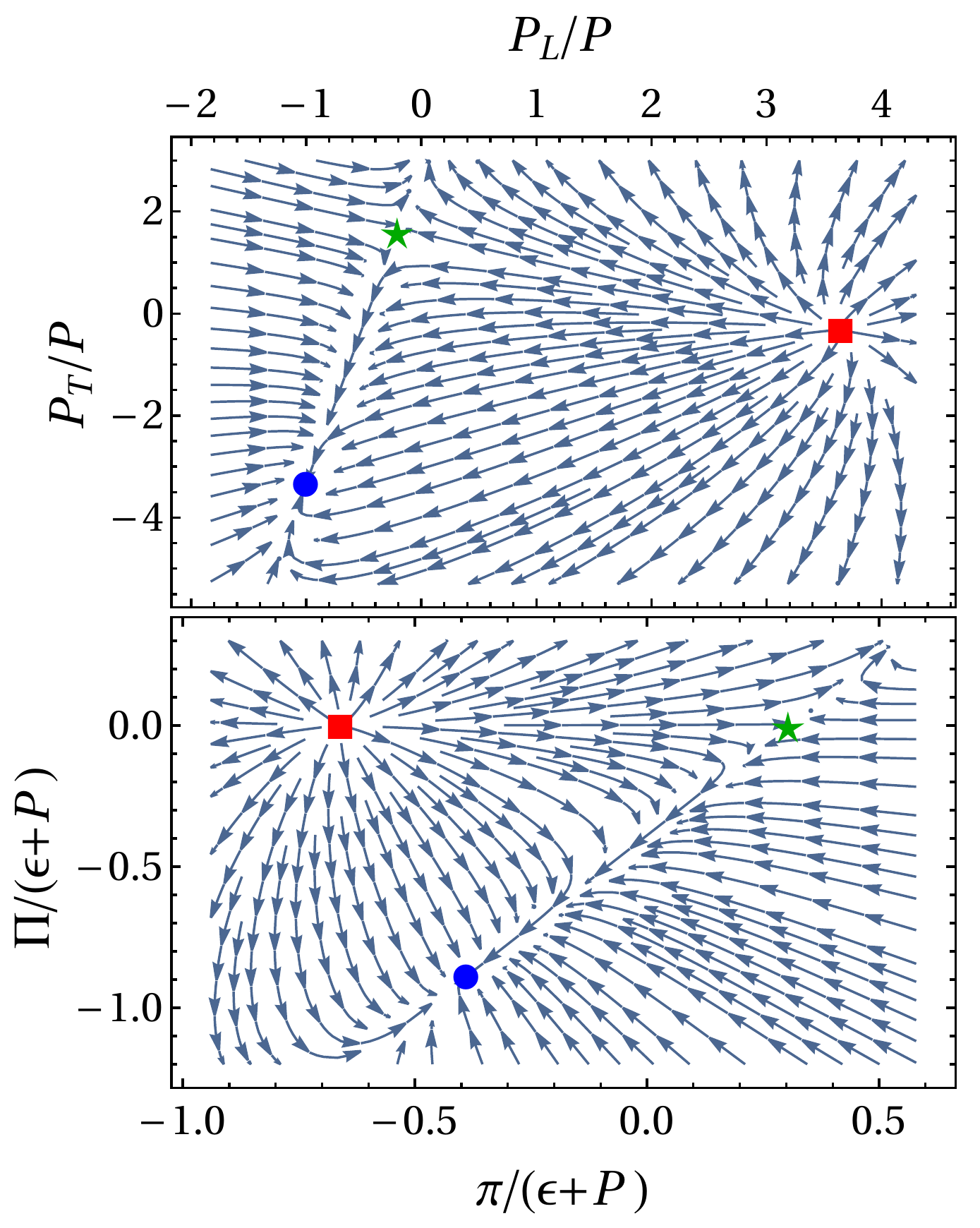}
\end{center}
\vspace*{-6mm}
  \caption{
  Stream lines showing the behavior of Eqs.~(\ref{plbar_limit1},\ref{ptbar_limit1}) (i.e. for transport coefficients evaluated in the $z{\,\to\,}0$ limit) near the first (green star), second (red square) and third (blue circle) of the three fixed points in Eqs.~(\ref{fp:non_conf},\ref{fp:non_conf_2}). The upper and lower panels correspond, respectively, to streamlines in $\PTbar-\PLbar$ and $\bPi-\bpi$ planes.
  \vspace*{-2mm}
  }
\label{fig:streamplot}
\end{figure} 

The upper panel of figure~\ref{fig:streamplot} illustrates the dynamical evolution of initial conditions in the neighborhood of these fixed points by showing streamlines of the trajectories of $\PLbar$ and $\PTbar$ obtained by solving Eqs.~(\ref{plbar_limit1}, \ref{ptbar_limit1}) using $\ln(\tau/\tau_0)$ as the time variable. The red square corresponds to the repelling fixed point, $\{\PLbar^* = 3.64, \PTbar^* = -0.32\}$. All trajectories starting in its vicinity are repelled. The green star is the saddle fixed point, $\{\PLbar^* = -0.212, \PTbar^* = 1.61\}$. Trajectories that lie exactly on the line joining the red and green points eventually merge with the green star. All other trajectories, however, are repelled by the green star. Trajectories that are repelled downward to the left from the green star eventually merge with the attracting fixed point at $\{\PLbar^* = -1, \PTbar^* = -3.33\}$, denoted by a blue circle. Before reaching this point, trajectories are attracted to a line joining the green star and blue circle. This line corresponds to the attracting fixed line of second-order hydrodynamics with transport coefficients evaluated in the massless limit. The line joining the red square and green star corresponds to the conformal condition $\Pi/P\equiv 0$ and is thus another fixed line of hydrodynamics. 

\begin{figure}[t]
\begin{center}
 \includegraphics[width=.85\linewidth]{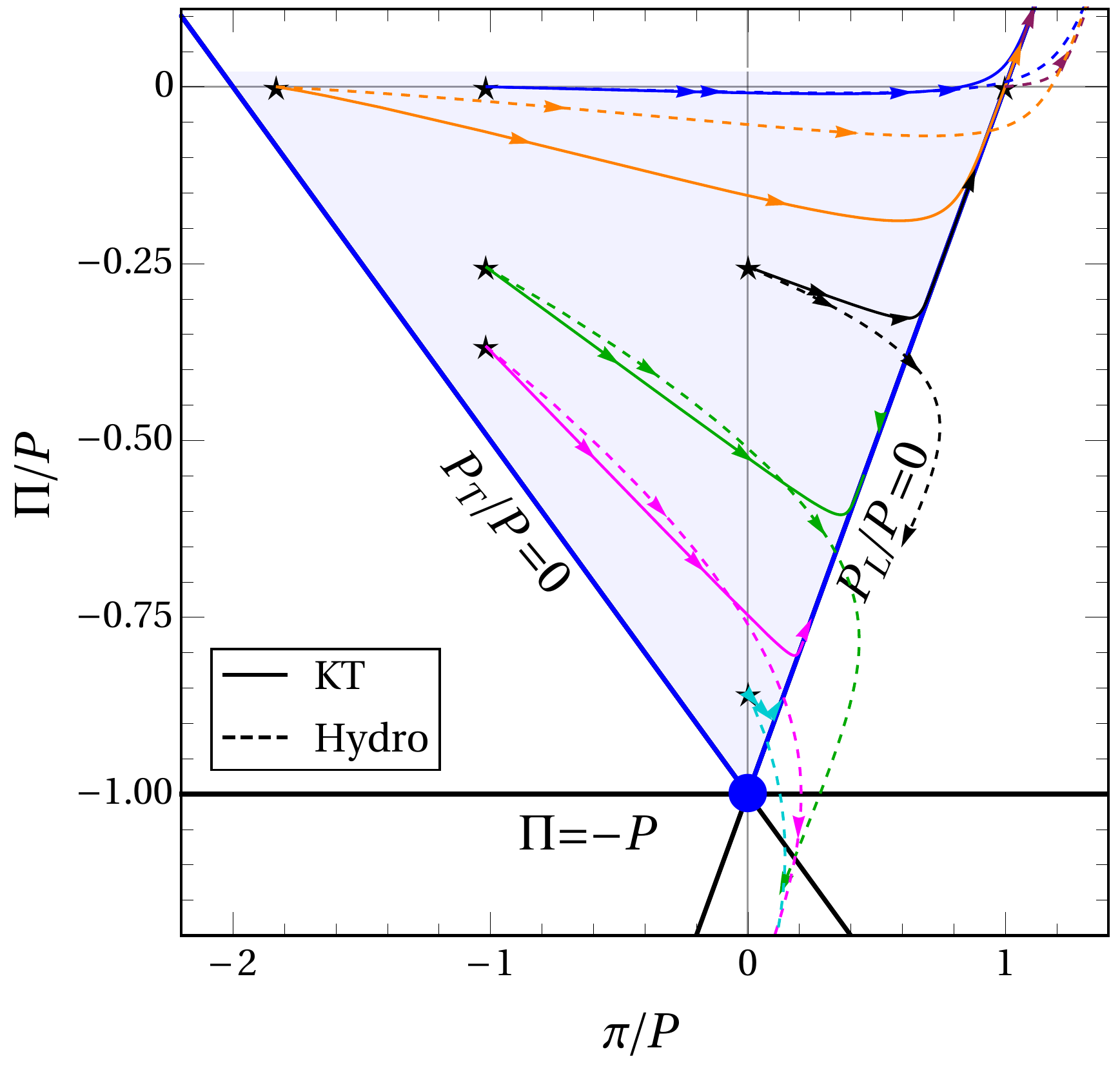}
\end{center}
\vspace*{-6mm}
  \caption{
  Comparison between the evolution of various free-streaming solutions of kinetic theory (solid lines) and hydrodynamics (dashed lines) for non-zero particle mass $m{\,=\,}0.2$\,GeV.}
  \vspace*{-2mm}  
  \label{Fig_kt_hydro_fs:1}
\end{figure}

To facilitate later discussions of Figs.~\ref{Fig_kt_hydro_fs:1}-\ref{fig:attrdisplay2}, we show in the lower panel of Fig.~\ref{fig:streamplot} the same streamlines of trajectories and the fixed points in the space $(\bpi,\bPi) \equiv \bigl(\pi/(\epsilon{+}P), \Pi/(\epsilon{+}P)\bigr)$. The red square $(\bpi^* \approx -0.66, \bPi^* = 0)$ and the blue circle $(\bpi^* \approx - 0.39, \bPi^* \approx -0.89)$ behave, respectively, as repulsive and attracting fixed points, whereas the green star $(\bpi^* \approx 0.3, \bPi^* = 0)$ is a saddle point. The evolution of linearized perturbations, $(\delta_{\bpi}, \delta_{\bPi})$, around a given fixed point can be simply obtained from Eqs.~(\ref{delta_sol}) by choosing the corresponding fixed point and then using $\delta_{\bpi} = (\delta_T{-}\delta_L)/6$ and $\delta_{\bPi} =  (\delta_L{+}2\delta_T)/12$.\footnote{%
    The resulting growth and decay rates are the same for either choice of variables.} 

In Fig.~\ref{Fig_kt_hydro_fs:1} we compare the free-streaming evolution for $\Pi/P$ and $\pi/P$ using kinetic theory and hydrodynamics for non-conformal systems. The solid lines from kinetic theory are copied from Fig.~\ref{CM_fs}. For the dashed hydrodynamic trajectories we solve Eqs.~(\ref{eq_z1}-\ref{eq_ptbar1}) numerically, by setting $\tau_R \to \infty$ but without taking the massless limit for the transport coefficients. We observe that, contrary to the massless case $m/T{\,=\,}0$, no fixed lines or fixed points are found for hydrodynamics. Both hydrodynamics and kinetic theory start by evolving initial conditions towards the right; but whereas the solid lines from kinetic theory eventually settle on the $\PL{\,=\,}0$ fixed line of zero longitudinal pressure, following it towards the upper right, the hydrodynamic curves cross this line into the kinetically forbidden region of negative longitudinal pressure, eventually bending left or right and following a direction roughly parallel to the $\PL{\,=\,}0$ line. The dashed lines bending downward are attracted by a (temperature dependent) attractive ``fixed'' point which itself presents a moving target as the temperature decreases. In the massless limit $z=0$ the hydrodynamic trajectories would lie on a fixed line obtained by joining the attractor and saddle fixed points. Note that, even in the $z\to 0$ case where an attracting fixed line exists in hydrodynamics, this line would lie entirely outside the kinetically allowed region of positive longitudinal and transverse pressures.

We remind the reader that Figs.~\ref{fig:streamplot}, \ref{Fig_kt_hydro_fs:1} were obtained by setting $\taur\to\infty$ and thus correspond to the expansion-dominated, effectively free-streaming dynamics at very early times. We proceed to the onset of thermalizing dynamics in the following subsection where we study the modifications arising from finite relaxation times.

\vspace*{-2mm}
\subsection{Hydrodynamics with finite $\tau_R$}
\label{sec3c}
\vspace*{-2mm}

We now include interaction effects in the hydrodynamic equations, via a finite relaxation time, $\tau_R = 5 C/T$, with $C = 10/4\pi$ as before. We discuss the evolution as a function of the scaled proper time $\bar\tau \equiv \tau/\tau_R$ (a.k.a. the inverse Knudsen number) such that Eqs.~(\ref{eq_z1}-\ref{eq_ptbar1}) take the form
\begin{align}
\label{eq_z}
    \left( k_1 + k_2 \PLbar \right) \frac{dz}{d\btau} &= - \frac{z \, k_2}{\btau}  \left( \frac{\epsilon}{P} + \PLbar \right),
\\ 
\label{eq_plbar}
    \left( k_1 + k_2 \PLbar \right) \frac{d\PLbar}{d\btau} 
    &+ \left(\PLbar{-}1 \right) 
\\\nonumber
    &= \frac{1}{\btau} \left( a_1 + a_2 \PLbar +c_s^2 \PLbar^2 + a_3 \PTbar \right),
\end{align}
\begin{align}
\label{eq_ptbar}
    \left( k_1 + k_2 \PLbar \right) \frac{d\PTbar}{d\btau} 
    &+ \left( \PTbar{-}1 \right) 
\\\nonumber
    &= \frac{1}{\btau} \left( b_1 + b_2 \PTbar + c_s^2 \PLbar \PTbar + b_3 \PLbar \right),
\end{align}
with Jacobian coefficients
\begin{equation}
    k_1 = 1+ \frac{c_s^2}{z} \frac{\epsilon}{dP/dz}\, , \qquad 
k_2 = \frac{c_s^2}{z} \frac{P}{dP/dz}.
\end{equation}
Before comparing results from these hydrodynamic equations with those from kinetic theory in full generality we discuss how a finite $\taur$ affects the evolution of the inverse shear and bulk Reynolds numbers, $\bpi \equiv \pi/(\epsilon{+}P)$ and $\bPi \equiv \Pi/(\epsilon{+}P)$, in the limit of vanishing $m/T$ for the transport coefficients that was also studied in Sec.~\ref{dim_eq}.

\begin{figure}[t]
\begin{center}
\hspace{-5mm}
\includegraphics[width=.9\linewidth]{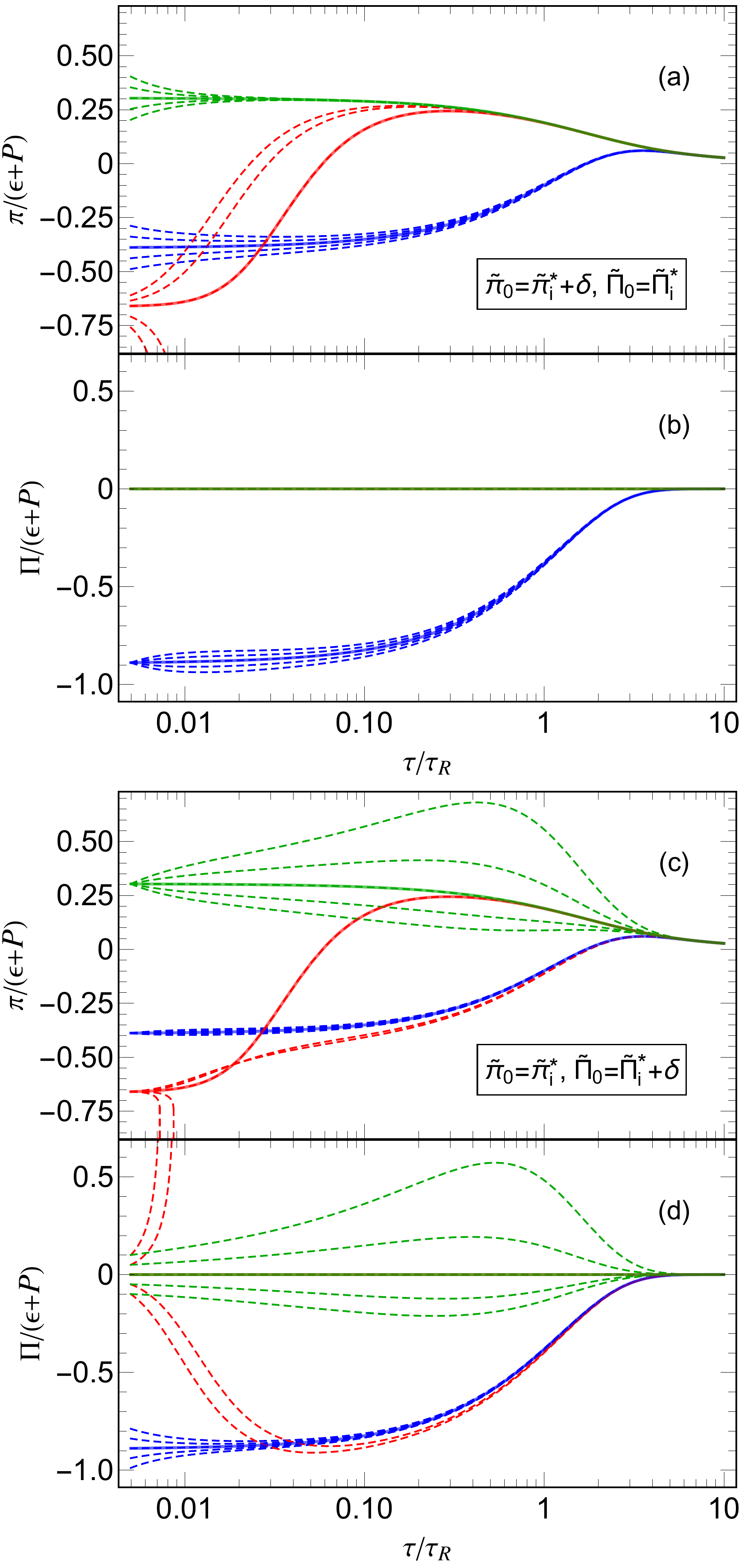} 
\end{center}
\vspace*{-6mm}
\caption{%
    Scaled time evolution of the normalised shear and bulk stresses $\bpi=\pi/(\epsilon{+}P)$ and $\bPi=\Pi/(\epsilon{+}P)$, using transport coefficients evaluated in the massless limit $z\to0$, initialised close to the three fixed points; see text for discussion.
    \label{fig:attrdisplay2}
    \vspace*{-2mm}}
\end{figure}

Figure~\ref{fig:attrdisplay2} shows solutions for $\bpi$ and $\bPi$ for a variety of initial conditions close to the three fixed points $\{ \bPi, \bpi \} = \{0,0.3\},\{-0.89,-0.39\},\{0,-0.66\}$ corresponding to Eqs.~(\ref{fp:non_conf},\ref{fp:non_conf_2}). In Figs.~\ref{fig:attrdisplay2}a,b we explore different initial conditions for $\bpi$ while always starting $\bPi$ from its corresponding fixed points, i.e. $0$, $-0.89$, and $0$, respectively. The green and blue dashed lines for $\bpi$ in Fig.~\ref{fig:attrdisplay2}a are seem to converge to the corresponding solid lines, whereas the red dashed lines diverge. In Fig.~\ref{fig:attrdisplay2}b the green and red dashed lines for $\bPi$ (which lie on top of each other) start with identical initial conditions, $\bPi = 0$, and stay there (since we used transport coefficients evaluated in the massless limit), whereas the blue dashed lines start from $\bPi = -0.89$. The blue dashed lines are seen to get initially repelled from the corresponding blue solid line, as a result of bulk-shear coupling, before merging with it after $\btau \approx 1$.

In Figs.~\ref{fig:attrdisplay2}c,d we switch the roles of $\bpi$ and $\bPi$ in the initial conditions: we vary the initial values for $\bPi$ while starting the normalized shear stress $\bpi$ at one of its fixed point values: green for $\bpi^*{\,=\,}0.3$, blue for  $\bpi^*{\,=\,}-0.39$, and red $\bpi^*{\,=\,}-0.66$. The green dashed lines in Fig.~\ref{fig:attrdisplay2}c show that that shear stress trajectories starting from the same fixed point $\bpi^*$ with different initial values for $\bPi$ diverge initially as a result of bulk-shear coupling; at late times $\btau > 4$ they converge again and join the Navier-Stokes attractor. For the blue curves the normalised shear $\bpi$ always stays close to the blue solid line which also joins the Navier-Stokes attractor at $\btau > 4$. For the red curves we observe a strong tendency for divergence at early times: the solid red line for $\bpi^*{\,=\,}-0.66$ shows that, when the system is initialized with zero bulk viscous pressure, the normalized shear stress very quickly (i.e. after $\btau\lesssim 0.5$) joins the corresponding green curve for $\bpi^*{\,=\,}0.3$ while for different initial bulk viscous pressures $\bPi\ne0$ they either join the blue trajectory long before both converge to the Navier-Stokes attractor, or they move to very large negative $\bpi$ values (moving even ``backwards'' in the inverse Knudsen number $\btau$). Looking at the evolution of the bulk viscous pressure $\bPi$ Fig.~\ref{fig:attrdisplay2}d, the divergence of the red curves emerging from the unstable red fixed point is even more obvious, with the zero bulk pressure curve as a singular case, remaining at zero at all times, while for negative initial bulk pressures $\bPi$ immediately crosses over to and merges at $\btau \ll 1$ with the blue set of curves, and for positive initial values it diverges to $+\infty$. 

These figures confirm what we previously obtained in the preceding subsection using a linearised analysis: for hydrodynamics in the $m/T \to 0$ limit, the red fixed point acts as a repellor, the blue fixed point behaves as an attractor, whereas the green one is a saddle point.

\vspace*{-2mm}
\section{Kinetic theory vs. Hydrodynamics}
\label{sec:hydrovskt}
\vspace*{-2mm}

\begin{figure}[b!]
\begin{center}
 \includegraphics[width=.85\linewidth]{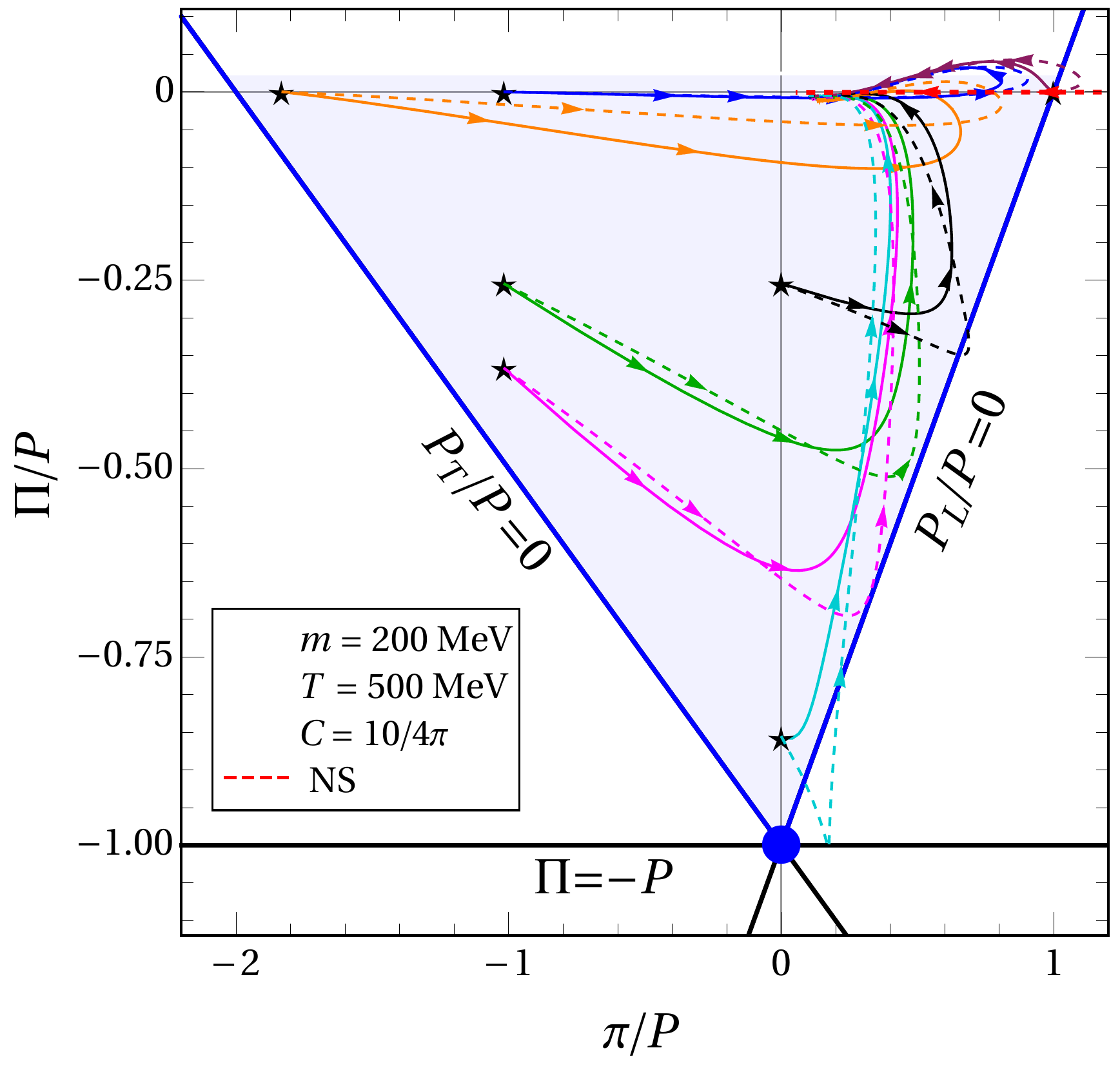}
\end{center}
\vspace*{-6mm}
  \caption{
  Comparison between evolution of various solutions of exact kinetic equation (solid lines) and hydrodynamic equations (dashed lines).
  \vspace*{-4mm}  }
  \label{Fig_kt_hydro:1}
\end{figure} 

\begin{figure}[t]
\begin{center}
 \includegraphics[width=.9\linewidth]{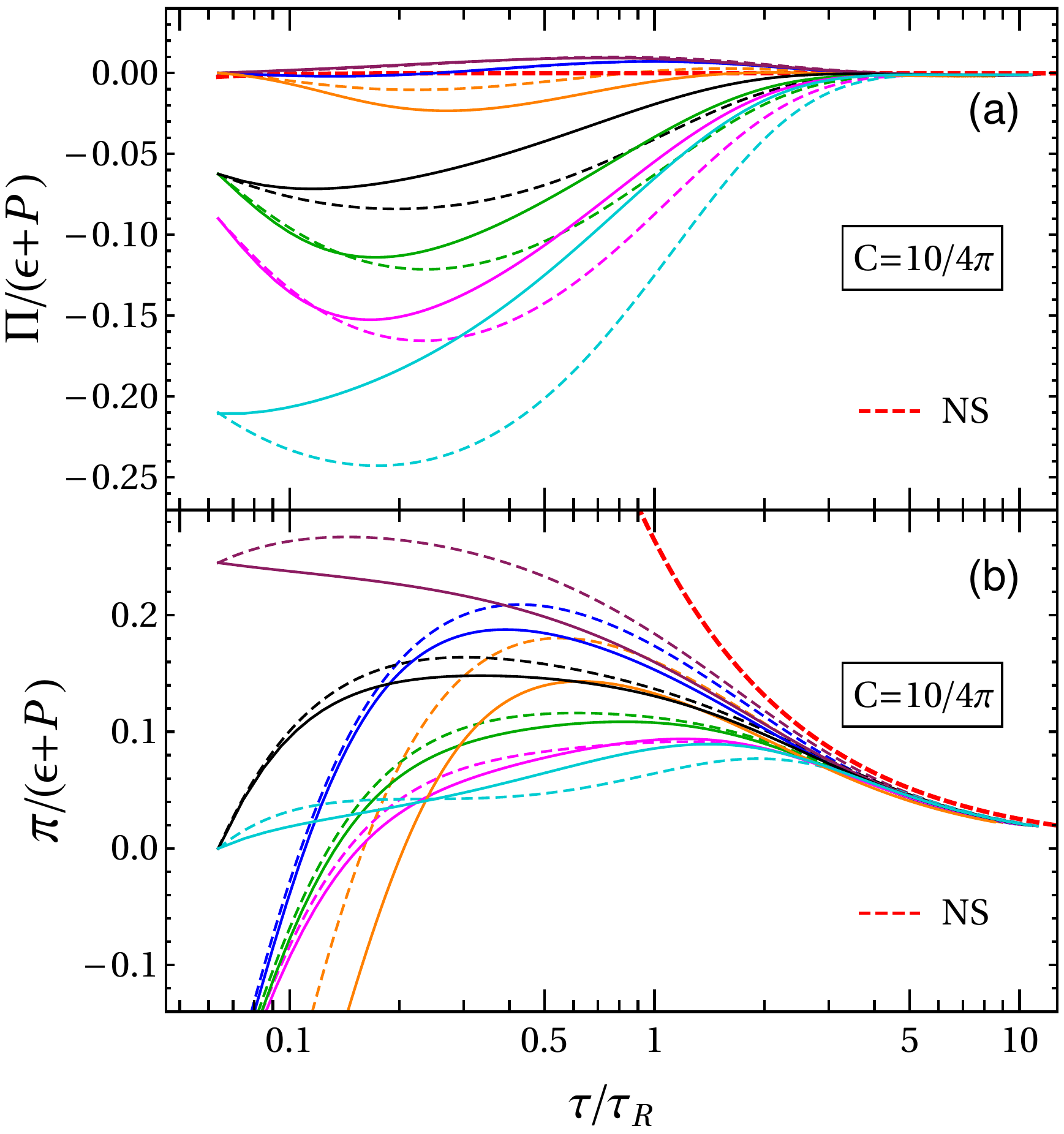}
\end{center}
\vspace*{-6mm}
  \caption{
  Comparison between hydrodynamic (dashed lines) and kinetic theory (solid lines) evolution of (a) scaled bulk viscous pressure and (b) scaled shear stress as a function of scaled time.
  \vspace*{-2mm}}
  \label{Fig_kt_hydro:2}
\end{figure} 

We now proceed to a numerical comparison of the hydrodynamic approach defined by Eqs.~(\ref{eq_z})-(\ref{eq_ptbar}) with  the results from kinetic theory obtained in Sec.~\ref{sec:kt}, for non-conformal systems outside the limits studied in the preceding section. Fig.~\ref{Fig_kt_hydro:1} shows the evolution of the scaled viscous pressures in the $(\Pi/P, \pi/P)$ plane. The solid lines from kinetic theory are the same as in Fig.~\ref{CM_exact}. They are here compared with hydrodynamic results for the same initial conditions, shown as dashed lines. Qualitatively, the hydrodynamic trajectories match those from kinetic theory. All curves starting with $\pi/P \leq 0$ initially move right, driven by rapid longitudinal expansion. The magenta and green curves come very close to the bound $P_L \geq 0$ whereas the cyan, black and maroon curves break the bound, moving into the kinetically forbidden region, before the onset of thermalizing interactions causes the trajectories to turn around and re-enter the kinetically allowed region. This is because the early-time (or non-interacting) hydrodynamic attracting fixed point and the saddle point corresponding to longitudinal kinetic free streaming (which for massless systems are located at $\{\Pi/P = - 3.56, \pi/P = - 1.56\}$ and $\{\Pi/P = 0, \pi/P = 1.214\}$, respectively) both violate the kinetic theory bounds. The cyan dashed curve, which is initialized close to the attracting (third) fixed point, exhibits the strongest violation of the allowed bound. This can be understood from Eq.~\eqref{delta_sol} where we see a power law decay of trajectories initialized near the third fixed point towards this fixed point (which lies outside the kinetically allowed region). For kinetic theory, the fixed point $\{\Pi/P = -1, \pi/P = 0\}$ is not an attracting one, and hence the solid cyan curve does not indicate any convergence towards this fixed point.

Figures~\ref{Fig_kt_hydro:2}a,b compare the (dashed) hydrodynamic and (solid) kinetic evolutions of the normalised bulk and shear stresses. In Fig.~\ref{Fig_kt_hydro:2}a, the magenta and green dashed lines follow the exact RTA curves for a while before deviating, whereas the cyan, black and orange dashed lines deviate from the exact solutions from the beginning. The shared feature between the magenta and green curves is that they are initially characterised by larger shear than bulk stress. For the cyan and black curves it is the opposite: they are initially driven predominantly by bulk stress. Although the orange trajectory has large scaled shear stress initially, its dynamics is initially rather close to the transverse free-streaming fixed point. As hydrodynamics does not accurately describe any of the fixed lines or fixed points that characterize the underlying kinetic theory in the limit $m/T\to 0$), it makes sense that differences with kinetic theory become particularly apparent whenever the system is driven close to these fixed lines/points. Note that all hydrodynamic trajectories merge with exact solutions around $\btau \approx 3$ as the thermal fixed point is approached, but this convergence appears to be delayed for the dashed hydrodynamic trajectories compared to the solid kinetic ones, especially for the dashed cyan and magenta curves which are initialized closest to the kinetic theory fixed point
(indicated by the solid blue dot in Fig~\ref{Fig_kt_hydro:1}).

\begin{figure}[t]
\begin{center}
 \includegraphics[width=.9\linewidth]{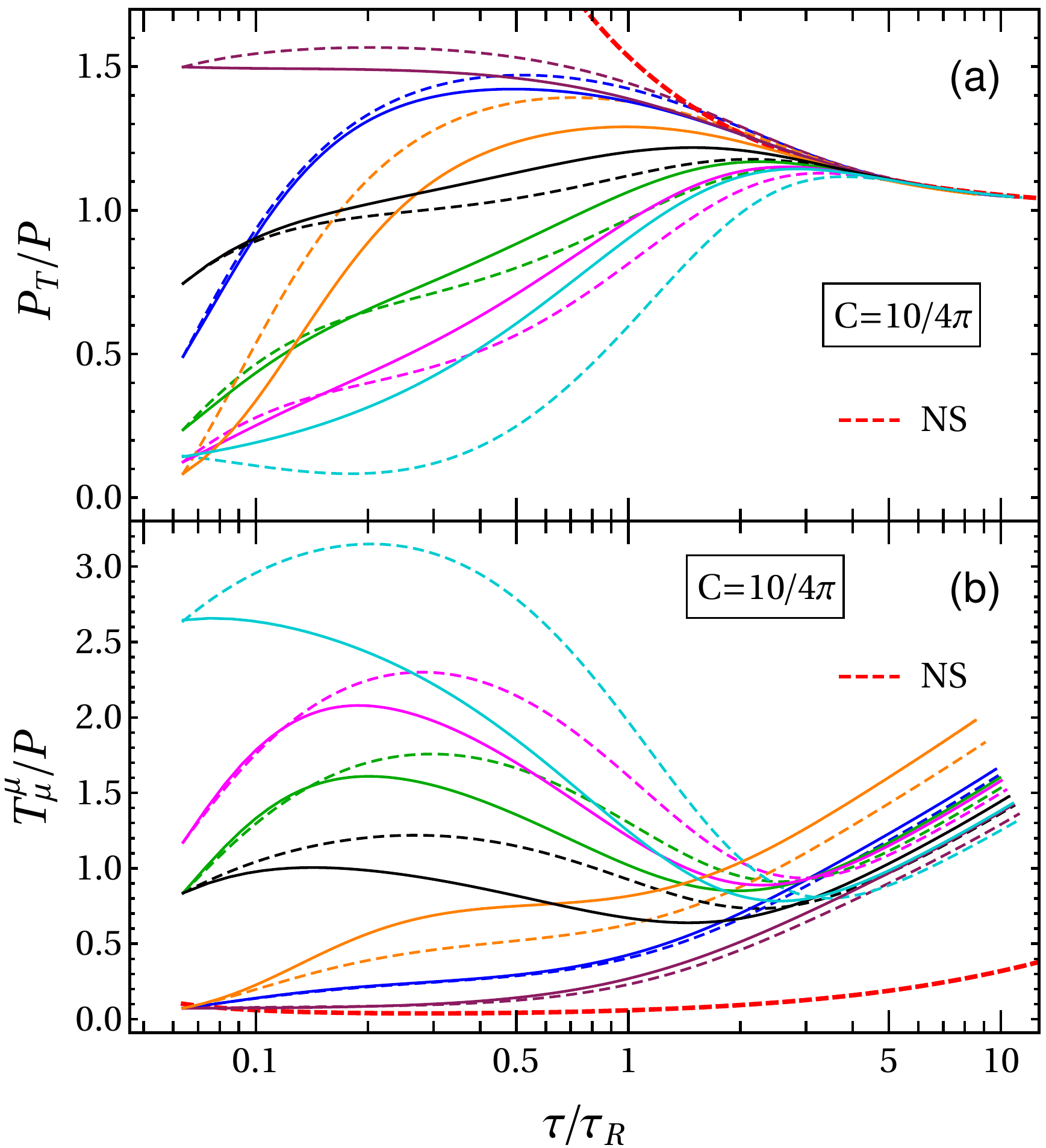}
\end{center}
\vspace*{-6mm}
  \caption{
  Comparison of hydrodynamic (dashed lines) and kinetic theory (solid lines) solutions for (a) scaled transverse pressure and (b) scaled trace of energy-momentum tensor.
  \vspace*{-2mm} }
  \label{Fig_kt_hydro:3}
\end{figure} 

For the normalized shear stress, shown in Fig.~\ref{Fig_kt_hydro:2}b, the largest differences between the hydrodynamic (dashed) and kinetic (solid) evolution are observed for the maroon and orange curves. These shear stress trajectories start at initial values particularly close to the two free-streaming fixed lines (i.e. the lines of vanishing transverse and longitudinal pressure). However, all the remaining dashed curves eventually show deviations from the exact solutions. After $\btau \sim 3$, the hydrodynamic, kinetic theory and Navier-Stokes results are indistinguishable.  

\begin{figure}[t!]
\begin{center}
 \includegraphics[width=0.9\linewidth]{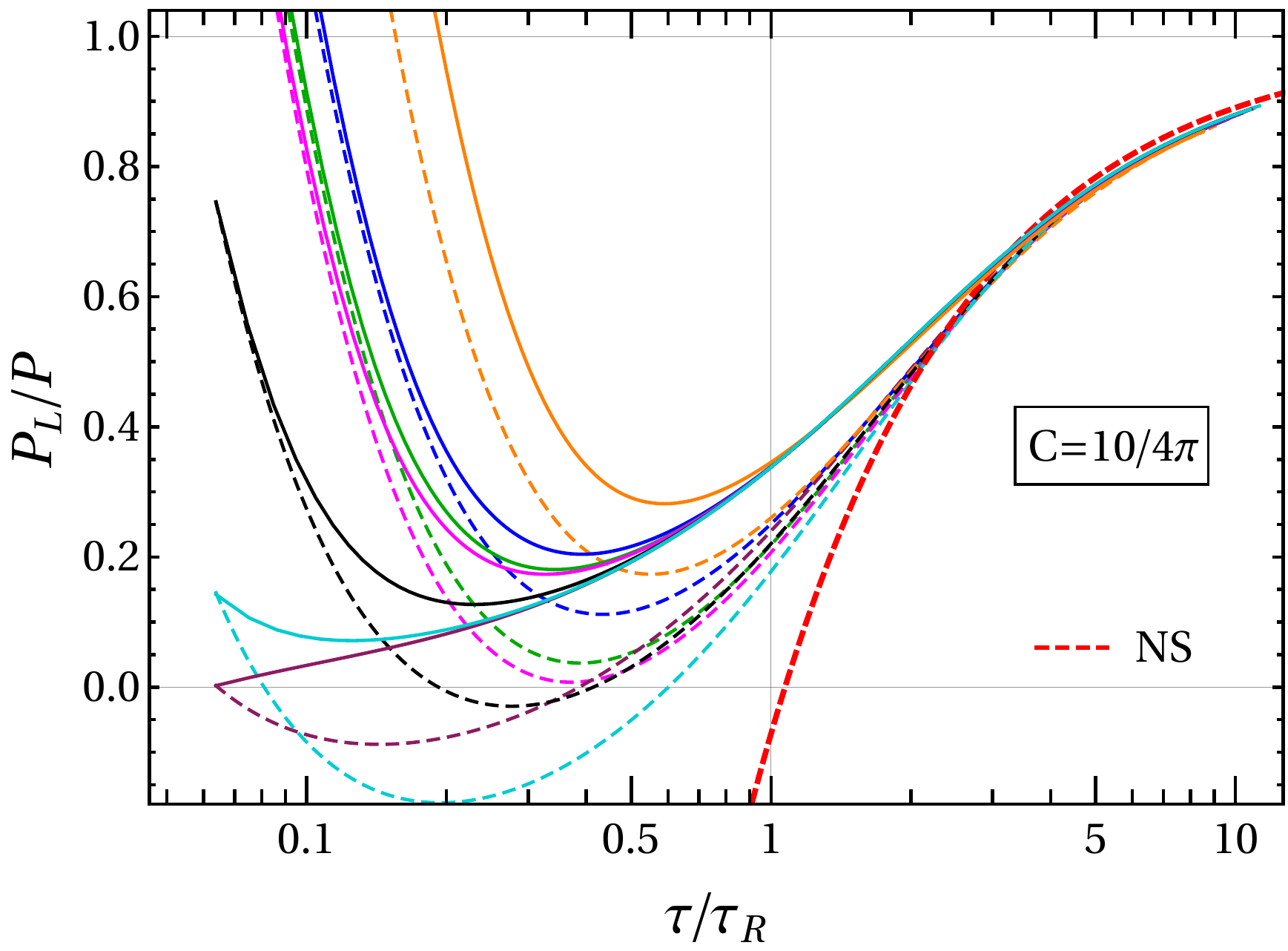}
\end{center}
\vspace*{-6mm}
  \caption{
 Evolution of scaled longitudinal pressure in hydrodynamics (dashed lines) and kinetic theory (solid lines).
  \vspace*{-2mm} }
  \label{Fig_kt_hydro:4}
\end{figure}

In Figs.~\ref{Fig_kt_hydro:3} and \ref{Fig_kt_hydro:4} we repackage the information contained in Figs.~\ref{Fig_kt_hydro:1} and \ref{Fig_kt_hydro:2} in terms of the transverse and longitudinal pressures and the trace anomaly of the energy momentum tensor (all scaled by the thermal pressure $P$). For the scaled transverse and longitudinal pressures, $\PT/P$ and $\PL/P$, the hydrodynamic curves (dashed) with large initial bulk (cyan) or shear stresses (maroon, orange) show again the largest deviations from kinetic theory (solid lines), whereas the agreement is slightly better for other initial conditions, at least at early times.  Convergence of the hydrodynamic trajectories with the exact kinetic theory curves occurs only after $\bar\tau \sim 3$ when both microscopic and macroscopic evolutions are well described by Navier-Stokes theory. The earlier convergence at $\bar\tau < 1$ of $\PL/P$ to a common attractor, that was observed for kinetic theory in Fig.~\ref{PL_P_KT} and more generally for both kinetic theory and hydrodynamics in conformal theories with zero bulk viscous pressure $\Pi=0$ \cite{Heller:2015dha, Romatschke:2017vte, Strickland:2017kux, Blaizot:2017ucy, Heller:2018qvh, Behtash:2018moe, Jaiswal:2019cju, Kurkela:2019set}, is not reproduced by the hydrodynamic approximation when conformal symmetry is broken, as illustrated in Fig.~\ref{Fig_kt_hydro:4}.

For the trace anomaly $T^\mu_\mu/P$, shown in Fig.~\ref{Fig_kt_hydro:3}b, the best agreement between hydrodynamics and kinetic theory is seen for the maroon and blue curves which are characterized by small values for the bulk viscous pressure $\bPi$ throughout the evolution. For the other curves, non-negligible bulk stresses induce significant non-zero $T^\mu_\mu$ values, and just as the evolution of $\bPi$ is not accurately captured by hydrodynamics, neither is that of $T^{\mu}_{\mu}$. At late times $T^\mu_\mu/P \approx \epsilon/P -3$ increases monotonically with time since the temperature decreases and thus the violation of conformal symmetry (characterized by the ratio $z{\,=\,}m/T$) increases. The different vertical offsets, combined with near-identical slopes, of the $T^{\mu}_{\mu}$ trajectories at large times can be understood in terms of different amounts of viscous heating at early times, caused by different initial conditions for the bulk and shear viscous stresses. This early-time production of additional entropy is not well captured by the Navier-Stokes (NS) solution which fails at early times by generating too much viscous heating. The late-time differences between the solid and dashed lines in Fig.~\ref{Fig_kt_hydro:3}b are rooted in the failure of the hydrodynamic models to correctly reproduce early-time entropy production in the kinetic theory, as previously discussed in Ref.~\cite{Chattopadhyay:2018apf} for a conformal gas but enhanced here by the nonzero bulk viscous pressure.

\vspace*{-2mm}
\section{Anisotropic hydrodynamics}
\label{aHydro}
\vspace*{-2mm}

The hydrodynamic equations introduced in Sec.~\ref{sec:hydro} and compared with the underlying kinetic theory in Sec.~\ref{sec:hydrovskt} were originally derived from the kinetic theory by expanding the phase-space distribution function $f(x,p)$ around a locally isotropic thermal equilibrium distribution (see, e.g., Ref.~\cite{Denicol:2012cn}). This locally isotropic leading-order distribution does not follow the rapid shrinking at early times of the $p_z$ distribution caused by the rapid longitudinal Bjorken expansion that was demonstrated in Sec.~\ref{sec2d}. As a result, dissipative corrections generated by deviations of the exact distribution function from its locally isotropic leading-order thermal equilibrium form grow large very quickly, and the hydrodynamic approach fails. In particular, this form of hydrodynamics cannot reproduce the early-time, far-off-equilibrium attractor for $\PL/P$ from kinetic theory shown in Fig.~\ref{PL_P_KT} \cite{Chattopadhyay:2021ive}.  

To address this shortcoming {\it anisotropic hydrodynamics} was introduced, in various degrees of refinement, in Refs.~\cite{Florkowski:2010cf, Martinez:2010sc, Martinez:2012tu, Bazow:2013ifa, Strickland:2014pga, Florkowski:2014bba, Molnar:2016vvu, Molnar:2016gwq, Alqahtani:2017mhy, McNelis:2018jho, Nopoush:2019vqc}. In this section we therefore explore to which extent anisotropic  hydrodynamics leads to an improved description of the underlying kinetic theory for the non-conformal systems presented in Sec.~\ref{sec:kt}.

The standard derivation of anisotropic hydrodynamics is based on an expansion of the distribution function around an ellipsoidally deformed leading-order distribution of Romatschke-Strickland form \cite{Romatschke:2003ms, Tinti:2015xwa} which for systems with Bjorken flow simplifies to
\begin{equation}
\label{f_a_postulate}
    f(\tau; p_T, w) \approx f_a \equiv \exp \left( - \frac{\sqrt{p_T^2/\alpha_T^2{+}(w/\tau)^2/\alpha_L^2{+}m^2}}{\Lambda} \right).
\end{equation}
Here the parameters $(\alpha_T, \alpha_L, \Lambda)$ are all functions of proper time. The deformation parameters $(\alpha_T, \alpha_L)$ allow $f_a$ to adjust its form to the longitudinally contracted form resulting from strong longitudinal Bjorken expansion at early times that is seen as the red core in the plots shown in the second column of Fig.~\ref{fig:Contour}. The parameters $(\alpha_T, \alpha_L, \Lambda)$ are Landau matched to the corresponding energy density $\epsilon$ and the transverse and longitudinal pressures $\PT$ and $\PL$ \cite{Tinti:2015xwa, McNelis:2018jho},
\begin{align}
    \epsilon = \left\langle (p^\tau)^2 \right\rangle_{a}, \quad
    \PL = \left\langle (w/\tau)^2 \right\rangle_{a}, \quad
    \PT = (1/2) \left\langle p_T^2 \right\rangle_{a},
\end{align}
where the expectation values are taken with the leading-order distribution (\ref{f_a_postulate}). These three quantities are also used (instead of the shear stress $\pi$ and the bulk viscous pressure $\Pi$ that we evolved in Sec.~\ref{sec:hydro} via Eqs.~(\ref{eng_bj})-(\ref{shear_bj})) as the hydrodynamic dynamical variables \cite{McNelis:2018jho}: 
\begin{align} 
    \frac{d\epsilon}{d\tau} &= -\frac{\epsilon + \PL}{\tau},
\label{e_sec_aH}
\\
    \frac{d\PL}{d\tau} &= - \frac{\PL - P}{\tau_R} + \frac{\bar{\zeta}^{L}_{z}}{\tau}, 
\label{PL_sec_aH}
\\
    \frac{d\PT}{d\tau} &= - \frac{\PT - P}{\tau_R} + \frac{\bar{\zeta}^{\perp}_{z}}{\tau}. 
\label{PT_sec_aH}
\end{align}
Here $\bar{\zeta}^{L}_{z}$ and $\bar{\zeta}^{\perp}_{z}$ are transport coefficients to which we will return shortly.

It turns out, however, that the Romatschke-Strickland parametrization (\ref{f_a_postulate}) of the leading-order anisotropic distribution is unable to accommodate the entire range of $\PT$ and $\PL$ values allowed according to Fig.~\ref{CM_fs} in kinetic theory for a weakly interacting gas of massive particles. In particular, it can never be matched to yield large negative values for the bulk viscous pressure $\Pi$, close to the lower corner in Fig.~\ref{CM_fs} for $m/T \ll 1$. The reason for this can be argued physically. As mentioned before, generating $\Pi/P \approx -1$ amounts to a distribution with almost vanishing isotropic pressure, such that $f \sim A \, \delta (|\bm{p}|)/|\bm{p}|^2$ where $A$ is a constant with dimensions of $[\mathrm{GeV}]^3$.\footnote{%
    The corresponding energy density is then $\epsilon = m \,A /(2 \pi^2)$.} 
Thus, to yield the given energy density $\epsilon$, the mean particle density at $|\bm{p}|{\,\approx\,}0$ must be greatly enhanced. With the ansatz (\ref{f_a_postulate}), however, the maximum number density at zero momenta is $f(\tau; 0, 0) = \exp(-m/\Lambda) \leq 1$. This prohibits $f_a$ from simultaneously generating large temperatures ($T \gg m$) and large negative bulk viscous pressures $\Pi/P\simeq-1$.\footnote{%
    For a more detailed analysis, please refer to Appendix \ref{RS_appendix}.}

This problem can be circumvented by replacing Eq.~(\ref{f_a_postulate}) by the following modified ansatz for the leading-order anisotropic distribution:
\begin{equation} 
\label{modifed_aHydro_ansatz}
    f \approx \tilde{f}_a = \frac{1}{\alpha(\tau)} \exp\left(- \frac{\sqrt{p_T^2  + (1+\xi(\tau)) w^2/\tau^2 + m^2 }}{\Lambda(\tau)} \right). 
\end{equation}
The reader will recognize it as the initial distribution given earlier in Eq.~(\ref{f_in}); here, however, the parameters $\alpha$, $\Lambda$, $\xi$ have been promoted to functions of proper time $\tau$.

Now we return to the transport coefficients $\bar{\zeta}^{L}_{z}$ and $\bar{\zeta}^{\perp}_{z}$ appearing in the evolution equations (\ref{e_sec_aH})-(\ref{PT_sec_aH}). Following the derivation of anisotropic hydrodynamics using the method of moments \cite{Molnar:2016gwq, Molnar:2016vvu} one finds
\begin{align}
\label{zeta_long}
   \bar{\zeta}^{L}_{z} = - 3 \PL + I^{\mathrm{exact}}_{240},
\\
\label{zeta_perp}
   \bar{\zeta}^{\perp}_{z} = - \PT + I^{\mathrm{exact}}_{221},
\end{align}
where $I^{\mathrm{exact}}_{nrq}$ are higher-order (non-hydrodynamic) moments of the exact solution $f(x,p)$ of the RTA Boltzmann equation:
\begin{equation}
\label{I_exact}
    I^{\mathrm{exact}}_{nrq} = \frac{1}{(2q)!!} \int dP \, E_{\mathrm{LRF}}^{n-r-2q} \, p_{z,\mathrm{LRF}}^r \, p_{T, \mathrm{LRF}}^{2q} \, f.
\end{equation}
Note that with this definition $\epsilon{\,=\,}I^{\mathrm{exact}}_{200}$, $\PL{\,=\,}I^{\mathrm{exact}}_{220}$, and $\PT{\,=\,}I^{\mathrm{exact}}_{201}$.

Although Eqs.~(\ref{e_sec_aH})-(\ref{PT_sec_aH}) are exact, they do not form a closed set for the hydrodynamic moments $(\epsilon, \PL, \PT)$ because Eqs.~(\ref{zeta_long}), (\ref{zeta_perp}) couple them to the higher-order non-hydrodynamic moments $I^{\mathrm{exact}}_{nrq}$ of the exact solution for $f$. To truncate the resulting infinite hierarchy of coupled moment equations one usually approximates the non-hydrodynamic moments (\ref{I_exact}) by replacing the full solution $f$ under the integral in (\ref{I_exact}) by the leading-order anisotropic distribution $f_a$, with parameters matched to the hydrodynamic moments $\epsilon$, $\PT$ and $\PL$. This standard procedure will be modified here by substituting $f_a$ from Eq.~(\ref{f_a_postulate}) by a similarly matched $\tilde{f}_a$ from Eq.~(\ref{modifed_aHydro_ansatz}).\footnote{%
    The temperature appearing in $\tau_R$ and in the equilibrium pressure $P$ are defined, as usual, via Landau matching: $\epsilon = \epsilon_\mathrm{eq}(T)$.}
The resulting approximation for the moments $I^{\mathrm{exact}}_{nrq}$ in Eq.~(\ref{I_exact}) we denote by $\tilde{I}_{nrq}$.

\begin{figure}[b!]
\begin{center}
 \includegraphics[width=0.85\linewidth]{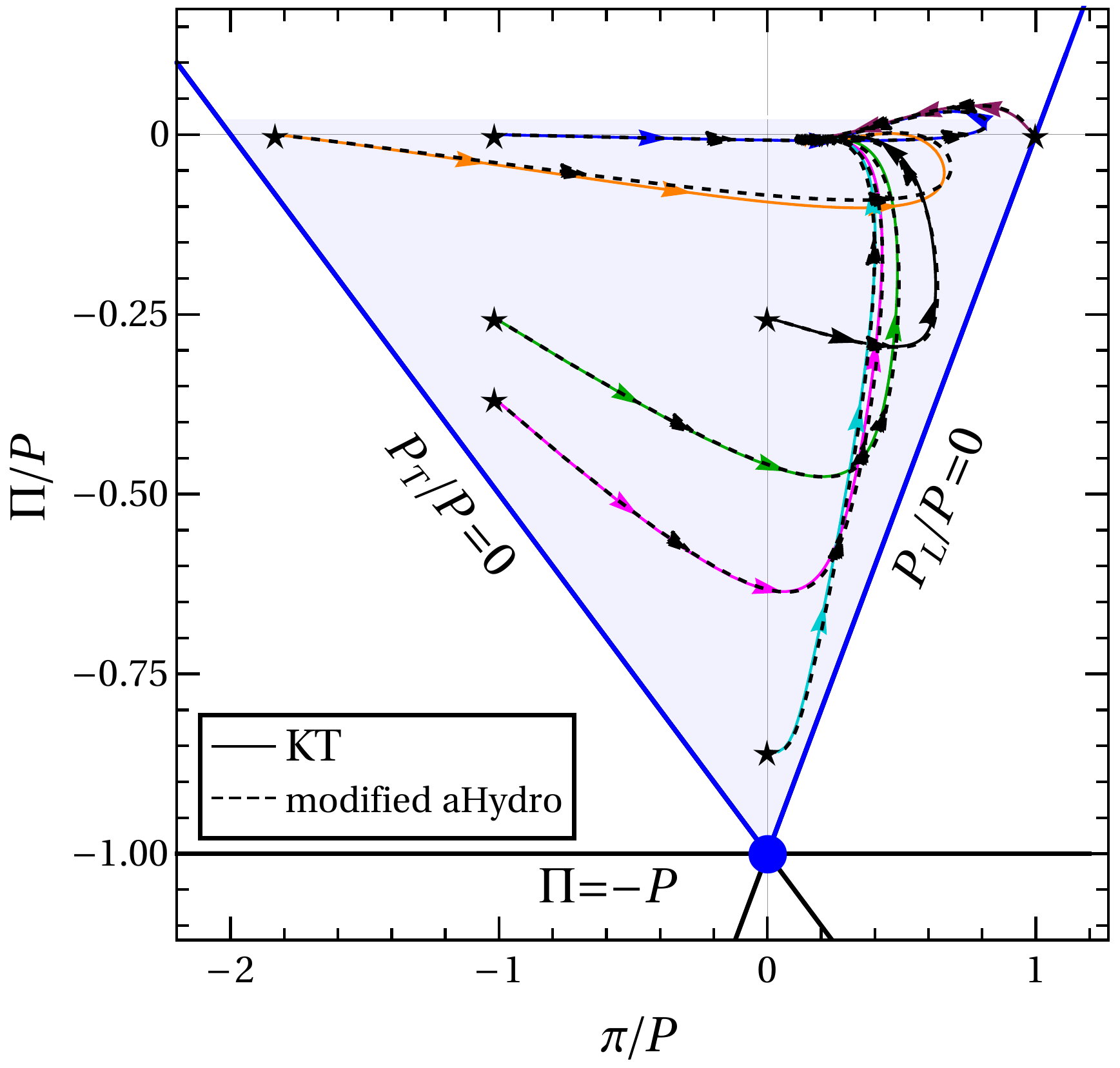}
\end{center}
\vspace*{-6mm}
  \caption{
  Selected evolution trajectories from kinetic theory (solid colored lines, copied from Fig.~\ref{CM_exact}) compared with those from anisotropic hydrodynamics (black dashed lines).
  \vspace*{-4mm}  }
  \label{Fig_kt_ahydro:1}
\end{figure} 

Note that the moments $\tilde{I}^{240}$ and $\tilde{I}^{201}$ are functions of $\alpha$, $\Lambda$, and $\xi$. To solve the above equations in terms of $(\epsilon, \PL, \PT)$ one therefore has to invert $(\alpha, \Lambda, \xi) \to (\epsilon, \PL, \PT)$ such that the moments $\tilde{I}_{nrq}$ are all expressed in terms of hydrodynamic moments. We sidestep the technical complications of this inversion procedure by switching variables from $X_a \equiv (\epsilon, \PL, \PT)$ to $x_a \equiv (\alpha,\Lambda, \xi)$. The differentials $dX_a$ can be expressed in terms of $dx_a$ as
\begin{equation}
    dX_a = M_{a}^{b} \, dx_b
\end{equation}
with the Jacobian $M$. Instead of the moments $\tilde{I}_{nrq}$ we switch to the moments used in \cite{McNelis:2018jho},
\begin{equation}
\label{Mike_moments}
    I_{nrqs} = \frac{1}{(2q)!!} \int dP \, E_{\mathrm{LRF}}^{n-r-2q} \, p_{z,\mathrm{LRF}}^r \, p_{T,\mathrm{LRF}}^{2q} \, E_a^s \, f_a,    
\end{equation}
where for Bjorken flow $E_{\mathrm{LRF}}{\,=\,}p^\tau$, $p_{z,\mathrm{LRF}}{\,=\,}w/\tau$, $p_{T,\mathrm{LRF}}=p_T$, and where $f_a = \exp(- E_a/\Lambda)$ with $E_a \equiv \sqrt{p_{T,\mathrm{LRF}}^2 + (1{+}\xi) p_{z,\mathrm{LRF}}^2 + m^2 }$. In terms of these moments $\epsilon{\,=\,}I_{2000}/\alpha$, $\PL = I_{2200}/\alpha$, and $\PT = I_{2010}/\alpha$. One finds
\begin{equation}
    M = \frac{1}{\alpha}
\begin{pmatrix}
  -I_{2000}/\alpha & I_{2001}/\Lambda^2 & -I_{420,-1}/(2\Lambda) \\
    - I_{2200}/\alpha & I_{2201}/\Lambda^2 &  - I_{440,-1}/(2\Lambda) \\
    -I_{2010}/\alpha & I_{2011}/\Lambda^2 & - I_{421,-1}/(2\Lambda) 
\end{pmatrix},
\end{equation}
from which it is straightforward to obtain the evolution equations for the parameters $(\alpha, \Lambda, \xi)$:
\begin{equation}
    \frac{dx_a}{d\tau} = (M^{-1})^{\ b}_{a}\, \frac{dX_b}{d\tau}.
\end{equation}
These ordinary differential equations are solved numerically using a fourth-order Runge-Kutta algorithm. Inserting the solutions $\bigl(\Lambda(\tau), \xi(\tau)\bigr)$ into $f_a$ in Eq.~(\ref{Mike_moments}) and evaluating $I_{2000}$, $I_{2200}$ and $I_{2010}$ by quadrature provides the time evolution of $\epsilon$, $\PL$ and $\PT$.

\begin{figure}[t!]
\begin{center}
\includegraphics[width=0.9\linewidth]{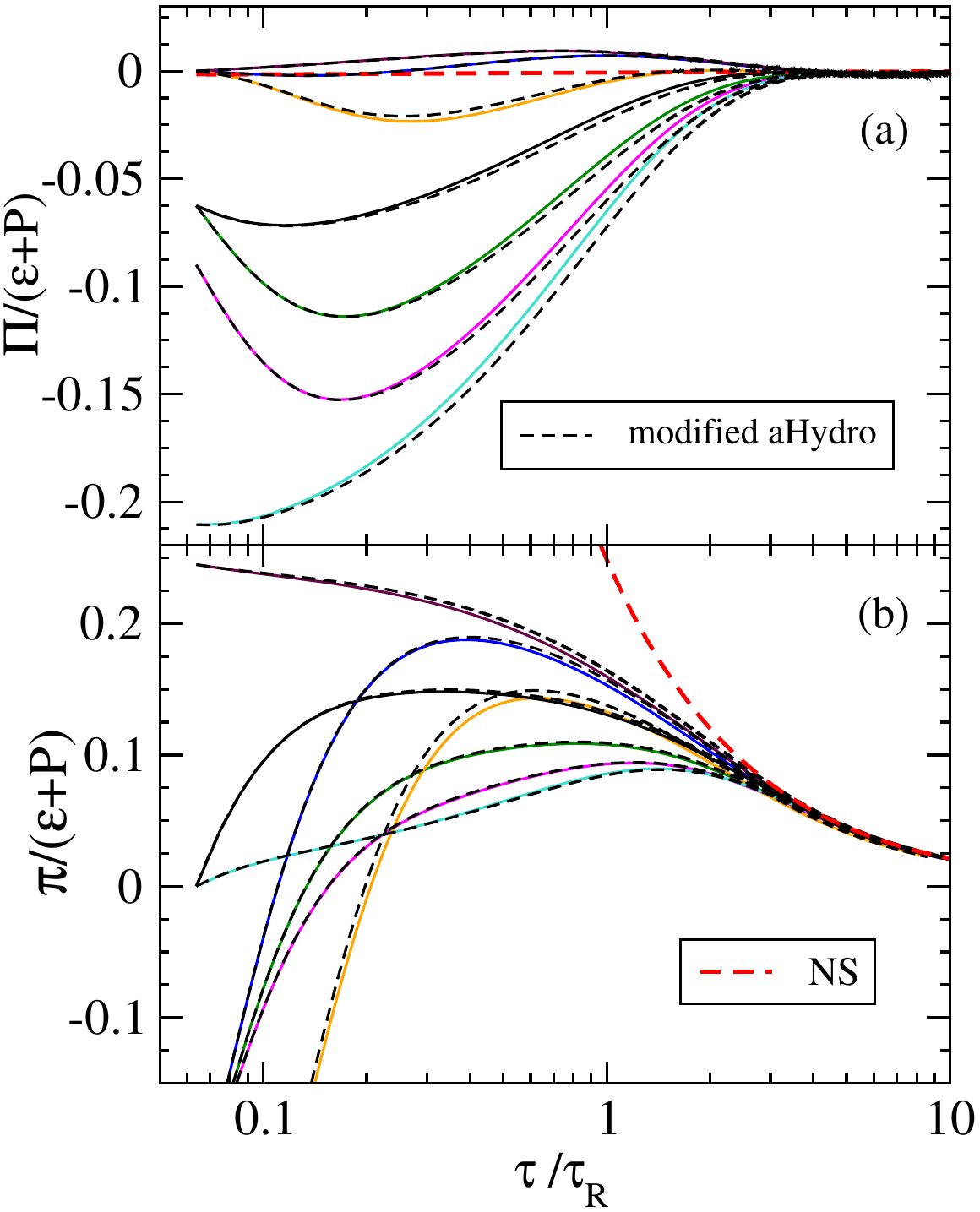}
\end{center}
\vspace*{-6mm}
\caption{
    Scaled time evolution of (a) $\Pi/(\epsilon{+}P)$ and (b) $\pi/(\epsilon{+}P)$. The colored curves showing exact solutions of the RTA Boltzmann equation are from Fig.~\ref{fig:bulk_shear}. Dashed black lines show modified anisotropic hydrodynamic evolution.}
  \vspace*{-2mm}  
\label{fig:bulk_shear_aHydro}
\end{figure} 

In Fig.~\ref{Fig_kt_ahydro:1} we compare the evolution of the normalized bulk and shear viscous stresses from the kinetic theory solutions (colored curves, copied from Fig.~\ref{CM_exact}) with those from the modified anisotropic hydrodynamic approximation just described (black dashed lines). The anisotropic hydrodynamic solutions are seen to be in excellent agreement with the exact RTA BE results. Different from the standard second-order hydrodynamic evolution studied in the preceding Section (see Fig.~\ref{Fig_kt_hydro:1}), all anisotropic hydrodynamic trajectories stay within the bounds allowed by kinetic theory, throughout their entire evolution. We checked and found that the parameter $\alpha(\tau)$ remains positive at all times such that the distribution function $\tilde{f}_a$ is never negative.\footnote{%
    Note that the out-of-equilibrium distribution function obtained using second-order CE hydrodynamics \cite{Jaiswal:2014isa} always turns negative (i.e. unphysical) at sufficiently large momentum, irrespective of the values of the shear and bulk viscous stresses.} 

Figure~\ref{fig:bulk_shear_aHydro} compares the anisotropic hydrodynamic evolution of the bulk (panel a) and shear (panel b) inverse Reynolds numbers with the exact solution from the underlying kinetic theory. The improvement of the macroscopic description from standard second-order viscous hydrodynamics shown in Fig.~\ref{Fig_kt_hydro:2} to modified anisotropic viscous hydrodynamics is stunning, especially at very early times and even for the largest possible negative bulk viscous pressures in the initial condition. The reason for this dramatic improvement is the choice of the leading-order distribution function around which we expand when deriving modified anisotropic hydrodynamics, which is custom-made to allow it to follow the rapid shrinking of the $p_z$ distribution caused by the rapid initial longitudinal expansion of Bjorken flow. The relatively largest deviations between solid and dashed lines are seen at intermediate times $\tau\sim\tau_R$ when the system exits the approximately free-streaming stage and begins to thermalize, eventually converging to the late-time first-order hydrodynamic Navier-Stokes attractor (red dashed line) at $\tau\gtrsim4\tau_R$.

\begin{figure}[t]
\begin{center}
 \includegraphics[width=0.9\linewidth]{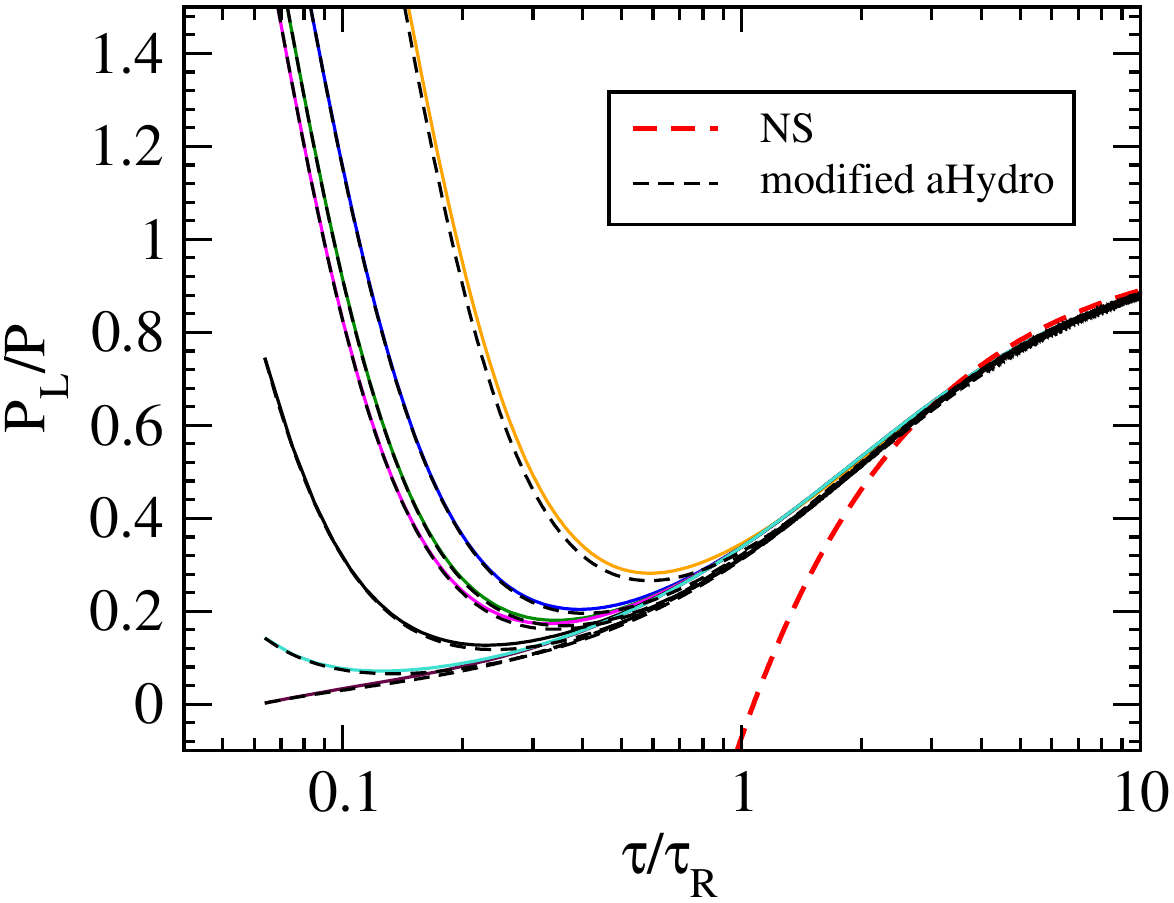}
\end{center}
\vspace*{-6mm}
  \caption{
  Scaled time evolution of $P_L/P$ from kinetic theory (colored lines, from Fig.~\ref{PL_P_KT}) compared with modified anisotropic hydrodynamics (black dashed lines).
  \vspace*{-2mm} }
  \label{fig:4_aHydro}
\end{figure} 

Given this good agreement of the modified anisotropic hydrodynamic evolution with the exact kinetic result for the bulk and shear stresses it is not surprising that this improved macroscopic theory is also able to quantitatively reproduce the universal attractor behavior of the scaled longitudinal pressure seen in Fig.~\ref{PL_P_KT}. This is shown in Fig.~\ref{fig:4_aHydro}. The anisotropic hydrodynamic solutions nicely reproduce the rapid convergence of solutions with arbitrary initial conditions onto an early-time attractor (the lowermost black dashed line) at $\tau/\tau_R < 1$, much before they merge with the late-time Navier-Stokes attractor at $\tau/\tau_R\gtrsim 3$. Most importantly, the anisotropic hydrodynamic attractor is in near-perfect agreement with the kinetic theory one, and so are the decay rates for different initial deviations from the attractor. 

The work presented in this Section suggests that an inclusion of shear and bulk inverse Reynolds numbers to all-orders is necessary in a macroscopic theory for it to accurately capture the universal behavior of $P_L$ predicted by a microscopic kinetic description of a weakly-coupled non-conformal gas. We close this section with the following remark: As discussed, our main reason for introducing the modified ansatz (\ref{modifed_aHydro_ansatz}) for the leading-order distribution when deriving anisotropic hydrodynamics was to repair the inability (see Appendix~\ref{RS_appendix}) of the Romatschke-Strickland ansatz (\ref{f_a_postulate}) to simultaneously describe systems with large negative bulk viscous pressure and small $m/T$. Many other choices for $\tilde{f}_a$ may achieve the same goal. Recent work offers a hope for solving this ambiguity in a systematic fashion, by choosing the so-called `maximum entropy' distribution \cite{Everett:2021ulz} as the leading-order distribution when deriving anisotropic hydrodynamics. Its functional form (given in Eq.~(\ref{ME_appendix}) in Appendix~\ref{Initial paramters}) represents the least biased choice (one that maximizes entropy) for a distribution function that can be constructed using only the information given by hydrodynamic energy-momentum tensor (i.e., in our case, the energy density and the shear and bulk viscous stresses). We leave the exploration of this idea to future work.

\section{Conclusions}
\label{sec:conclusion}
%
In this work we performed a detailed analysis of the non-conformal dynamics of a gas of massive Boltzmann particles undergoing boost-invariant Bjorken expansion, using kinetic theory based on the RTA Boltzmann equation, second-order Chapman-Enskog hydrodynamics, and anisotropic hydrodynamics. 
In all three approaches we found that neither the bulk nor the shear viscous stress exhibit early-time attractor behavior. Instead, we showed that in kinetic theory the scaled effective longitudinal pressure, $\PL/P{\,=\,}1+(\Pi{-}\pi)/P$, exhibits universal early-time behavior where solutions $(\PL/P)(\tau/\tau_R)$ with different initial conditions converge within $\tau/\tau_R < 1$, i.e. before microscopic collisions become significant, onto a universal attractor. This attractor starts from $\PL\approx0$ at $\tau\approx0$ and smoothly joins the late-time Navier-Stokes solution at $\tau/\tau_R \gtrsim 3$ when thermalizing dynamics gains control over free-streaming expansion. The origin of this early-time, far-off-equilibrium attractor for $\PL/P$ can be attributed to the existence of a universal attracting fixed line at $\PL = 0$ in any free-streaming gas affected by longitudinal expansion. For a weakly coupled gas undergoing Bjorken flow, where the longitudinal expansion rate diverges like $1/\tau$ as $\tau\to0$ and thus dominates over the microscopic scattering rate $1/\tau_R$ until $\tau \sim \cal{O}(\tau_R)$, this free-streaming dynamics dominates any collision-induced thermalizing effects at early times $\tau/\tau_R\ll1$, moving the system rapidly towards this $\PL=0$ fixed line. For other types of flows with a finite expansion rate at $\tau\to0$, or for strongly coupled systems that cannot be described by the RTA Boltzmann equation, a similar early-time dominance of free-streaming dynamics over thermalization is not expected, and this early-time, far-off-equilibrium attractor is likely lost. 

While during the early free-streaming-dominated stage the strong longitudinal expansion rapidly shrinks the width of the longitudinal momentum distribution in the comoving frame, moving the system towards zero longitudinal pressure, this does not entail any universal consequences for the shear and bulk viscous stresses which depend on other characteristics of the LRF momentum distribution. Hence $\pi$ and $\Pi$ separately do not exhibit any universal early-time far-off-equilibrium attractive behaviour --- only their scaled difference $(\Pi{-}\pi)/P$, which is equivalent to $\PL/P$, does. In the conformal limit where $\Pi\equiv0$, the shear stress $\pi$ and longitudinal pressure $\PL$ become interchangeable, and the far-off-equilibrium early-time attractor for $\PL$ manifests itself as a similar attractor for $\pi/P$ and for the shear inverse Reynolds number Re$_{\pi}^{-1}=\big(1{-}\PL/P\big)/4$, as observed in many prior works.

Studying these phenomena at the macroscopic level with second-order Chapman-Enskog hydrodynamics, which is based on a perturbative gradient expansion around local thermal equilibrium, we found that it does not correctly reproduce the early-time evolution of the bulk and shear viscous stresses predicted by the underlying kinetic theory, nor does it recover the universal early-time attractor for the longitudinal pressure $\PL$. The reason for this failure can be traced back to the inability of this hydrodynamic scheme to correctly reproduce the free-streaming fixed lines and fixed points of kinetic theory, including their attractive, repulsive or saddle point properties. All three early-time hydrodynamic fixed points of second-order CE-hydrodynamics were found to lie outside the region in the $\pi{-}\Pi$ plane allowed by the underlying kinetic theory. Universality only emerges late in this theory, around $\tau \gtrsim 3 \tau_R$, when the viscous dynamics enters a stage governed by the first-order Navier-Stokes equations. 

A more efficient macroscopic model of the kinetic theory results is provided by anisotropic hydrodynamics. Unlike standard second-order viscous fluid dynamics, this theory involves transport coefficients that include contributions at all orders in the inverse bulk and shear Reynolds numbers. Anisotropic hydrodynamics generalizes the standard perturbative gradient expansion to one around an ellipsoidally deformed local momentum distribution. For our purposes here we had to modify this momentum-anisotropic leading-order distribution from its usual Romatschke-Strickland form to be able to accommodate large bulk viscous pressures. This modified anisotropic hydrodynamic framework was found to reproduce the macroscopic properties of the RTA Boltzmann solutions, including its early-time far-off-equilibrium universal attractor for the scaled longitudinal pressure, with excellent accuracy.

Several questions remain: (i) The present work suggests that different expansion geometries may lead to universal early-time attractive behavior in different physical observables. An interesting case to study this hypothesis might be a Friedman-Lema\^{i}tre-Robertson-Walker cosmology with a diverging initial scalar expansion rate \cite{Bazow:2016oky, Du:2021fok}. This system is spatially homogeneous and undergoes isotropic expansion in three dimensions, with vanishing shear stress. We expect that in this profile, for weakly-coupled systems that can be described by the Boltzmann equation, the scaled effective {\it radial} pressure (which in this case is interchangeable with the scaled bulk viscous pressure) would show early-time attractor behavior.  
(ii) The Bjorken flow studied in this work is a well-motivated approximation for the early-stage flow pattern of the hot and dense matter created in ultra-relativistic nucleus-nucleus collisions. If the strong 1-dimensional longitudinal expansion during the pre-hydrodynamic stage leads to the emergence of a free-streaming attractor, rapid far-off-equilibrium convergence of expansion trajectories towards this attractor could generate a certain degree of universality in the initial conditions for the subsequent 3-dimensional hydrodynamic expansion phase. It would be interesting to explore to what extent such a uniformity in initial conditions manifests itself through universal features in final-state observables, even in the absence of a (3+1)-dimensional far-off-equilibrium hydrodynamic attractor.   
(iii) Finally, the analysis in this work assumes that the fluid is composed of a weakly coupled massive gas whose dynamics is well described by the Boltzmann equation. Whether the expanding matter formed in ultra-relativistic nucleus-nucleus collisions before the quark-gluon plasma stage is sufficiently weakly coupled so as to warrant the use of the Boltzmann equation requires additional study. We leave these exciting questions to future research.   

\section*{Acknowledgements}
%
The authors thank A. Jaiswal for his involvement during the initial stages of this project. We gratefully acknowledge helpful discussions with J. Noronha, M. Heller, M. Strickland, J. P. Blaizot, K. Ingles, M. McNelis, D. Liyanage, and D. Everett. C.C., L.D. and U.H. were supported by the U.S. Department of Energy (DOE), Office of Science, Office for Nuclear Physics under Award No.~\rm{DE-SC0004286}. S.J. and S.P. acknowledge financial support by the Department of Atomic Energy (Government of India) under Project Identification No. RTI 4002. \\

\appendix

\section{Initial parameters} 
\label{Initial paramters}

In Table \ref{table_ic_appendix} we list the values of parameters $(\Lambda_0, \alpha_0, \xi_0)$ used in the initial distribution function \eqref{f_in},
\begin{equation}
\label{f_in_appendix}
    f_\mathrm{in}(\tau_0; p_T, w) = \frac{1}{\alpha_0} \exp\left(- \frac{\sqrt{p_T^2  + (1+\xi_0) w^2/\tau_0^2 + m^2 }}{\Lambda_0} \right),
\end{equation}
to generate the initial conditions for scaled bulk and shear stresses summarized in Table \ref{table:IC}. As mentioned in the main text, the mass is taken to be $m = 200$ MeV, and the initial temperature has been held fixed at $T_0 = 500$ MeV.

\begin{widetext}

\begin{table}[h!]
 \begin{center}
	\resizebox{0.75\columnwidth}{!}{
	\begin{tabular}{|c|c|c|c|c|c|c|c|}
   		\hline
   		&  Blue &  Green  &  Magenta &  Maroon &  Orange &  Black &  Cyan \\
   		\hline
        $m/\Lambda_0$ & 0.616 & 4.808 & 10.89 & 0.294 & 1.818 & 2.023 & 20  \\
    	\hline
    	$\alpha_0$ & 0.655 & $4\times 10^{-5}$ & $2.5\times 10^{-8}$ & 0.078 & 0.0632 & $1.06\times 10^{-3}$ & $1.48\times 10^{-13}$ \\
    	\hline
    	$\xi_0$ &  -0.832 & -0.908 & -0.949 & 1208.05 & -0.987 & 0 & 0 \\
   		\hline  \hline
   		$e^{-(m/\Lambda_0)}/\alpha_0$ & 0.82 & 199.13 & 736.9 & 9.55 & 2.57 & 124.2 & 13945.7 \\
   		\hline  \hline
   		$e^{-(m\, \Lambda)}|_{\rm ME}$ & 1.37 & 821.05 & 2385.89 & 77.34 & 15 & 495.38 & 15450.4 \\
   		\hline
  	\end{tabular}
  	}
  \caption{Values of initial parameters $\Lambda_0, \alpha_0, \xi_0$ considered for different colors of the curves given in Table \ref{table:IC}. The last two rows denote number of particles at zero momenta for the ansatz (\ref{f_in_appendix}) and (\ref{ME_appendix}), respectively.}
  \label{table_ic_appendix}
 \end{center}
 \vspace*{-.6cm}
\end{table}
\end{widetext}

%
\begin{figure*}[bht!]
\begin{center}
 \includegraphics[width=0.9\linewidth]{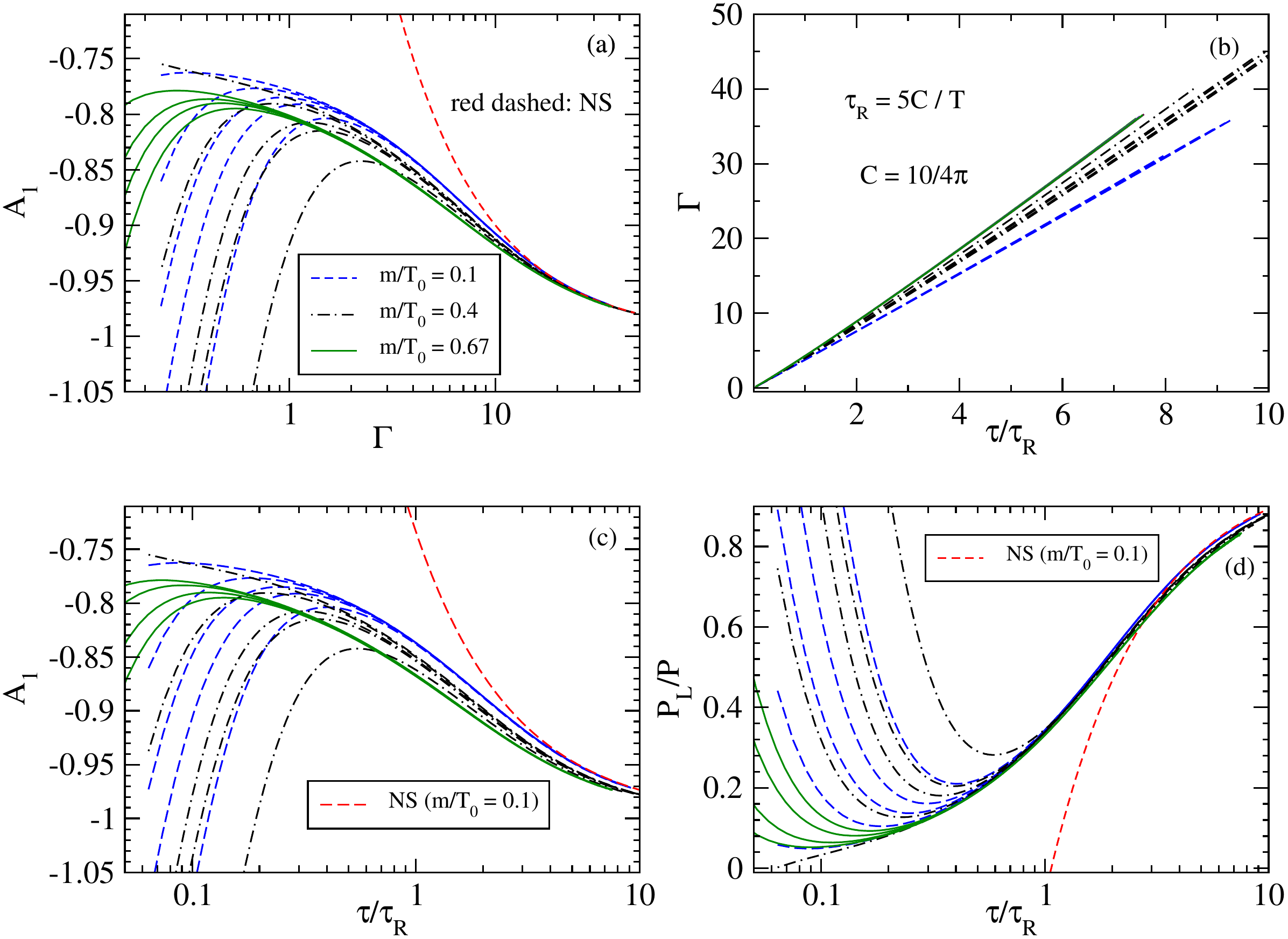}
\end{center}
\vspace*{-6mm}
  \caption{
  {Evolution of $A_1 \equiv -1 + Re_{\mathrm{\pi}}^{-1} - Re_{\mathrm{\Pi}}^{-1}$ as a function of gradient strength $\Gamma$ and scaled time in panels (a) and (c) respectively. Evolution of $\Gamma$ and scaled effective longitudinal pressure versus $\tau/\tau_R$ in panels (b) and (d) respectively. Different colors indicate solutions with different ratio's of $m/T_0$. The red dashed curve in panel (a) corresponds to the Navier-Stokes value, $A_{1,NS} = -1 + 1/\Gamma$, whereas those in panels (c) and (d) are obtained using first-order hydro with $m = 50$ MeV and $T_0 = 500$ MeV.}
  \vspace*{-2mm}}%
  \label{fig:A1_appendix}
\end{figure*}%
%

The second-last row of Table \ref{table_ic_appendix} denotes the initial mean occupation density at zero momenta, $f(\tau_0; 0, 0) = \exp(-m/\Lambda_0)/\alpha_0$. For most of the curves (black, green, magenta, and cyan) the initial particle number at low momentum regions of phase space are seen to be enhanced (the enhancement ranging from ${\cal O}(10^2) - {\cal O}(10^5)$), such that the distribution ansatz (\ref{f_in_appendix}) generates the desired scaled bulk and shear stresses. The presence of large numbers of particles at small momenta is reminiscent of a Bose condensate and invalidates the use of Boltzmann statistics, at least during the early evolution stage (small $m/T$) near the lower corner of the allowed region in Fig.~\ref{CM_fs}. In this work we continue, however, to use Boltzmann statistics even in regions of high phase-space occupancy, to enable comparison of kinetic theory results with second-order hydrodynamics whose transport coefficients have been obtained assuming that the fluid is microscopically constituted of a Boltzmann gas. 

This feature is not unique to our ansatz (\ref{f_in_appendix}) for the initial distribution. The same enhancement at small momenta appears when one generates initial conditions for $(\Pi/P, \pi/P)$ from Table~\ref{table:IC} using the so-called `maximum-entropy' distribution for Boltzmann statistics~\cite{Everett:2021ulz}:
\begin{align}
\label{ME_appendix}
    f_{\rm ME}= \exp\Bigl[-\Big(\Lambda\, p^2_0 + \lambda_\Pi p^2 + \gamma\bigl(\textstyle{\frac{1}{2}}p_T^2{+}p_z^2\bigr) \Bigr)\Big/p_0 \Bigr].
\end{align}
As mentioned at the end of Sec.~\ref{aHydro}, the maximum entropy distribution function is the minimally biased phase-space distribution that can be constructed using only the information contained in conserved hydrodynamic charge currents. Eq.~\eqref{ME_appendix} is obtained by expressing the original definition of the maximum-entropy distribution function \cite{Everett:2021ulz} in the local rest frame for Bjorken flow. By computing the Lagrange parameters $(\Lambda, \lambda_\Pi, \gamma)$ that yield the required initial $(T, \Pi/P, \pi/P)$, we show in the last row of Table \ref{table_ic_appendix} that for all the curves, the initial enhancement of particles at zero momenta, $f_{\rm ME}(\tau_0; 0, 0) = \exp(-\Lambda \, m)$, is in fact slightly larger than what is obtained using the ansatz (\ref{f_in_appendix}). This additional enhancement arises because the parameter $\Lambda$ is negative for all of the initial conditions for bulk and shear stresses considered here. This is not a problem since an exponential fall-off of $f_{\rm ME}$ at large momenta is guaranteed as long as $\Lambda + \lambda_\Pi > |\mathrm{min}(\gamma/2, \gamma)|$ \cite{Everett:2021ulz}; this was always found to be the case here. 

\section{Early- and late-time attractors}
\label{A1_Gamma_appendix}

In this Appendix we contrast features of the attractor demonstrated in this paper, namely, in the quantity $P_L/P$ vs $\tau/\tau_R$, to the attractor found by Romatschke in \cite{Romatschke:2017acs}, i.e., in the quantity $A_1 \equiv -1 + \pi/(\epsilon+P) - \Pi/(\epsilon+P)$ vs the `inverse gradient strength' $\Gamma \equiv \tau/\gamma_s$, where $\gamma_s \equiv (4\eta/3 + \zeta)/(\epsilon+P)$. Note that when a Bjorken system is close to its Navier-Stokes limit, $\pi_{NS} = 4\eta/3\tau$ and $\Pi_{NS} = - \zeta/\tau$, the quantity $A_1$ approaches $A_{1,{\rm NS} } = -1 + 1/\Gamma$. Hence, $A_1$ plotted as a function of $\Gamma$ is expected to show universality, at least at late times. However, at early times the effective longitudinal pressure $\PL$ rapidly approaches 0 such that the quantity $A_1 = - (\epsilon + \PL)/(\epsilon+P) \approx - \epsilon/(\epsilon + P)$. Thus, the early-time dynamics of $A_1$ is not universal (unlike that of $\PL/P$) and depends on the ratio $m/T$. It is thus reasonable to expect that the quantity $A_1$ will \textit{not} exhibit early-time universality once we choose initial conditions that give rise to substantially different evolutions of $m/T$.

This is, in fact, observed in Fig. \ref{fig:A1_appendix}a where we plot $A_1$ vs $\Gamma$ for different ratios of particle mass $m$ to the initial temperature $T_0$. The blue dashed and black dashed dotted curves both correspond to same initial temperature $T_0 = 500$ MeV, but with masses $m = 50$ MeV and $200$ MeV respectively. The green curves are obtained using $T_0 = 300$ MeV and $m = 200$ MeV. The green, blue, and black solutions are found to converge to three different curves, which in turn merges with each other around $\Gamma \approx 20$. The green and blue solutions converge to their respective `universal' curves at large gradient strength ($\Gamma \lesssim 1$). However, the black dashed dotted curves merge with each other at much smaller gradient strengths of $\Gamma \approx 10$. These features can be understood from the following: As mentioned above, early-time universality in $A_1$ is disrupted by substantial differences in $m/T$ evolution. However, for the green curves the temperature evolutions are near identical as their initial $\PL$'s are close to each other.
Although for the blue curves the temperature evolutions are different, the ratio $m/T$ is small enough to make the early time limit of $A_1$ become near universal $\approx -4/3$ (i.e., independent of $m/T$). For the black curves the evolution of $m/T$ is rather different leading to delayed convergence with one another. We emphasize that the blue, black, and green solutions show crossings at small $\Gamma$ and merge to a universal curve only when the system's dynamics is essentially described by Navier-Stokes equations (red dashed curve).

We plot $\Gamma$ vs $\tau/\tau_R$ in Fig. \ref{fig:A1_appendix}b to explore how the inverse gradient strength varies as a function of the scaled time corresponding to the different solutions shown in panel (a). Owing to complicated $m/T$ dependence of $\eta/s$ and $\zeta/s$, the quantity $\Gamma$ is not simply a function of $\tau/\tau_R = \tau T/(5C)$; hence the splittings in panel (b). Moreover, $\Gamma$ is seen to increase with increasing $m/T$. This is because for small $m/T$, $\zeta/s \approx 0$ such that $\Gamma \propto 1/(\eta/s)$, and $\eta/s$ decreases with increasing $m/T$ (see Fig. \ref{Fig_etas_zetas}). For the blue dashed curves, $m/T$ is small. As a result, $\Gamma \approx 3 \tau T/(4C) = (15/4) \tau/\tau_R$. Whereas all the green lines for $\Gamma$ are almost on top of each other because of their similar temperature evolutions, the black dashed dotted lines show substantial $m/T$ induced splitting.

In panel (c) we plot the evolution of $A_1$ as a function of scaled proper time. The universality seen in panel (a) at $\Gamma \approx 20$ is disrupted while using the variable $\tau/\tau_R$ because a fixed $\Gamma$ does not correspond to a unique $\tau/\tau_R$, but instead, depends on the temperature evolution, as manifested in the splittings shown in panel (b). Finally, in panel (d) we plot the  evolution of $\PL/P$. In contrast to $A_1$ evolution, the scaled effective longitudinal pressure continues showing universal behaviour at $\tau/\tau_R < 1$ for different choices of $m/T_0$. Interestingly, the crossing of curves of different colours as seen in panels (a) and (c) have disappeared in panel (d). However, small splittings appear at late-times because the NS limit of $P_L/P$ is not merely a function of $\tau/\tau_R$ but also of $m/T$.

\section{Scaled bulk viscous pressure using the Romatschke-Strickland ansatz}
\label{RS_appendix}

\begin{figure}[b!]
\begin{center}
\includegraphics[width=0.8\linewidth]{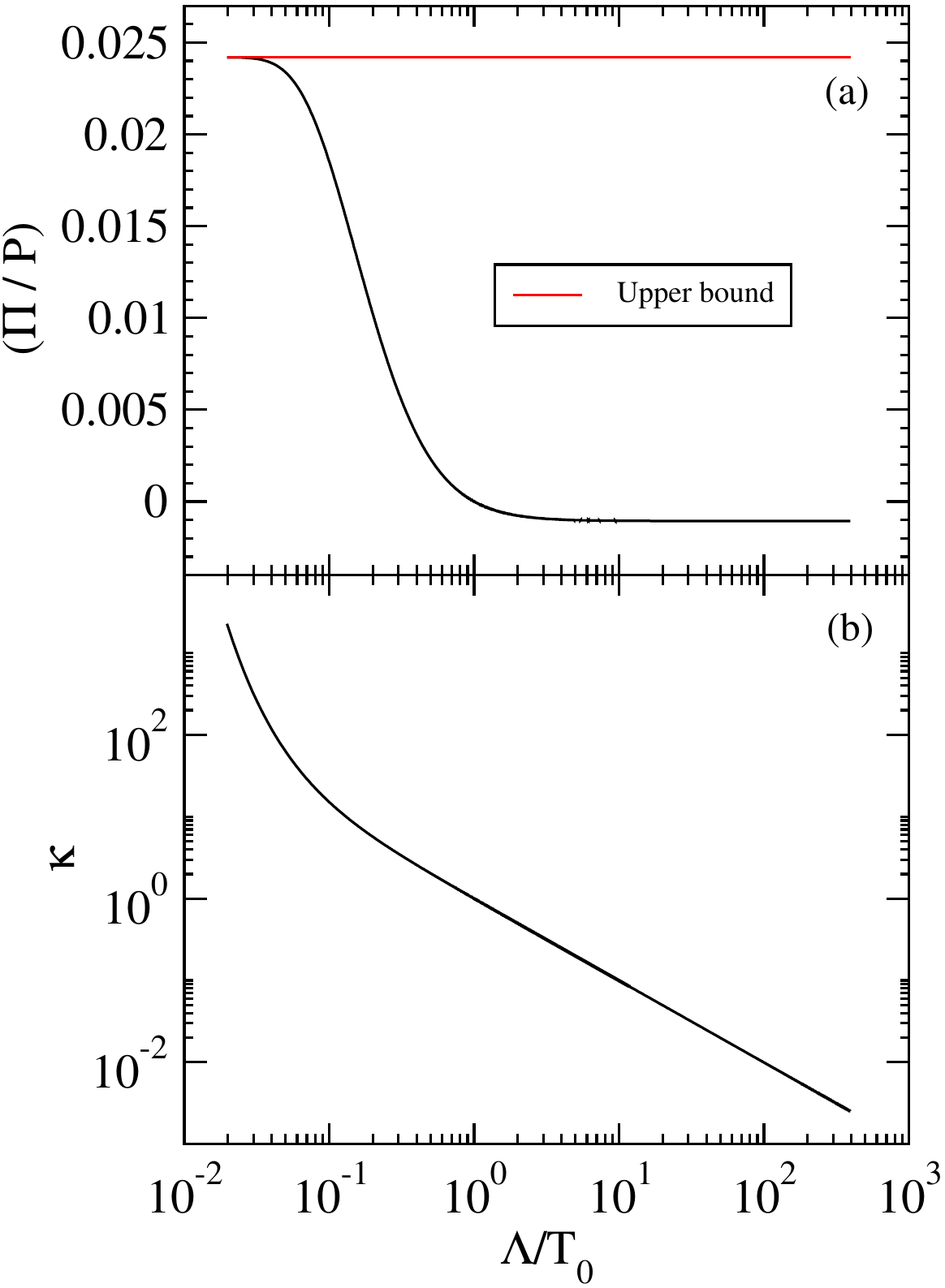}
\end{center}
\vspace*{-6mm}
\caption{
    Variation of (a) scaled bulk viscous pressure $(\Pi/P)$ and (b) parameter $\kappa$ in Eq. \eqref{f_a2_appendix} with respect to $\Lambda/T_0$. The solid red line in panel (a) denotes the maximum kinetically allowed $(\Pi/P)$ for $m/T = 0.4$.}
  \vspace*{-2mm}  
\label{bulk_RS_appendix}
\end{figure} 

Let us generate initial conditions having finite bulk viscous pressure, vanishing shear stress, and fixed temperature $T_0 = 500$ MeV using the Romatschke-Strickland distribution:
\begin{equation}
\label{f_a_appendix}
    f_a = \exp \left( - \frac{\sqrt{p_T^2/\alpha_T^2{+} p_z^2/\alpha_L^2{+}m^2}}{\Lambda} \right),
\end{equation}
with mass $m = 200$ MeV as considered in the main text of the paper. Without the requirement of momentum space anisotropy, it suffices to set $\alpha_L = \alpha_T \equiv \kappa$, thus reducing the number of parameters in the RS ansatz from three to two, $(\Lambda,\kappa)$:
\begin{equation}
\label{f_a2_appendix}
    f_a = \exp \left( - \frac{\sqrt{{\bf p}^2/\kappa^2 {+}m^2}}{\Lambda} \right),
\end{equation}
We choose a wide range of values for the parameter $\Lambda$ in units of the initial temperature such that $\Lambda/T_0 \in (0.02, 10^3)$. For each value of $\Lambda$ we solve for $\kappa$ that yields a temperature of $500$ MeV. Using these sets of $(\Lambda, \kappa)$ we compute the scaled bulk viscous pressure $\Pi/P$, which is plotted in Fig. \ref{bulk_RS_appendix}a, as a function of $\Lambda/T_0$. Clearly, the scaled bulk is seen to asymptotically approach a small constant negative value as $\Lambda/T_0$ becomes large. Fig. \ref{bulk_RS_appendix}b shows that with increasing $\Lambda/T_0$ the parameter $\kappa$ rapidly approaches zero with a constant slope at large $\Lambda$ depicting power-law decay (note both axes use logarithmic spacing). Fig. \ref{bulk_RS_appendix} indicates that the minimum scaled bulk is generated in the limit $\kappa \to 0$, $\Lambda \to \infty$ such that 
\begin{equation}
\label{f_approx_appendix}
    f_a \approx \exp \left( - \frac{|{\bf p}|}{\Lambda'} \right),
\end{equation}
where $\Lambda' \equiv \kappa \, \Lambda$ sets the scale of momentum in the distribution. We have checked that $\Lambda'$ is finite and slightly less than the temperature: $\Lambda' \approx 497$ MeV. 

To get an analytical estimate of how negative $\Pi/P$ can become, we compute the total isotropic pressure using \eqref{f_approx_appendix}:
\begin{equation}
\label{PplusPI_appendix}
    P + \Pi = \frac{1}{3} \int \frac{d^3p}{(2\pi)^3 \, E_p} \, |{\bf p}|^2 \, \exp \left( - \frac{|{\bf p}|}{\Lambda'} \right).
\end{equation}
Expanding the above equation in powers of $m/\Lambda'$ and also noting that for $z \equiv m/T \ll 1$,
\begin{equation}
    P \approx \frac{T^4}{\pi^2} \left( 1 - \frac{z^2}{4} \right),
\end{equation}
we obtain,
\begin{equation}
\label{PplusPi_appendix2}
    1 + \frac{\Pi}{P} \approx \frac{\Lambda'^4}{T^4} \left( 1 + \frac{z^2}{4} - \frac{z^2}{12} \,  \frac{T^2}{\Lambda'^2} \right).
\end{equation}
In order to compute $\Lambda'/T$ we use the Landau matching condition,
\begin{equation}
    \epsilon_{\mathrm{eq}}(T,m) =  \int \frac{d^3p}{(2\pi)^3} \, E_p \, \exp \left( - \frac{|{\bf p}|}{\Lambda'} \right) 
\end{equation}
Expanding the l.h.s. and r.h.s. of the above equation in powers of $m/T$ and $m/\Lambda'$, respectively, we obtain,
\begin{equation}
    \frac{3T^4}{\pi^2} \left( 1 - \frac{z^2}{12} \right) \approx \frac{3 \Lambda'^4}{\pi^2} \left( 1 + \frac{1}{12} \, \frac{m^2}{\Lambda'^2} \right),
\end{equation}
which yields the desired ratio of $\Lambda'$ and T to be,
\begin{equation}
\label{Lambda_T_appendix}
 \frac{\Lambda'}{T} \approx 1 - \frac{z^2}{24}.
\end{equation}
Using $z = m/T = 0.4$, we get $\Lambda' \approx 497$ MeV, in perfect agreement with the numerical result mentioned above. Inserting Eq.~\eqref{Lambda_T_appendix} into Eq.~\eqref{PplusPi_appendix2} we find that the ${\cal O}(z^2)$ terms cancel each other and $\Pi/P$ vanishes. Thus, for small $m/T$, the largest magnitude of a negative $\Pi/P$ that can be generated by the Romatschke-Strickland ansatz (at least for isotropic momentum distributions) is smaller than ${\cal O}(z^2)$. 

\bibliography{reference}

\end{document}